\documentclass[pdflatex,sn-aps]{sn-jnl}% American Physical Society (APS) Reference Style

\usepackage[english]{babel}
\usepackage{csquotes}

\usepackage{graphicx} % Required for inserting images
\usepackage{indentfirst}% Allows indents after section heading
\usepackage{multirow}%
\usepackage{amsmath,amssymb,amsfonts}%
\usepackage{amsthm}%
\usepackage{mathrsfs}%
\usepackage[title]{appendix}%
\usepackage{xcolor}%
\usepackage{textcomp}%
\usepackage{manyfoot}%
\usepackage{booktabs}%
\usepackage{algorithm}%
\usepackage{algorithmicx}%
\usepackage{algpseudocode}%
\usepackage{listings}%
\usepackage{upgreek} % lettere greche in roman
\usepackage{xspace} % comando \xpace
\usepackage{caption}
\usepackage{subcaption} % per \subcaptionbox
\usepackage{siunitx} % nice unit usage with \SI{3}{\micro\meter}
\usepackage{paralist} % per avere compactitem e compactenum
\usepackage{hyperref} % Making nice hyperlinks
\usepackage{soul}

\providecommand*{\un}[1]{\ensuremath{\mathrm{\,#1}}} 
\providecommand{\rwell}[0]{\mbox{$\upmu$-RWELL}\xspace}
\providecommand{\rwellPCB}[0]{\mbox{$\upmu$-RWELL\_PCB}\xspace}

\begin{document}

\title{The IDEA detector concept for FCC-ee}
\author{The IDEA Study Group\footnote{Full list of authors is shown at the end of the paper}}
\date{February 28, 2025}

%\linenumbers

\abstract{

\indent A detector concept, named IDEA, optimized for the physics and running conditions at the FCC-ee is presented. After discussing the expected running conditions and the main physics drivers, a detailed description of the individual sub-detectors is given. These include: a very light tracking system with a powerful vertex detector inside a large drift chamber surrounded by a silicon wrapper, a high resolution dual readout crystal electromagnetic calorimeter, an HTS based superconducting solenoid, a dual readout fiber calorimeter and three layers of muon chambers embedded in the magnet flux return yoke. Some examples of the expected detector performance, based on fast and full simulation, are also given. }

\maketitle

\setlength{\parindent}{15pt}
\section{Introduction}

\indent The FCC-ee is a proposed e$^+$e$^-$ collider that would give access to a large variety of unique physics measurements. In the following a full detector concept, named IDEA, is described. It attempts to provide optimal performance on most of the physics accessible, while dealing with the constraints required by the accelerator. 

In section \ref{Req} the physics drivers on the detector design are reviewed after discussing the specific features of the accelerator. A detector overview is given in section \ref{Over} followed by a  description of all subdetectors and their expected performance.

\begin{figure}[!htbp]
    \centering
    \includegraphics[width=0.99\linewidth]{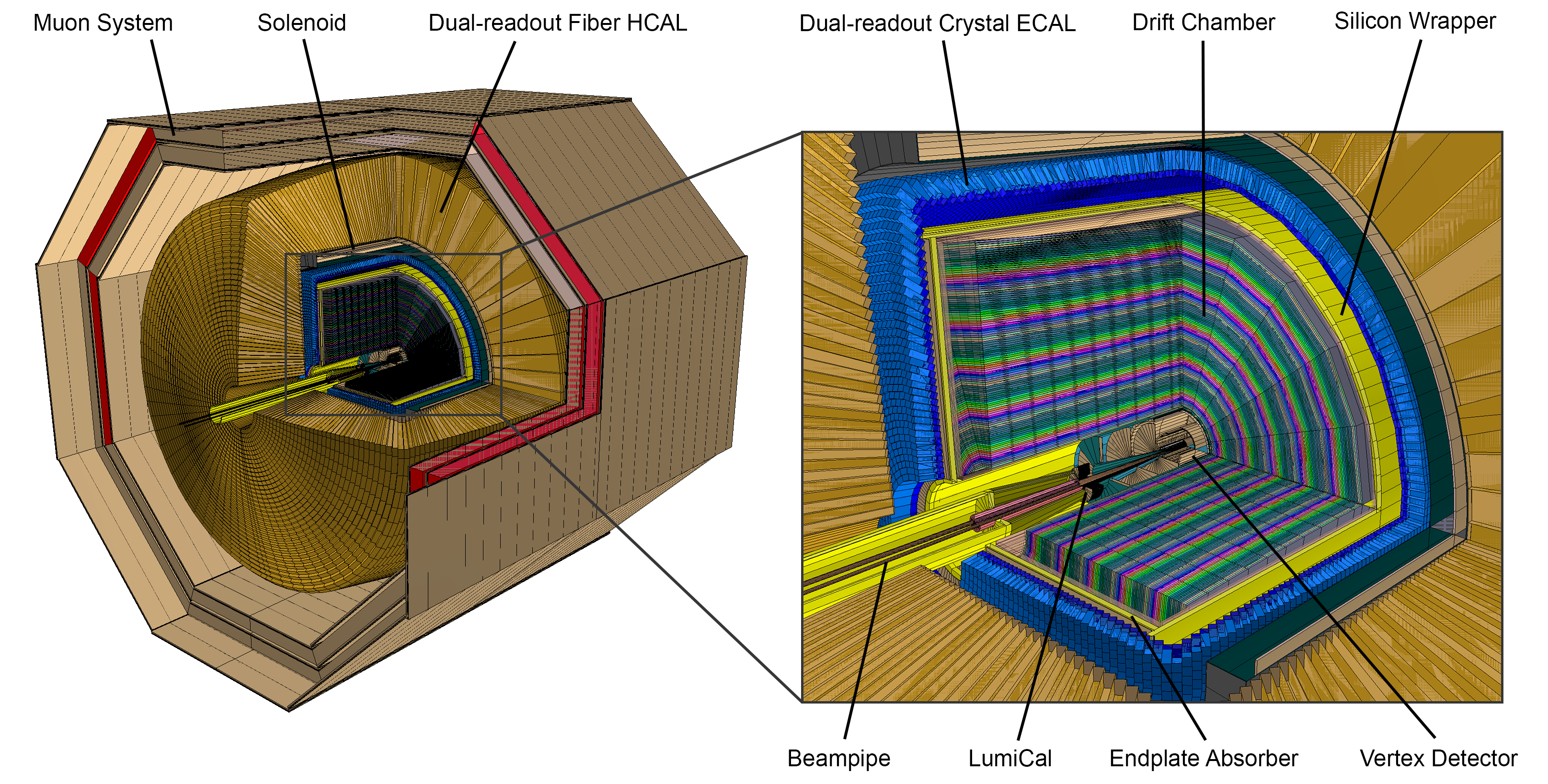}
    \caption{3D cutout view of the IDEA baseline detector design with subdetector labels. }
    \label{fig:idea_full}
\end{figure}
\begin{figure}[!htbp]
    \centering
    \includegraphics[width=0.99\linewidth]{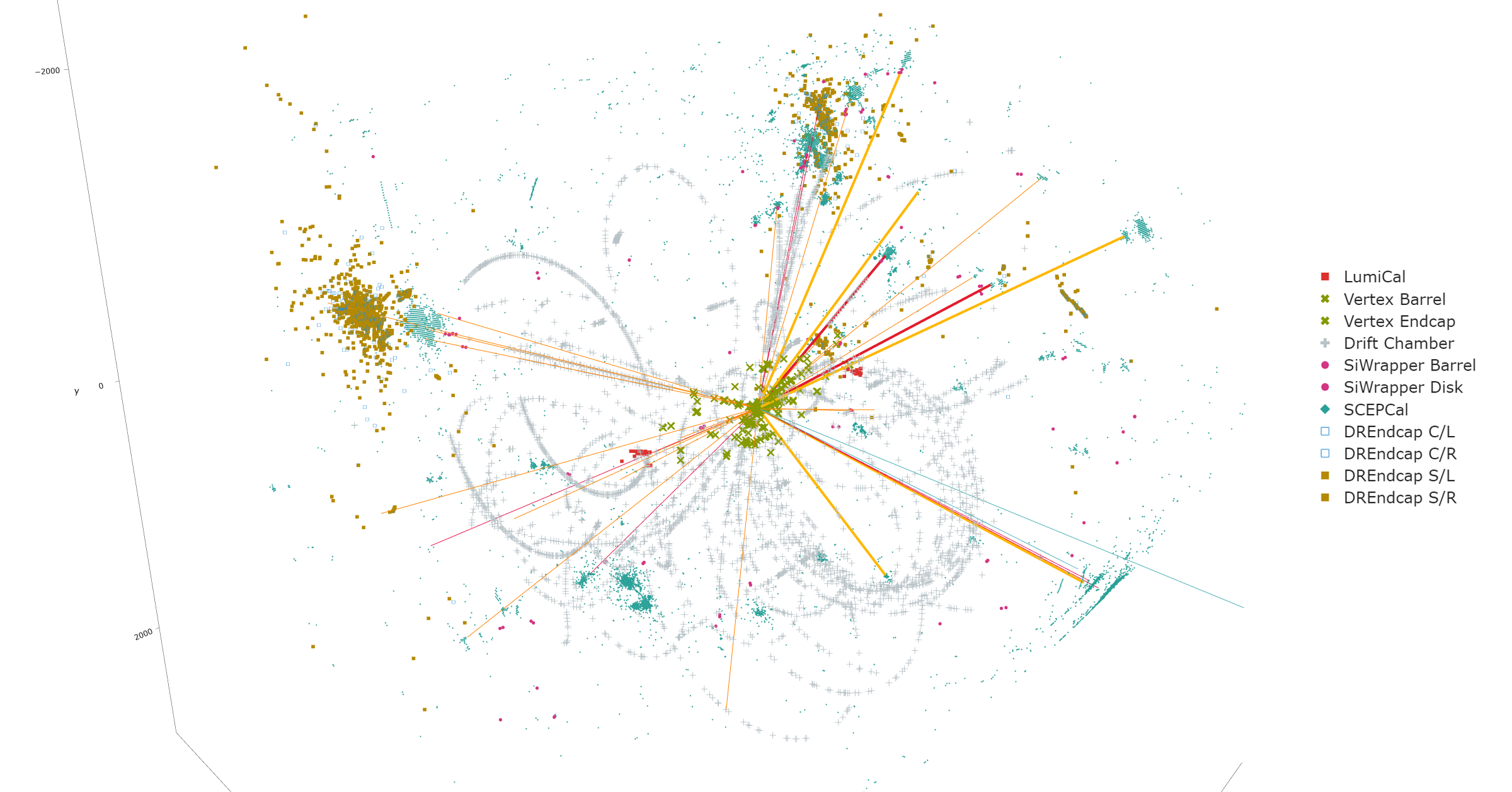}
    \caption{Event display of a sample ZZ  event at an $e^+e^-$ center of mass energy of 240\,GeV with the IDEA baseline design in DD4hep differentiable full simulation. Subdetector hit markers are coded by color and shape. The first 100 simulated tracks by generation time are shown.}
    \label{fig:idea_event_display}
\end{figure}

\section{Requirements \label{Req}}
    \subsection{Running conditions \label{ReqAcc}}
    The FCC-ee is planned to operate at or around four center of mass energies: the Z boson mass, the WW production threshold, the Z-Higgs associated production and the top quark pair production threshold. The luminosity, in excess of $10^{36}\,cm^{-2}s^{-1} $, is highest at the Z resonance with a bunch spacing of only 20 ns. It then drops rapidly while  the energy increases, as the beam currents are reduced to keep the maximum energy radiated by the beams below 50 MW/beam. The bunch spacing grows as the energy increases, reaching about  1 $\mu$s at the ZH associated production energy.

    These running conditions are particularly challenging for the detectors at the Z energy, due to the fast beam-related backgrounds and the rather high rate of interesting physics, which is expected to be in the order of 100 kHz. Limitations also exist in the current final focus configuration due to the magnetic field of the detector solenoid, which contributes to increase the beam emittances. A field of 2 T is currently considered to be the maximum acceptable to avoid degrading the luminosity when running at the Z pole. This restriction becomes much weaker at higher energies, so magnets that can provide up to 3 T when running above the Z resonance are being considered.

    The beam structure at FCC-ee does not have significant gaps in time with no beam as foreseen at ILC or CLIC. This has a large impact on the detector design, because powering on and off the detectors to reduce power consumption is not an option, thus increasing cooling requirements.

    Relative to linear colliders the beamstrahlung contribution~\cite{Datta:2005gm, Jeans:2024nba} to the center-of-mass energy uncertainty is much smaller, being in the order of 1\textperthousand$\;$  around the  ZH energy scale, thus allowing tighter constraints from the beam energy. A notable example where this helps significantly is the determination of the Higgs recoil mass.
    
    \subsection{Physics drivers \label{ReqPhys}}
    The physics accessible at FCC-ee falls into a few major categories: detailed measurement of the Higgs boson properties, high precision electro-weak and heavy flavor studies at the Z boson center of mass energy, precision measurements of the W boson and top quark parameters, and searches for new physics.

    The main objective for the tracking systems is to provide good momentum, angle and impact parameter resolution for tracks with $p_t$ below 100\,GeV/c. At higher momenta the requirement on the momentum resolution is driven mostly by the measurement of the Higgs recoil mass in HZ events with Z$\rightarrow\,\mu^+\mu^-$. The precision of this measurement depends on the center of mass energy spread due to beamstrahlung, that is $\sim$ 0.1\% at FCC-ee, and the experimental error on the momentum measurement of the two muons from the decay of the Z boson. A track momentum resolution, $\sigma_{p_t}/p_t$, of a few times $10^{-5}$ is required to preserve the precision driven by the beam energy spread. Given the constraints on the magnetic field of the detector solenoid this leads to a rather large tracking volume.

    Heavy flavor physics and flavor tagging, all profit from high accuracy in secondary vertex measurements implying the need for powerful vertex detectors with hit resolutions at the few micron level. Typical momenta are generally lower for these types of physics, 
    necessitating low mass  tracking detectors. Particle identification, in particular separation of pions from kaons up to a few tens of GeV, is also important to enhance the signal over background for many specific rare processes and to allow some level of strange flavor tagging.

    Jet energy and angular resolution is of key importance to exploit final states involving decays to high energy quarks. Typical examples are Z, W or Higgs boson decays to two jets or hadronic decays of top quarks. A useful figure of merit is the separation of Z from W decays using the di-jet invariant mass, leading to a required jet energy resolution $\sim$30-40\%/$\sqrt{E}$. A new generation of calorimeters, based on particle flow and/or dual readout, is needed to achieve such precisions. 

    Extreme electromagnetic resolutions, attainable only with crystals, can provide high quality neutral pion identification that is important for the measurement of many heavy flavor quark and tau final states. Resolved neutral pions also play a role in improving the jet resolution. A good  photon energy resolution is relevant in improving the signal over noise of the process $H\rightarrow\gamma\gamma$ and in correcting the electron track energy for bremsstrahlung radiation.

\section{Detector overview \label{Over}}

The Innovative Detector for E$^+$e$^-$ Accelerator (IDEA) is a new, innovative
%and possibly cost-effective 
detector concept, specifically designed for FCC-ee. 
It is based on a sophisticated tracker and a crystal dual-readout (DR) electromagnetic calorimeter within a superconducting solenoid magnet, followed by 
%a preshower detector and 
a DR fiber calorimeter and then completed by a muon detection system placed within the iron yoke that closes the magnetic field.

The tracker is composed of a high resolution silicon pixel vertex detector, a large-volume and extremely low-mass drift chamber for central tracking and particle identification, and a surrounding wrapper made from silicon sensors to ensure the best overall momentum resolution. 
The wire chamber provides up to 112 space-point measurements along a charged particle trajectory with excellent particle identification capabilities provided by the cluster counting technique. 
An outstanding energy resolution for electrons and photons is provided by a finely segmented crystal electromagnetic calorimeter. The particle identification capabilities of the drift chamber are complemented by a time-of-flight measurement provided either by LGAD technologies in the Si wrapper or by a first layer of LYSO crystals in the ECAL. Outside an ultra-thin, low-mass superconducting solenoid the calorimeter system is completed by the dual-readout fiber calorimeter. The hybrid dual-readout approach  promises an outstanding energy resolution for hadronic showers without loss of EM performance. 

The muon detection system is based on \mbox{\textmu{}-Rwell} detectors, a new micro pattern gas detector that will provide a space resolution of a few hundred of microns, giving the possibility of selecting muons with high precision and also reconstructing secondary vertices at a large distance from the primary interaction point. 

\begin{figure}[!htbp]
    \centering
    \includegraphics[width=1.1\linewidth]{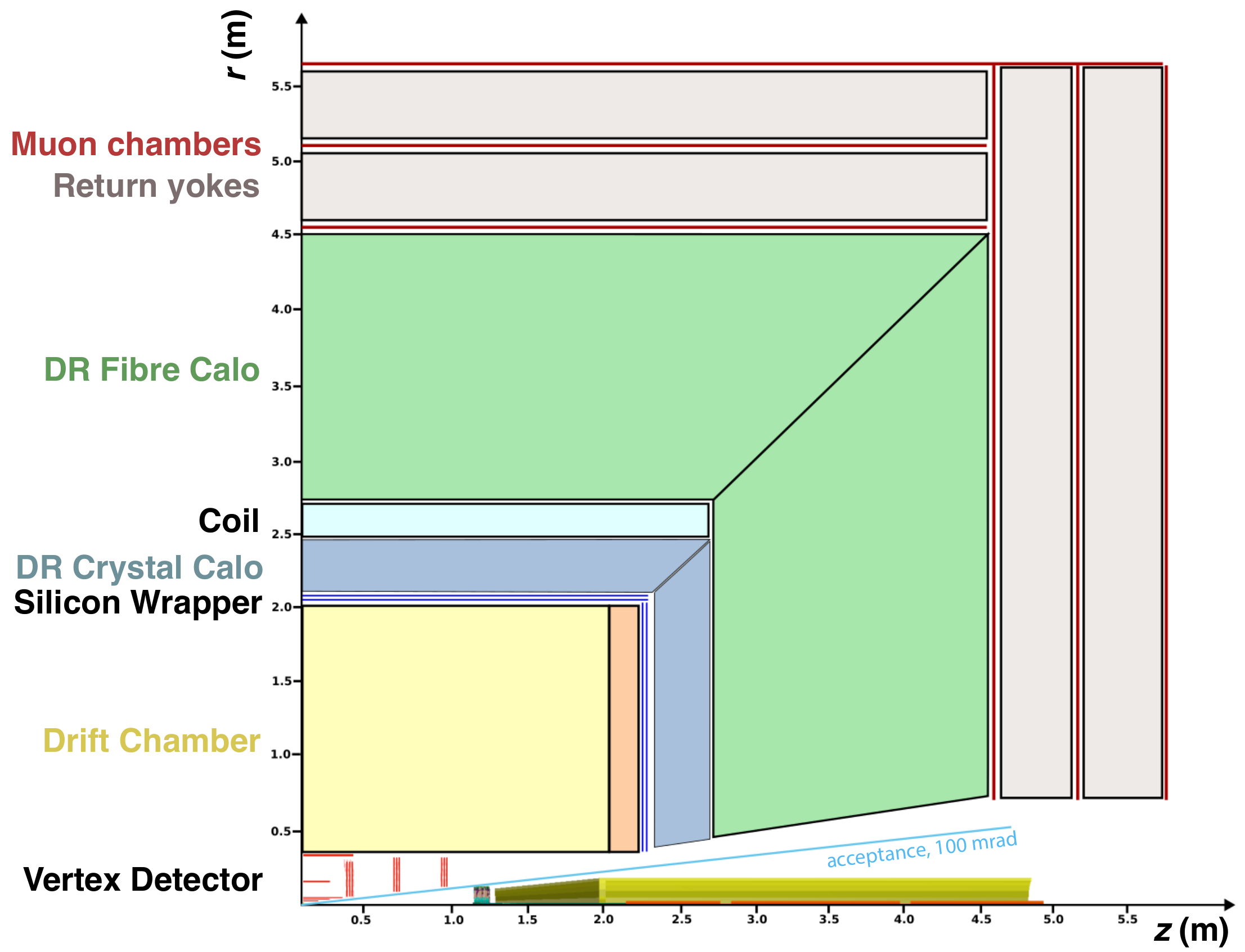}
    \caption{Overview of the IDEA detector layout.}
    \label{fig:idea_radenv}
\end{figure}

A 3-D view of the IDEA detector is shown in \mbox{Figure\ \ref{fig:idea_full}} and its cross-section is shown in \mbox{Figure\ \ref{fig:idea_radenv}}. An event display from a GEANT4 simulation of the detector is displayed in \mbox{Figure\ \ref{fig:idea_event_display}}. 

\clearpage\newpage

\section{Machine Detector Interface (MDI)\label{MDI}}

The FCC-ee interaction region (IR) design is based on the crab-waist collision scheme, whose main ingredients to reach the high luminosity are nano-beams at the interaction point (IP), large crossing angle, and crab-waist sextupoles~\cite{Raimondi:2007vi}. 
%To implement these requirements the IR turns out to be compact and complex, with relevant constraints to the Machine Detector Interface (MDI) of the FCC-ee. 

The IR magnet system includes, in addition to the detector solenoid, a compensating solenoid placed at 1.23\,m from the interaction point (IP) that cancels the $\int B_z dz$ between the IP and the faces of the final quadrupole, and a  screening solenoid around the final focus quadrupoles to produce an opposite field to that of the detector, thereby canceling the detector field. The final superconducting quadrupole sits inside the detector, at 2.2\,m  
where the distance between the  magnetic centers of the two beams is only a few centimeters.\par

\begin{figure}[!htbp]
    \centering
    \includegraphics[width=1.0\linewidth]{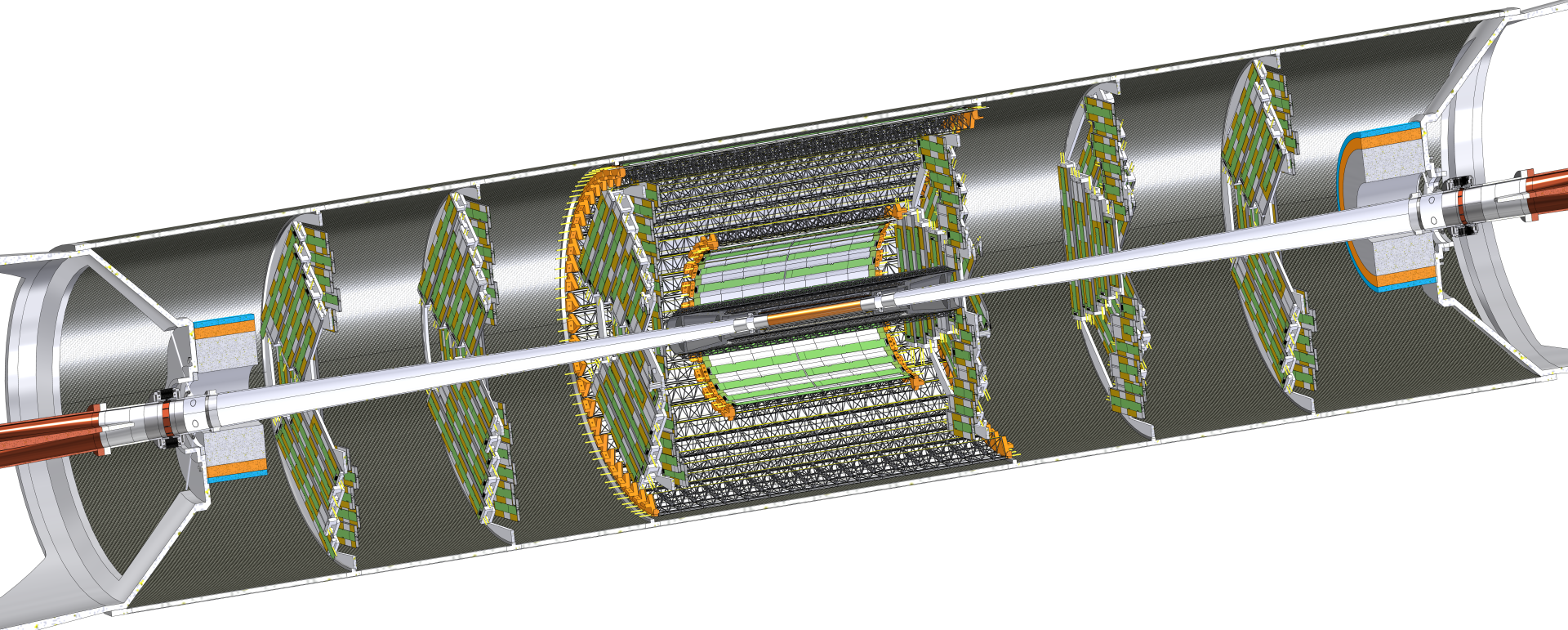}
    \caption{MDI layout with beam pipes, vertex detector, and LumiCal with the support tube.}
    \label{fig:mdi_layout}
\end{figure}

The central vacuum chamber extends for $\pm$\,90\,mm from the IP.
It will be composed of a double layer of  AlBeMet162, an alloy of 62\,\% of Beryllium and 38\,\% Aluminum,  with active liquid paraffin cooling inside through four inlets and outlets.
The geometry is optimized to minimize the material budget, which is calculated  to be 0.68\%~$X_0$ at normal incidence, and to guarantee a proper coolant flow to remove the heat generated by wakefields, estimated to be on the order of 60\,W. 
The vacuum chamber, that extends  between 90\,mm and 1154.5\,mm from the IP, has an ellipto-conical shape to minimize the material budget in front of  the LumiCal (the calorimeter that monitors the machine luminosity). This  beam pipe is supported by a lightweight carbon-fiber tube that also holds the vertex detector. The engineered MDI layout is shown in Fig.~\ref{fig:mdi_layout}.

\subsection{Beam related backgrounds\label{BeamBck}}
The most relevant background source for the detector is Incoherent Pair Creation (IPC)~\cite{Rimbault:2006ik}. The occupancies due to IPC were studied by applying a safety factor of 3 to the simulated rates. The largest effect occurs at the Z pole, and the resulting hit rate in the innermost layer of the IDEA vertex detector is about 200\,MHz/cm$^2$.
%, whereas the IDEA drift chamber occupancy is about 7\% every 400\,ns (with the safety factor of 1).

The Total Ionisation Dose (TID) and the 1 MeV n$_{eq}$ fluence in the IDEA interaction region have been estimated using FLUKA, considering  the main radiation sources at the Z pole operational mode, namely radiative Bhabha and IPC.
The peak annual dose in the inner vertex detector 
is at the level of few tens of kGy, while the peak fluence is of a few 10$^{13}$\,cm$^{-2}$ for the innermost layers.

The single beam backgrounds are mostly intercepted by the collimation system and absorbers which remove the bulk of the particles that would eventually hit the detectors, however some effects may remain due to 
radiation showers initiated by particle interactions with these devices upstream of the detectors.
Simulations of the beam-gas interactions have been done with a residual gas pressure profile resulting from 1~hr beam conditioning at a full nominal current of 1.27~A at the Z-pole. In the region of $\pm$ 500\,m from the IP  the gas composition was assumed to be 85\% Hydrogen, 10\%  $CO$ and 5\% $CO_2$ with peak  pressure in the order of $10^{-11}$mbar and spikes of $10^{-8}$ at the synchrotron radiation (SR) absorbers. The results indicate about 1\,Gy/year in the tunnel downstream of the IP and 1\,kGy/year in the SR absorbers and around the central vacuum chamber. 

Further studies are needed to estimate the injection backgrounds, as well as other possible sources of background from the incoming beams..

\section{Tracking system\label{Trk}}
The tracking system of IDEA consists of a silicon vertex detector, a lightweight drift chamber, and an enclosing silicon detector layer (\textit{Silicon Wrapper}), that could also provide timing information. The high granularity of the vertex detector enables the precise measurement of the vertices. At the same time, the drift chamber allows the tracks to be extended up to large radii to measure the charged particle momenta accurately. The Silicon Wrapper provides a last precise measurement before the Crystal ECAL. A time resolution of about $ \SI{100}{ps}$ in the Si wrapper or a LYSO layer would complement the $\text{d}N_\text{clusters}/\text{d}x$ information from the drift chamber to enable efficient particle identification for particle momenta up to about 30\,GeV/c.

    \subsection{Vertex detector \label{Vtx}}
    
The IDEA silicon vertex detector features two main subsystems, whose active elements are based on \SI{50}{\micro\meter} thick monolithic active pixel sensors (MAPS):
\begin{itemize}
    \item an inner vertex detector, located close to the beam pipe, at radii between 13.7 and 35.6 mm, covering an angular acceptance of about $\left|\cos(\theta)\right|<0.99$;
    \item an outer vertex detector, located at larger radii between 130 and 315\,mm, composed of a barrel section and forward disks.
\end{itemize}

Two alternative layouts are being explored for the inner vertex detector.
The first layout uses a traditional approach, in which modules are mounted on a carbon fiber support, with overlapping structures to allow full coverage and alignment.
The second, less advanced, layout is based on curved detectors, using a concept similar to the ALICE ITS3 \cite{its3_tdr}.

\begin{figure}[ht]
    \centering
\includegraphics[width=0.995\linewidth]{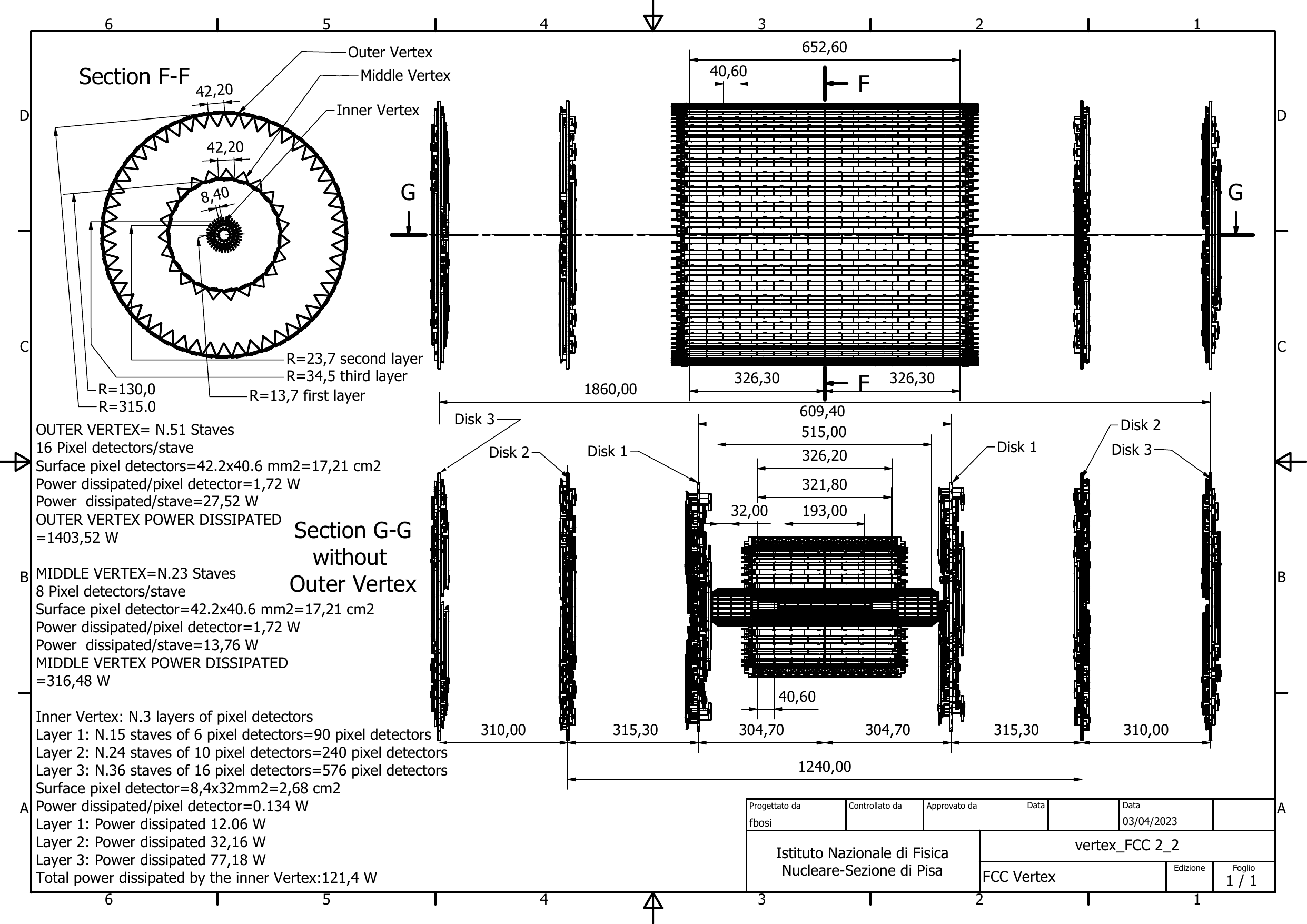}
    \caption{Engineered layout and main characteristics of the baseline vertex detector. Dimensions are given in mm. Top left: cross sectional view showing the barrel layers; top right: longitudinal view showing the outer barrel and disks; bottom right: longitudinal view with the inner and middle barrels and disks. The left panel reports the main elements for the outer, middle and inner barrel detectors together with the power dissipated.}
    \label{fig:Vertex_all}
\end{figure}

The baseline vertex detector layout is illustrated in Figure~\ref{fig:Vertex_all}. 

\subsubsection{Baseline inner vertex}

\begin{figure}
    \centering
    \includegraphics[width=0.8\linewidth]{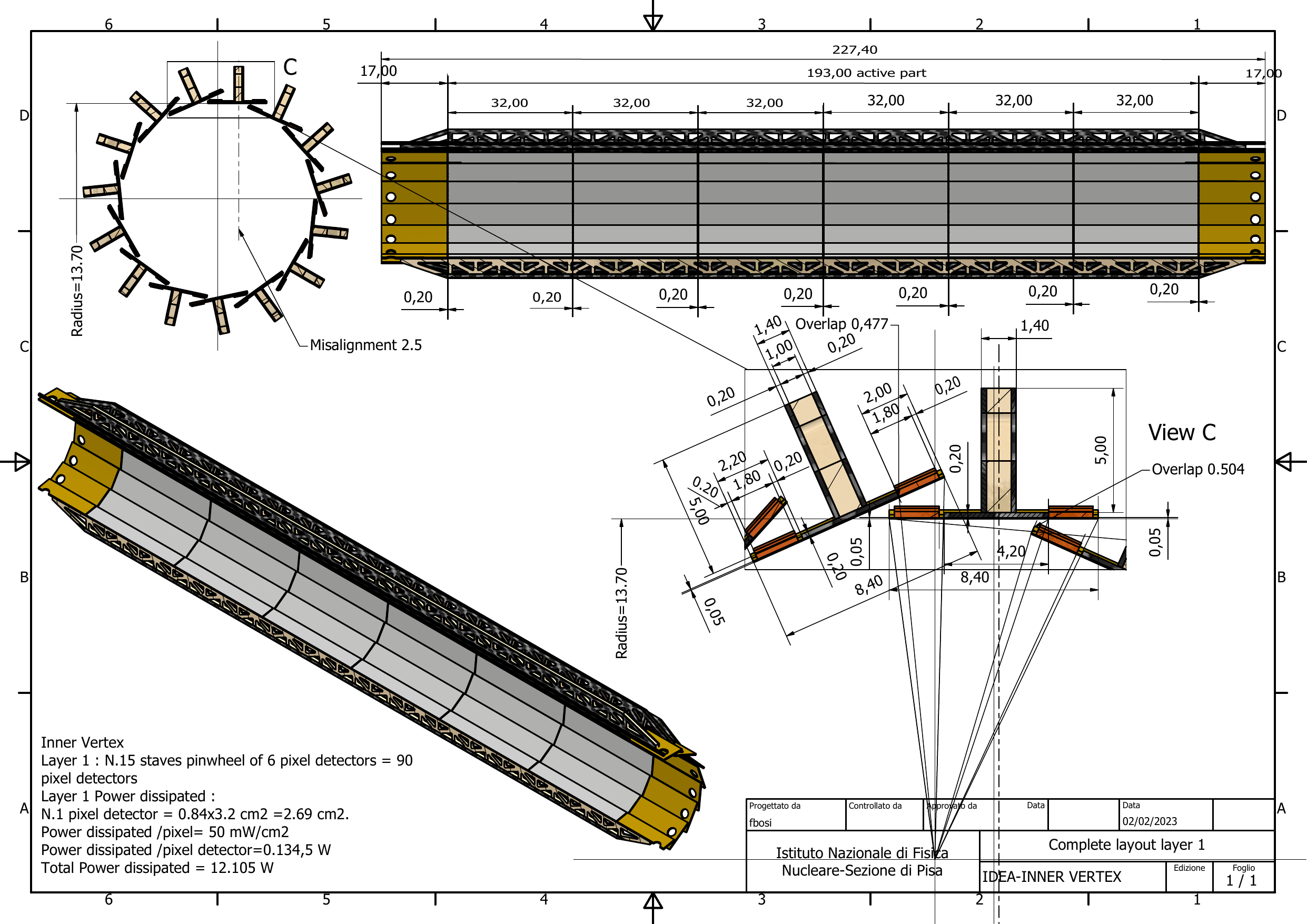}
    \caption{First layer of the baseline inner vertex detector. The top drawings show the transverse (left) and longitudinal (right) cross sections of the assembly, shown in three dimensions on the bottom left drawing. An enlarged view showing the overlaps and the different structures is given in the bottom right drawing. All measures are in mm.
    MAPS sensors of 50\,\textmu{}m thickness are facing the centre of the structure. The brown structures are the buses, while the gold parts are the electronic hybrid circuits for readout.}
    \label{fig:layer1}
\end{figure}

The inner vertex detector is composed of three concentric barrel layers mounted on carbon fiber support structures. The elementary unit is a module of dimensions $32\, (z) \times 8.4\, (r$--$\phi)\,\rm{mm}^2$. Each module has two chips abutted in $z$ inspired by the ARCADIA INFN R\&D program~\cite{ARCADIA}. 
The active area is made of $640\,(z) \times 256\,(r$--$\phi)$ pixels of $25\times \SI{25}{\micro\meter\squared}$ size. A 2\,mm inactive space is envisaged in $r$--$\phi$, which contains the periphery of the chip. The power consumption of the current ARCADIA prototype, read out at 100\,MHz/cm$^2$, is measured to be about 30\,mW/cm$^2$, and  considering the higher expected FCC-ee data rate and time resolution, 50\,mW/cm$^2$ has been conservatively assumed. 
To allow 2\,mm radial clearance for its insertion, 
the first layer is located at a radius of 13.7\,mm. 
The length is constrained by the central beam pipe cooling manifolds. The first layer comprises 15 staves of 6 modules each along $z$, as shown in Figure~\ref{fig:layer1}. The staves overlap in $\phi$, to allow internal alignment.
A lightweight support on each ladder provides rigidity and allows the mounting of the MAPS. The structure is made of thin carbon fiber walls interleaved with Rohacell, which holds the sensors (facing the beamline) and two buses (one for data and another for power) 1.8\,mm wide each on the opposite side. The thickness of the bus comprises \SI{200}{\micro\meter} kapton and \SI{50}{\micro\meter} aluminum for a total of 0.09\,\% of X$_0$. 
The ladders are arranged in a pinwheel geometry.

 The second layer, with a structure similar to the first, is made of 24 ladders of 10 modules each, at a radius of 23.7\,mm, arranged in a pinwheel geometry opposite in orientation to that of the first layer, in order to mitigate possible charge-dependent effects on charged-particle track reconstruction.
The third layer has 36 ladders each composed of 16 modules. The ladders are arranged in a $\phi$-symmetric lampshade fashion: half of the ladders are located at 34\,mm radius, the other half at 35.6\,mm. 

Table~\ref{tab:vertex_details} lists the properties of all the inner vertex layers.
\begin{table}[htbp]
    \caption{Main parameters of the baseline vertex detector.}
    \centering
    \small
    \begin{tabular}{|c|c|c|c|c|c|c|}\hline
        \bf Subsystem & \bf Layer ID & \bf Radius [mm] & \bf $|z|$ [mm] & \bf Staves & \bf Modules/staves & \bf Power [W]\\ \hline \hline
        Inner barrel & 1 & 13.7 &  $<96.5$ & 15 & 6 & 12\\
        Inner barrel & 2 & 23.7 &  $<160.9$ & 24 & 10 & 32\\
        Inner barrel & 3 & 34 and 35.6 & $<257.5$ & 36 & 16 & 77 \\ \hline
        Middle barrel & 1 & 130 & $<163.1$ & 23 & 8 & 316\\
        Outer barrel & 2 & 315 &  $<326.3$ &  51 & 16  & 1400\\ \hline

        Disks & 1 & $34.5 < r < 275$ & 304.7 & 56 & 2--6 & 135\\
        Disks & 2 & $70 < r < 315$ & 620 &   48 & 3--7 & 420\\
        Disks & 3 & $105 < r < 315$ & 930 &  40 & 4--7 & 370\\\hline
    \end{tabular}
    \label{tab:vertex_details}
\end{table}
Each layer has a radiation length of $0.25\%$ at normal incidence. The overlap between same-layer staves increases the average material budget by about 20\%.
The vertex detector layers are mounted on the conical support using two rings of PEEK material to thermally isolate them during bakeout.
% The material budget of the inner vertex is shown in Fig.~\ref{fig:vtx_material_budget_layer1}.%, and the impact parameter resolutions are shown in Figure~\ref{fig:idea_vertex_performance}.

\begin{figure}
    \centering
    \includegraphics[width=0.75\linewidth]{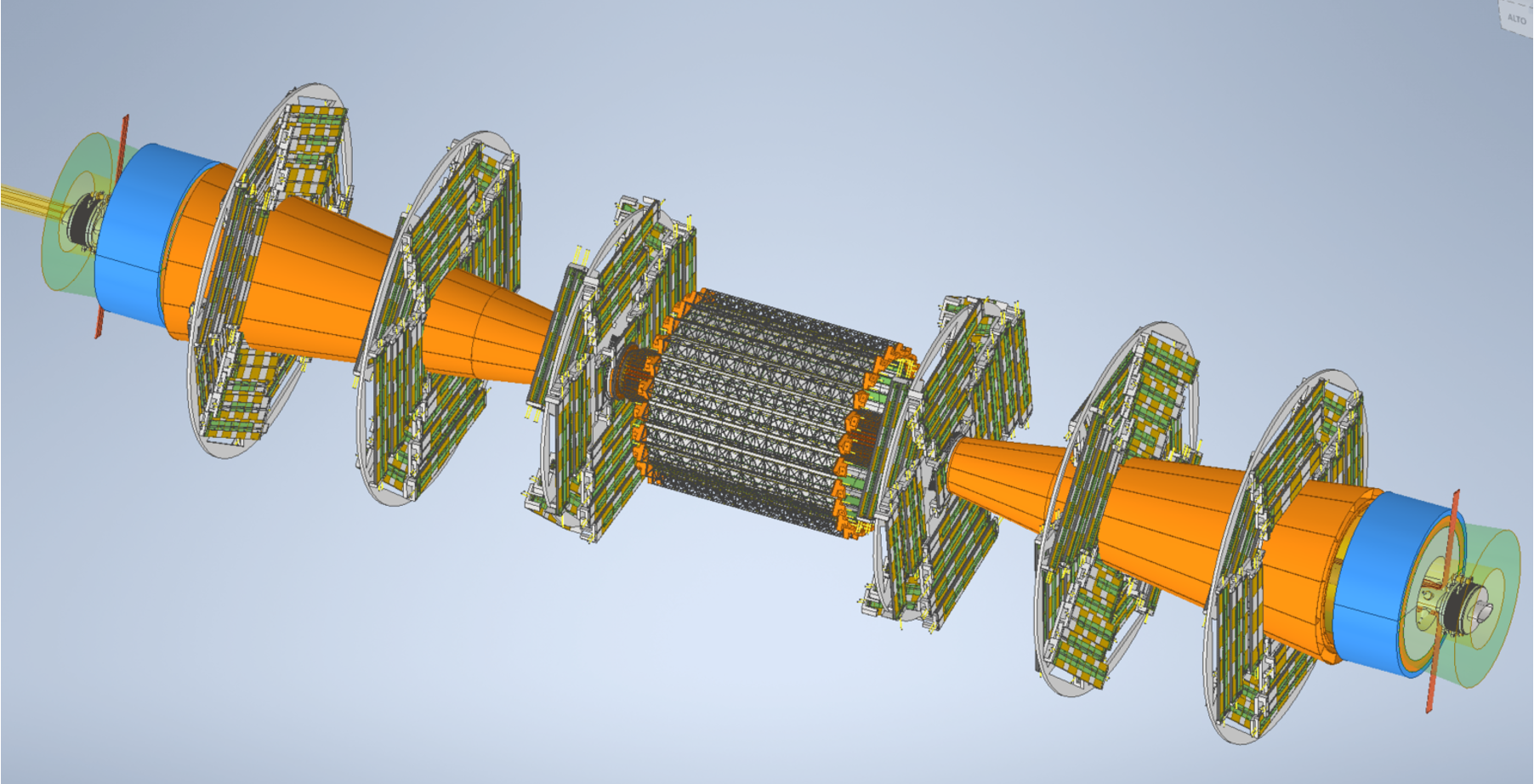}
    \caption{View of the vertex detector middle barrel and disks together with the cooling cones. The blue parts are the two luminosity detector calorimeters.}
    \label{fig:cooling_air}
\end{figure}

The power dissipated by each layer is reported in Table~\ref{tab:vertex_details}. The inner vertex detector will be cooled by gas passing through channels embedded in the conical carbon fiber support structure. Both atmospheric air and helium are considered as cooling gases.
 A system of carbon fiber cones is used to force gas convection inside the detector volume, and to support power and readout circuits, as shown in Figure~\ref{fig:cooling_air}. The cooling performance is analysed with a  Computational Fluid Dynamics (CFD) simulation, resulting in the largest temperature difference between the modules along the stave of the third layer of less than \SI{15}{\celsius}. All other layers dissipate less power (see Table~\ref{tab:vertex_details}). A mechanical vibration analysis has been performed using a Finite Element Analysis (FEA) simulation in ANSYS, resulting in the magnitude of the maximum displacement of about \SI{1.5}{\micro\meter} for a nominal airflow of \SI{0.7}{g/s}.

% \begin{figure}
%     \centering
%     \begin{subfigure}{a}
%    \includegraphics[width=0.45\linewidth]{Physics/DetectorConcepts/Figs/materialBudget_VTX.pdf}
%    \end{subfigure}
%   \begin{subfigure}{b}
%    \includegraphics[width=0.45\linewidth]{Physics/DetectorConcepts/Figs/VTXIB_curved_1D.pdf}
%    \end{subfigure}
   
%     \caption{Vertex Material Budget}
%     \label{fig:vertex_MB}
% \end{figure}

% \begin{figure}
% \centering
% \begin{subfigure}{a}
%     \includegraphics[width=0.4\textwidth]{Physics/DetectorConcepts/Figs/d0.pdf}
%     %\caption{}
%     %\label{fig:a}
% \end{subfigure}
% \begin{subfigure}{b}
%     \includegraphics[width=0.4\textwidth]{Physics/DetectorConcepts/Figs/z0.pdf}
%     %\caption{}
%     %\label{fig:b}
% \end{subfigure}
% \hfill
% \caption{Impact parameter resolution for different muon momenta in the IDEA vertex }
% \label{fig:idea_vertex_performance}
% \end{figure}

\subsubsection{Ultra-light inner vertex}
An alternative layout for the inner vertex is also explored, based on a similar concept of the ALICE ITS3 curved sensor technology \cite{its3_tdr}, which allows a self-supporting structure with almost only silicon sensors in the acceptance, thus resulting in an ultra-light vertex layout. ITS3
uses the stitching technique to form wafer-scale sensors from multiple repeated
sensor units (RSUs). Each ITS3 layer comprises two half-cylindrical sensors featuring ten RSUs in $z$ and three, four, or five in $\phi$ for the first, second, and third layers.
Unlike ITS3, the vertex detector at FCC-ee must cover the largest polar angle possible for all detection layers. The layers cover more or less the same angular acceptance to fulfill this requirement. To reach the small radius of \SI{13.7}{mm}, the first layer only uses two rows in $\phi$ of 10 RSU in z as listed in Table~\ref{tab:vertex_details_ultra-light}. It is supported only by two carbon foam longerons and rings. The spacing between the two half-shells is \SI{1.25}{mm}, thus leaving a gap in the $\phi$ acceptance, which is partially compensated by the second layer which is rotated in $\phi$ with respect to the first layer. It features one more row of RSUs in $\phi$ and 13 RSUs along $z$. The first two layers are read out and powered by either side. For the third and fourth layers, using two sensors in $z$ per half-layer is foreseen to circumvent the limitation of the 12-inch wafer diameter. The third (fourth) layer uses 8 (10) layers on the $-z$ side and 10 (8) on the $+z$ side.
In this manner, the gap in acceptance in $z$ in one layer is covered by the other. For these layers, the readout only occurs on the far $z$ sides.

\begin{table}[ht] 
      \caption{Main parameters of the ultra-light inner vertex detector.}
    \label{tab:vertex_details_ultra-light}
    \centering
    \small
    \begin{tabular}{|c|c|c|c|c|c|c|c|}\hline
         \bf Layer & \bf Sensors & \bf Radius [mm]  & \bf RSUs in $\phi$ & \bf RSUs in $z$ & \bf Length [mm] & \bf Coverage
         \\\hline\hline
         1 & 2 & 13.7   & 2 & 10                & 108.3 & $\left| \cos(\theta)\right| < 0.992$ \\\hline
         2 & 2 & 20.35  & 3 & 13                & 140.8 & $\left| \cos(\theta)\right| < 0.990$ \\\hline
         3 & 4 & 27     & 4 & \phantom{1}8 ($-z$)/10 ($+z$)    & 199.5 & $\left| \cos(\theta)\right| < 0.990$ \\\hline
         4 & 4 & 33.65  & 5 & 10 ($-z$)/\phantom{1}8 ($+z$)    & 199.5 & $\left| \cos(\theta)\right| < 0.986$ \\\hline
    \end{tabular}
\end{table}

\begin{figure}
    \centering
     \includegraphics[width=\linewidth]{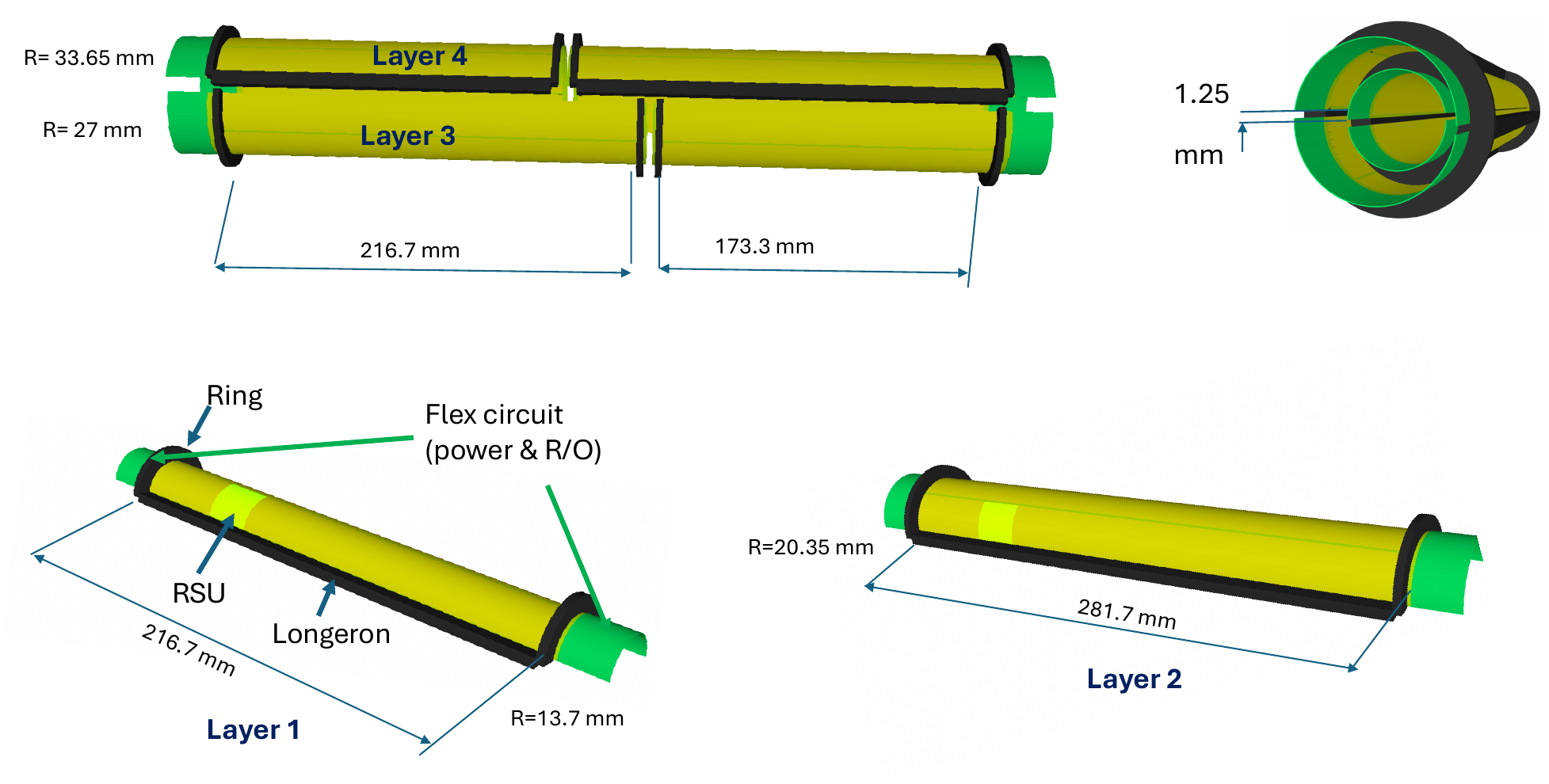}
    \caption{Ultra-light inner vertex detector layout (see text for detailed explanation). The four half-layers are shown separately. On the top right hand side an assembled view of the first two layers is shown.}
    \label{fig:curved_vertex}
\end{figure}

Figure~\ref{fig:curved_vertex} shows the layout of this ultra-light vertex detector concept. 
Integration studies are ongoing to validate the technical feasibility of such a layout.

\begin{figure}[htbp]
    \centering
    \subfloat[Flat inner vertex]{
        \includegraphics[width=0.5\textwidth]{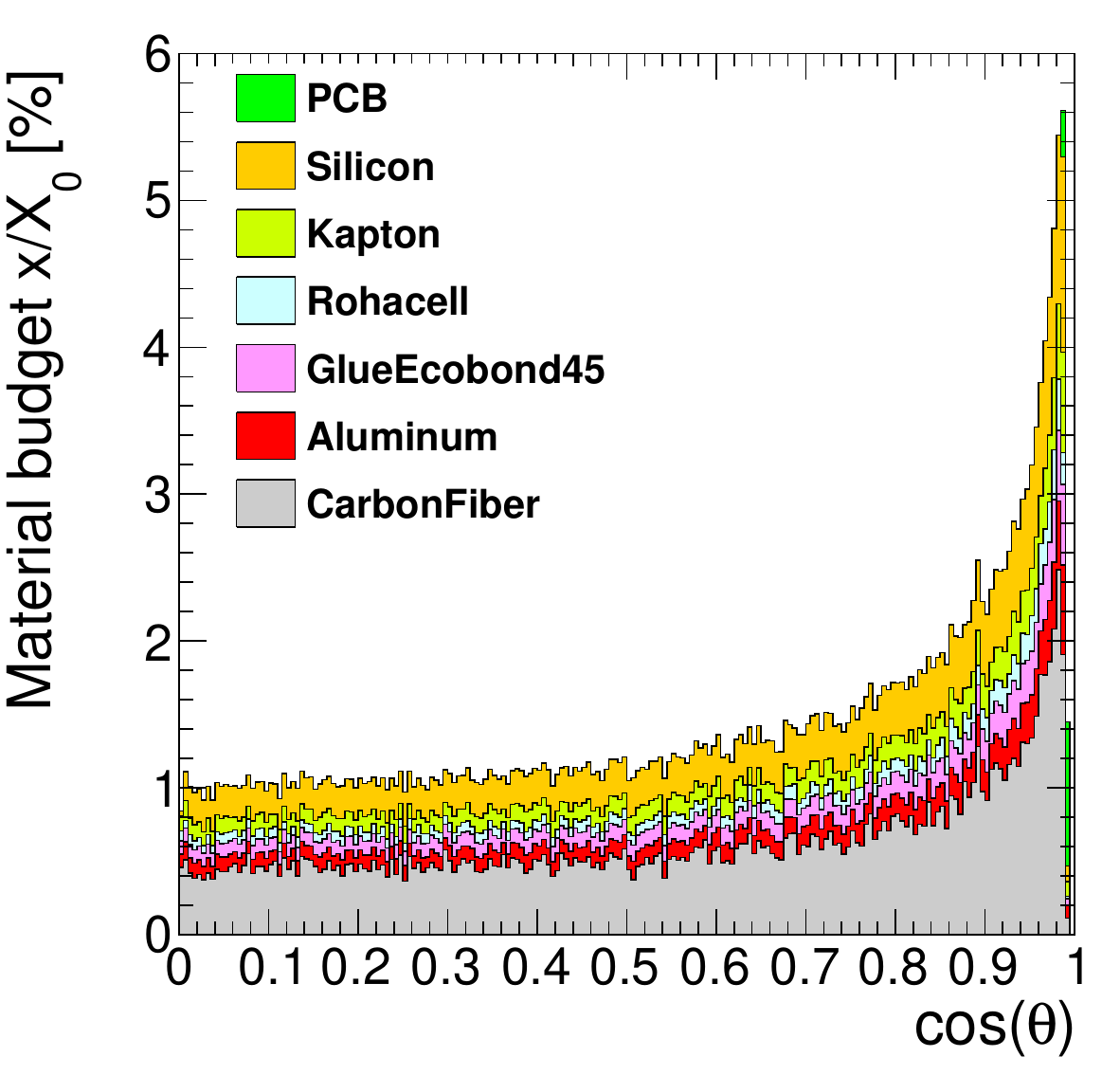}
        \label{fig:vtx_material_budget_layer1}
    }
    \subfloat[Complete flat vertex detector]{
        \includegraphics[width=0.5\textwidth]{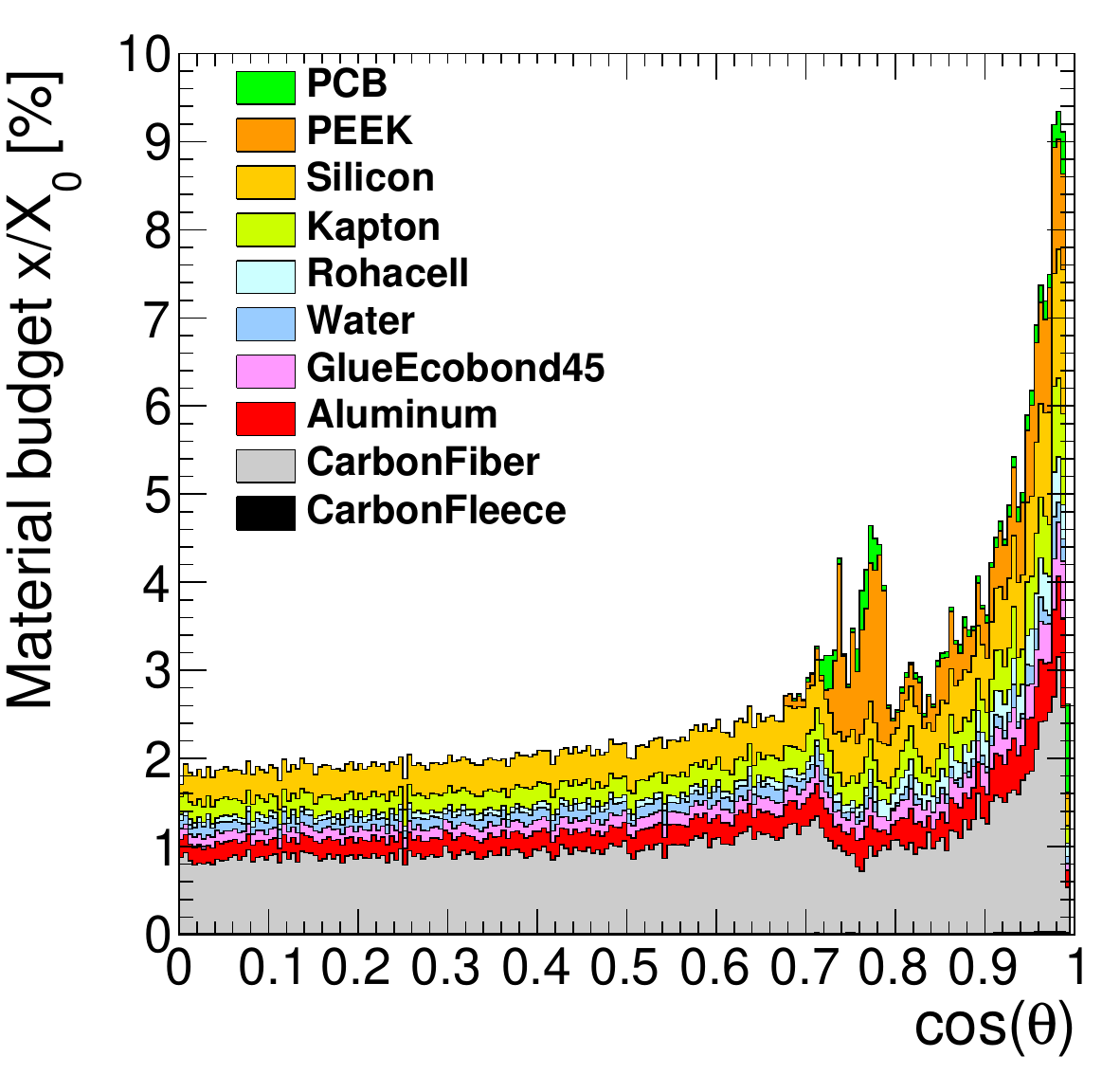}
        \label{fig:VTX_material_budget_all}
    }\\
    \subfloat[Ultra-light inner vertex detector]{
        \includegraphics[width=0.5\textwidth]{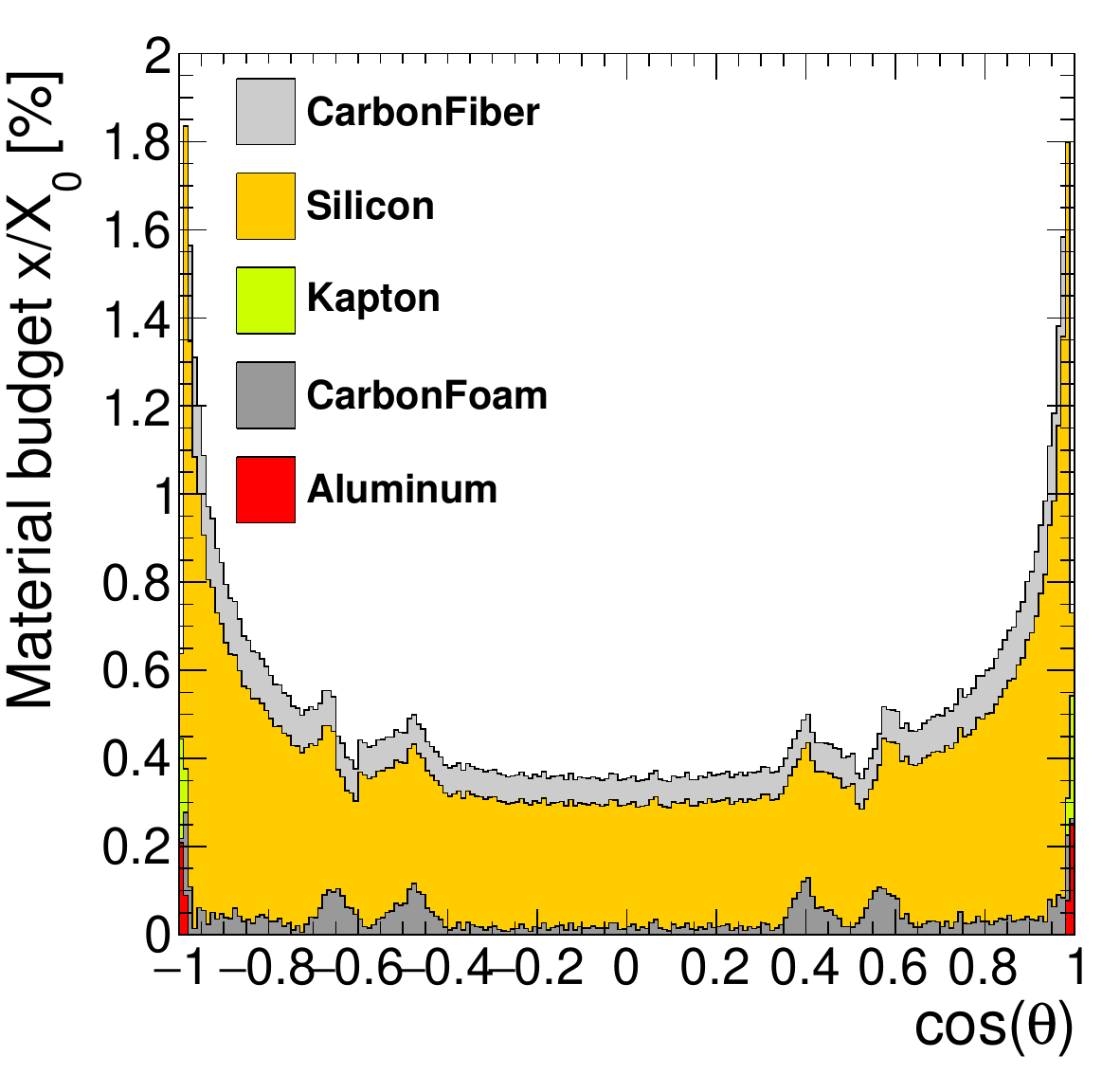}
    }
    \subfloat[2D ultra-light inner vertex detector ]{
        \includegraphics[width=0.5\textwidth]{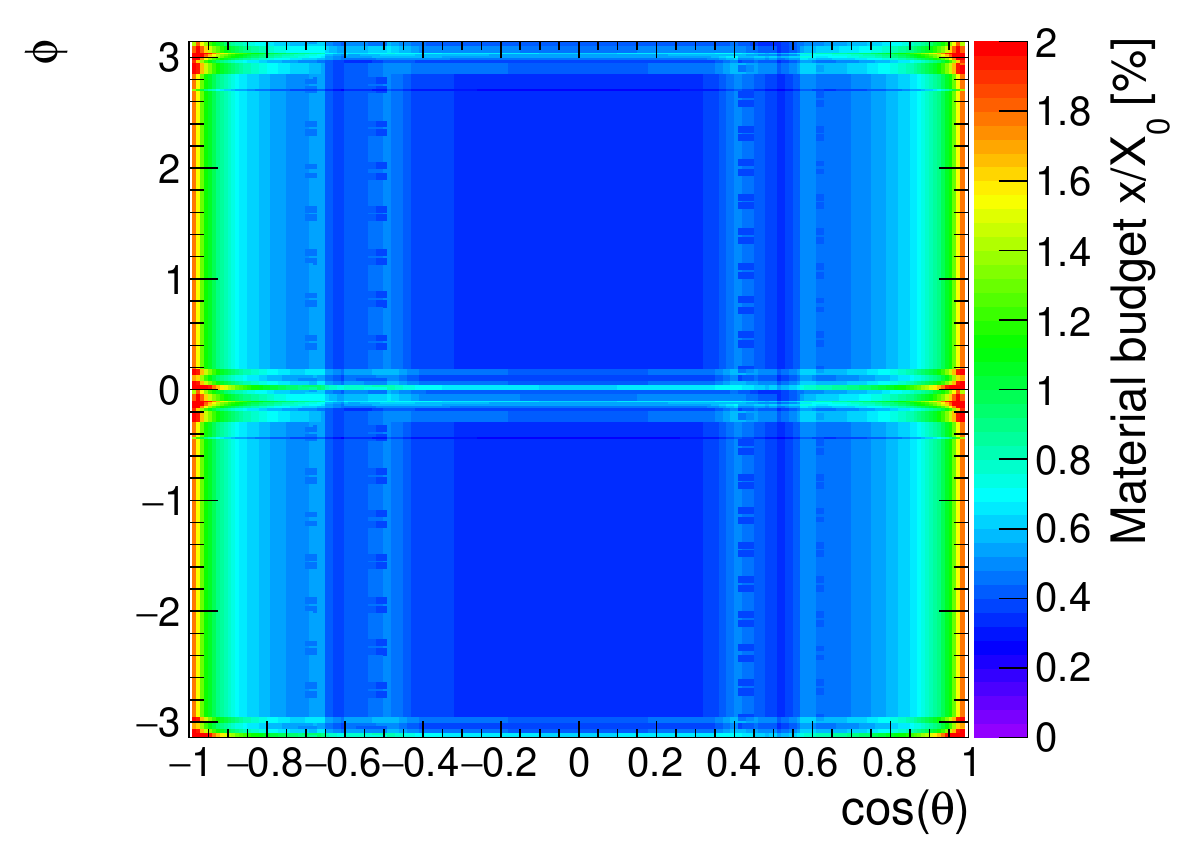}
\label{fig:UL_vtx_material_budget_2D}
    }
    \caption{Material budget for the vertex detector. On the top row for the flat inner vertex (a) and complete vertex detector (b). On the bottom row for the curved inner vertex detector (c) and two-dimensional distribution (d).}
    \label{fig:vtx_ultra-light_material_budget}
\end{figure}

The material budget is very much reduced with respect to using a classic design, as shown in Fig.~\ref{fig:vtx_ultra-light_material_budget}.
A single layer contributes about 0.075\% of $X_0$ at normal incidence, and the material budget is more uniformly distributed in $\phi$ due to the absence of overlapping structures in the same layer as shown in Fig.~\ref{fig:UL_vtx_material_budget_2D}. 

\subsubsection{Outer vertex and disks}
The outer vertex detector is composed of two barrel layers and three disks on either side of the IP. The elementary unit is a module of dimensions $40.6 (z) \times 42.2 (r-\phi) ~\rm{mm}^2$. Each module has four hybrid pixel sensors inspired by the ATLASpix3 design~\cite{ATLASPIX3}. The active area of the chip consists of 132 columns of 372 pixels each, with square pixels of $150 \times \SI{50}{\micro\meter\squared}$.
The sensor thickness of the modules is \SI{50}{\micro\meter} and their power consumption is assumed to be \SI{100}{mW\per\centi\meter\squared},
half of the level observed at the current stage of development, 
but still too high (by at least a factor of two) to be handled by air cooling.

\begin{figure}[H]
 \centering
    \begin{subfigure}[b]{0.8\textwidth}
        \centering
        \caption{Outer vertex barrel}
        \includegraphics[width=\linewidth]{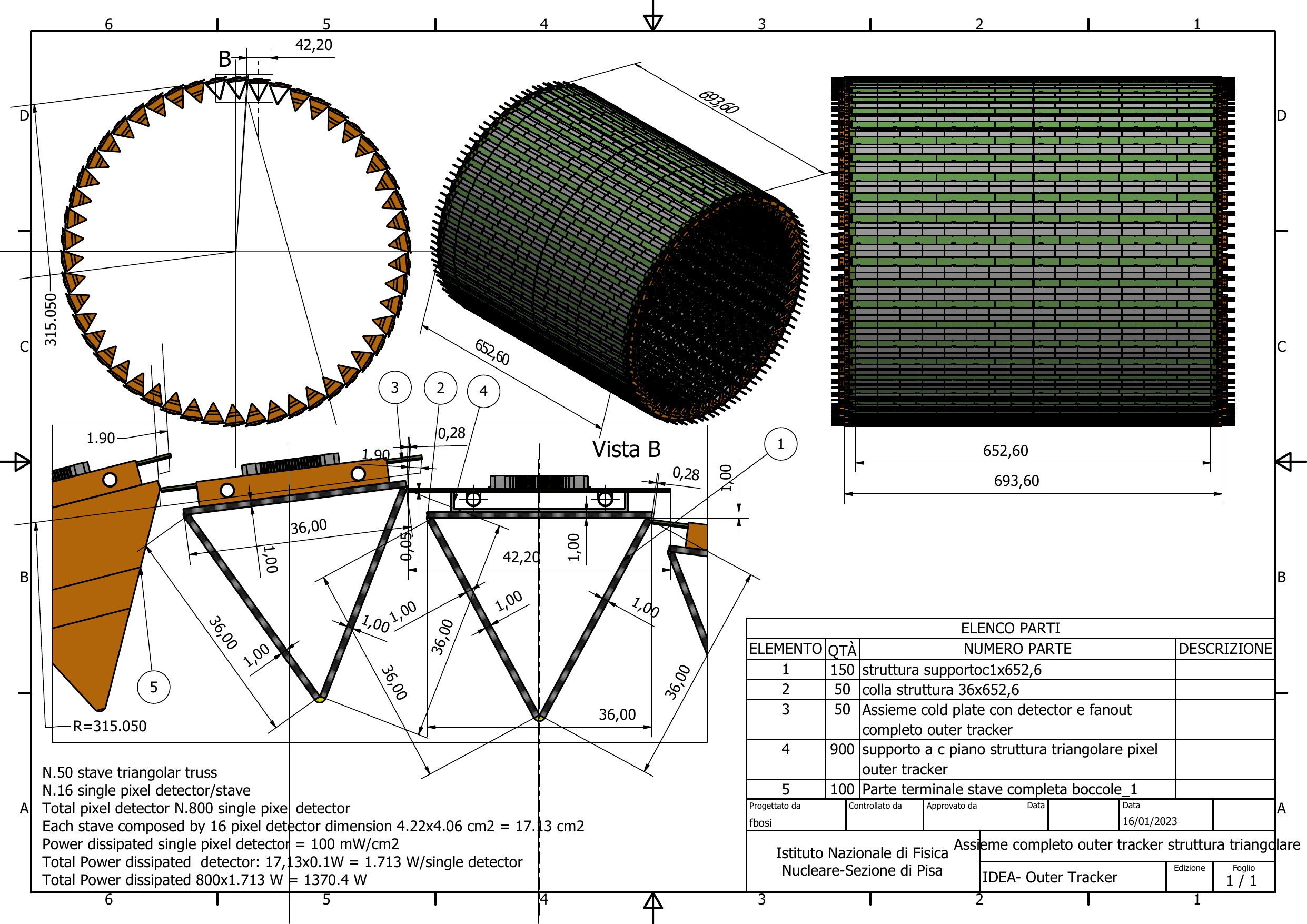}
        \label{fig:Outer_vertex}
    \end{subfigure}\hfill
    \begin{subfigure}[b]{0.8\textwidth}
        \centering
        \caption{Disk 1}
        \includegraphics[width=\linewidth]{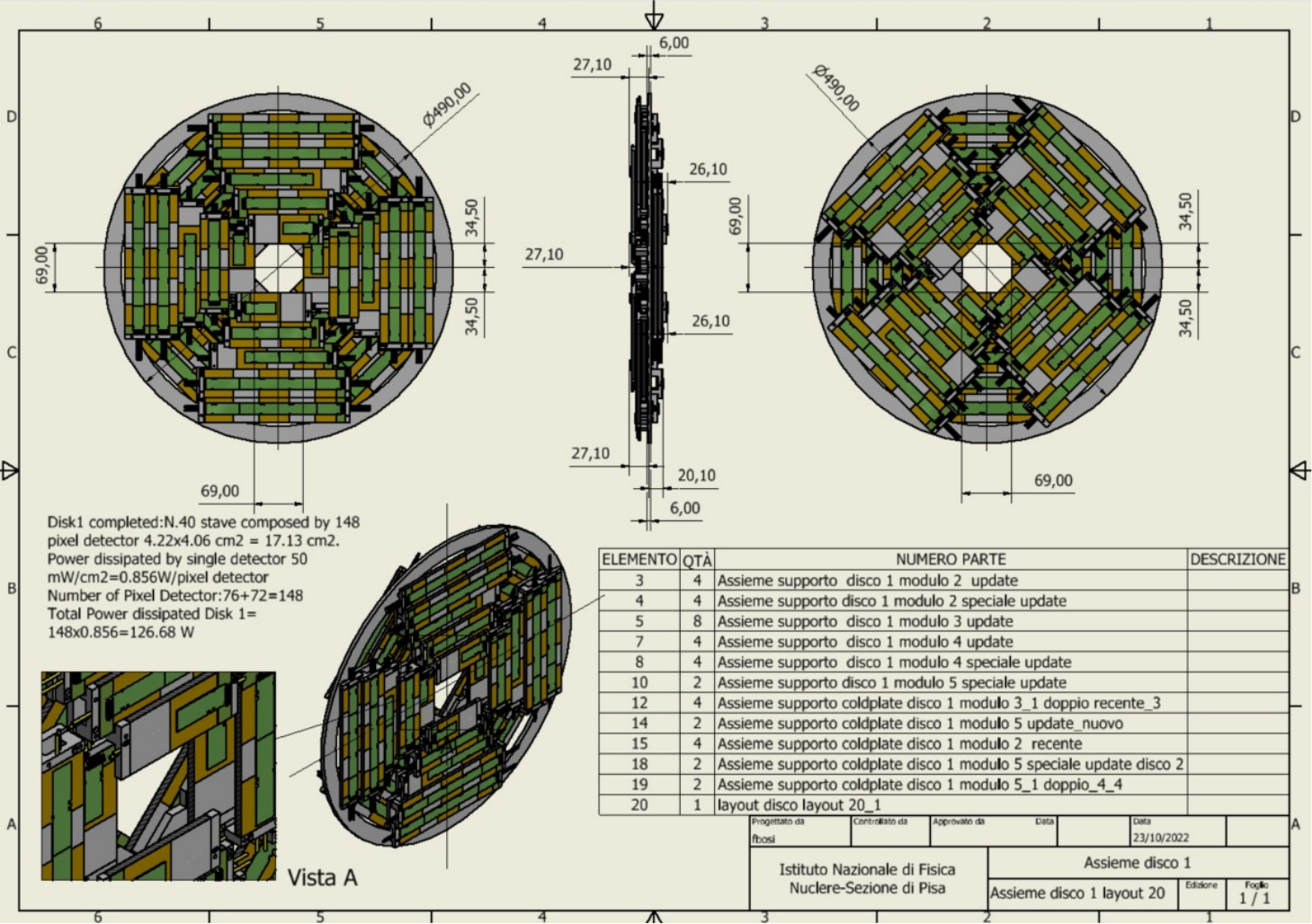}
        \label{fig:Disk1}
    \end{subfigure}
    \caption{Details of the outer vertex barrel  (a) and first disk (b).} 
\end{figure}

The same module type is used for the barrel layers and the disks. One barrel layer is placed at 13\,cm  radius (middle barrel) and is made of 22 ladders, each with 8 modules, as shown in 
Fig.~\ref{fig:Outer_vertex}.

The outermost layer (outer barrel) is placed at 31.5\,cm and is composed of 51 ladders of 16 modules each.
The outer layer is supported by a flange that is attached to the external support tube. The flange also supports the middle barrel as well as the first disks.

The modules are placed on top of a lightweight triangular truss structure. The staves are mechanical structures holding the modules.
Each stave is composed of a carbon fibre multilayered structure, comprising \SI{120}{\micro\meter} thick carbon fibre KDU13, and two carbon fleeces  of \SI{65}{\micro\meter} in total, to support two polyamide tubes of 2.2\,mm diameter and \SI{90}{\micro\meter} thick, in which demineralised water will be circulating.\\ An electronic bus,  bringing the power distribution and the readout and control signals, runs along the entire stave length and is put on top of the modules. It is terminated at the end of both sides of the stave by a hybrid circuit.
Three disks per side, located at z=$\pm 29.25, \pm 62  {\rm ~and }\pm 93$\,cm, complete the outer vertex tracker. The inner disk is located inside the barrel.
% and  extends from an innermost radius of $ 3.45$ cm to an outer radius of $24.5$ cm. The other two pairs of disks have an inner radius of $ 7 {\rm ~and ~} 10.5$ cm, and extend up to a radius of $28.5 $ cm to allow an hermetic coverage. \\
Each disk comprises four front and backward petals, made of modules of the same type as those of the barrels, as shown in Fig.~\ref{fig:Disk1}. The support structure of each disk is made of a sandwich of thin carbon fibre walls (each of 0.3\,mm thickness) interleaved with Rohacell (5.4\,mm thick).

\clearpage\newpage

    \subsection{Drift chamber \label{Dch}}
    
The drift chamber for the IDEA detector concept is designed to provide precise tracking, high-precision momentum measurement and excellent particle identification by adopting the ``cluster counting" technique \cite{Clucou}. The main peculiarity of this drift chamber is its high transparency, which is a crucial feature for the charged particle momentum range, from several tens of GeV/c to a few hundred of MeV/c, where the Multiple Scattering (MS) contribution is far from being negligible, but also to limit $\gamma$ conversions and hadronic interactions. This high transparency is obtained mainly thanks to a novel approach adopted for the wiring and assembly procedures~\cite{CHIARELLO2016512}.
The total amount of material in terms of radiation lengths is about 1.6\% $X_0$ in the radial direction towards the barrel calorimeter, mostly due to the outer wall. In the forward direction it is about 5.0\% $X_0$, with 75\% of this located in the end plates, which is instrumented with front-end electronics. 

The chamber design inherits some aspects of drift chambers  previously built and operated, like the ones of the KLOE experiment~\cite{ADINOLFI200251} and the MEG2 experiment~\cite{CHIAPPINI2023167740}. However it also presents fully innovative features, aimed to combine granularity and transparency needs. 

\begin{figure}[ht]
    \centering
\includegraphics[width=0.7\linewidth]{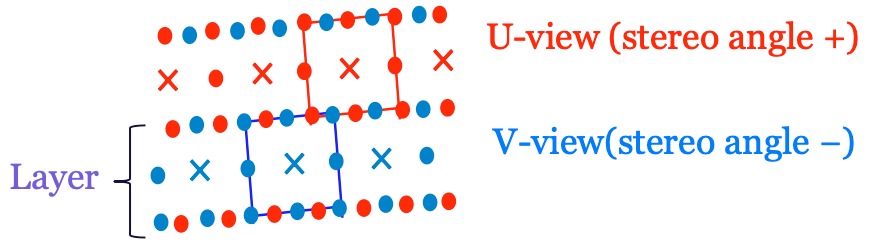}
    \caption{Detail of the cross-section at the middle of the drift chamber (z=0), showing the cell layout. Field wires are indicated with full dots, blue or red depending on the sign of the stereo angle, while sense wires are indicated with the symbol ``x". The cells are approximately square, and the field to sense wires ratio is 5:1.}
    \label{fig:celllayout}
\end{figure}

The IDEA drift chamber is a single-volume, cylindrical wire chamber, co-axial with the 2 T solenoidal field and operated with a very light gas mixture, 90\% He–10\% C$_{4}$H$_{10}$ (in addition to the excellent quenching properties, relatively low cost and ease of production, isobutane has a low global warming potential, GWP). It extends from an inner radius $R_{in} = 350$\,mm to an outer radius $R_{out} = 2000$\,mm, for a length of $L=4000$\,mm. Coverage in polar angle extends down to $\sim$$13^{\circ}$. The chamber is a full-stereo wire device, and it consists of 14 co-axial super-layers with 8 layers each, for a total of 112 layers, strung at alternating sign stereo angles ranging from 50 to 250 mrad. Wires are arranged azimuthally in 24 identical angular sectors, 15${^\circ}$ wide. The cells are approximately square, since this geometry ensures a more uniform cell distribution inside the chamber volume for a full stereo configuration. The cell size ranges between 12.0 and 14.5\,mm, thus implying a maximum drift time of about 400 ns, assuming a drift velocity of 2.2\,cm$/\mu$s in a typical 90\% He–10\% C$_{4}$H$_{10}$ gas mixture~\cite{ADINOLFI200251}. The total number of drift cells is 56,448, and the field to sense wires ratio is 5:1 (see Fig. \ref{fig:celllayout}). 

\begin{figure}[ht]
    \centering
\includegraphics[width=0.70\linewidth]{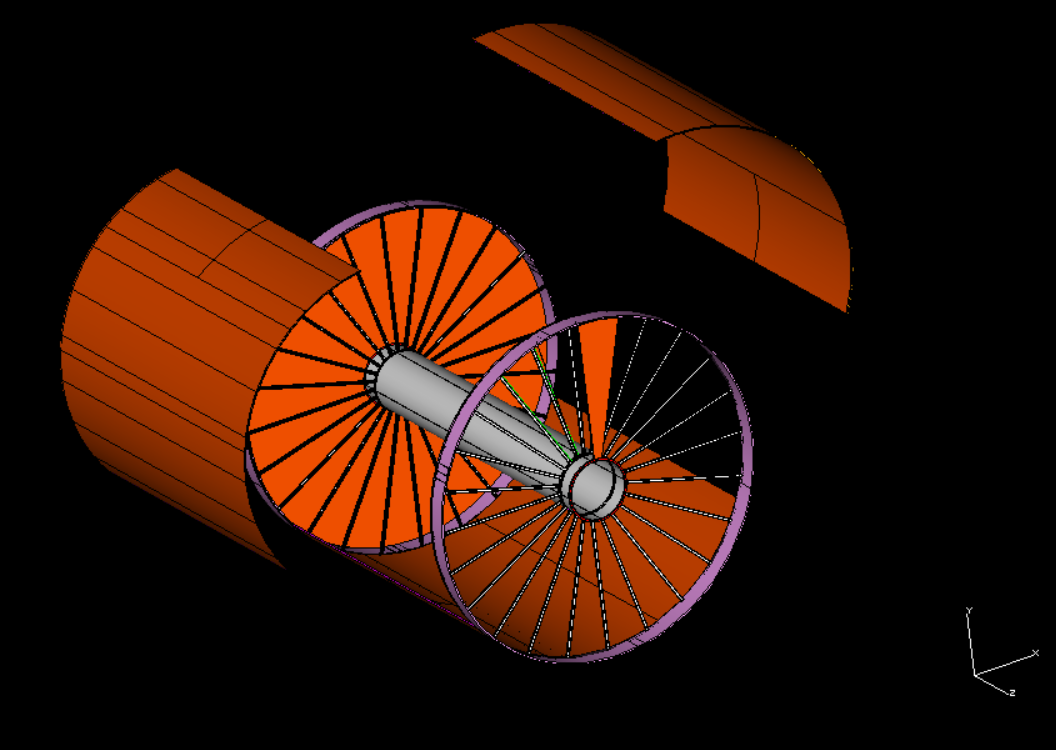}
    \caption{Schematic drawing of the drift chamber, showing the mechanical structure, consisting of the gas envelope and the wire cage.
}
    \label{fig:wirecage}
\end{figure}

The larger number of field wires per sense wire with respect to other large drift chambers (e.g. the KLOE drift chamber) allows the use of thinner field wires (40-50\,$\mu$ m diameter, in silver plated Al), benefiting the chamber granularity, the electric field isotropy inside the cell and the total load of the wires on the chamber end plates, at the price of wiring about 350,000 wires in total in the chamber.
This large number of wires requires a nonstandard wiring procedure and needs a feed-through-less wiring system. A novel wiring procedure, successfully exploited during the recent construction of the MEG2 drift chamber~\cite{BALDINI2020162152}, relies on anchoring the wires on carbon fiber supports placed on a cage structure mainly based on carbon fiber as well (shown in Fig. \ref{fig:wirecage}). 

\begin{figure}[ht]
    \centering
\includegraphics[width=0.7\linewidth]{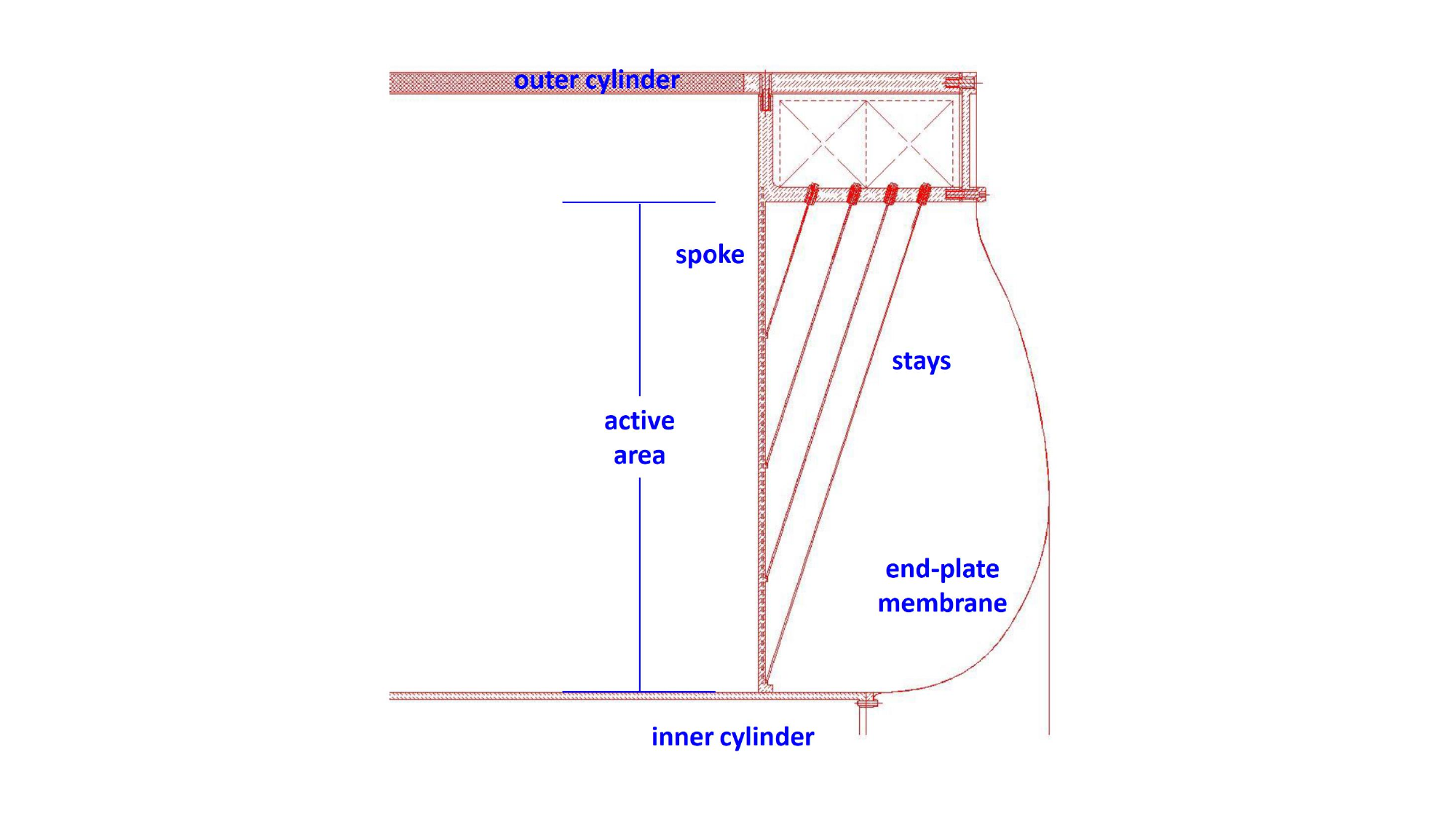}
    \caption{Schematic layout (not in scale) of the tension recovery system of the drift chamber spokes, obtained through a set of 15 stays per each spoke.}
    \label{fig:spokestays}
\end{figure}

The wire support boards and the spacers, both defining the cell geometry, are placed in between each pair of the 24 spokes, 1650\,mm long, at each chamber end. The deformation of the spokes due to the wire load is constrained to be below 200\,$\mu$m by applying a tension recovery system based on a system of 15 stays per spoke (as schematically shown in Fig. \ref{fig:spokestays}). 

\begin{figure}[ht]
    \centering
\includegraphics[width=0.47\linewidth]{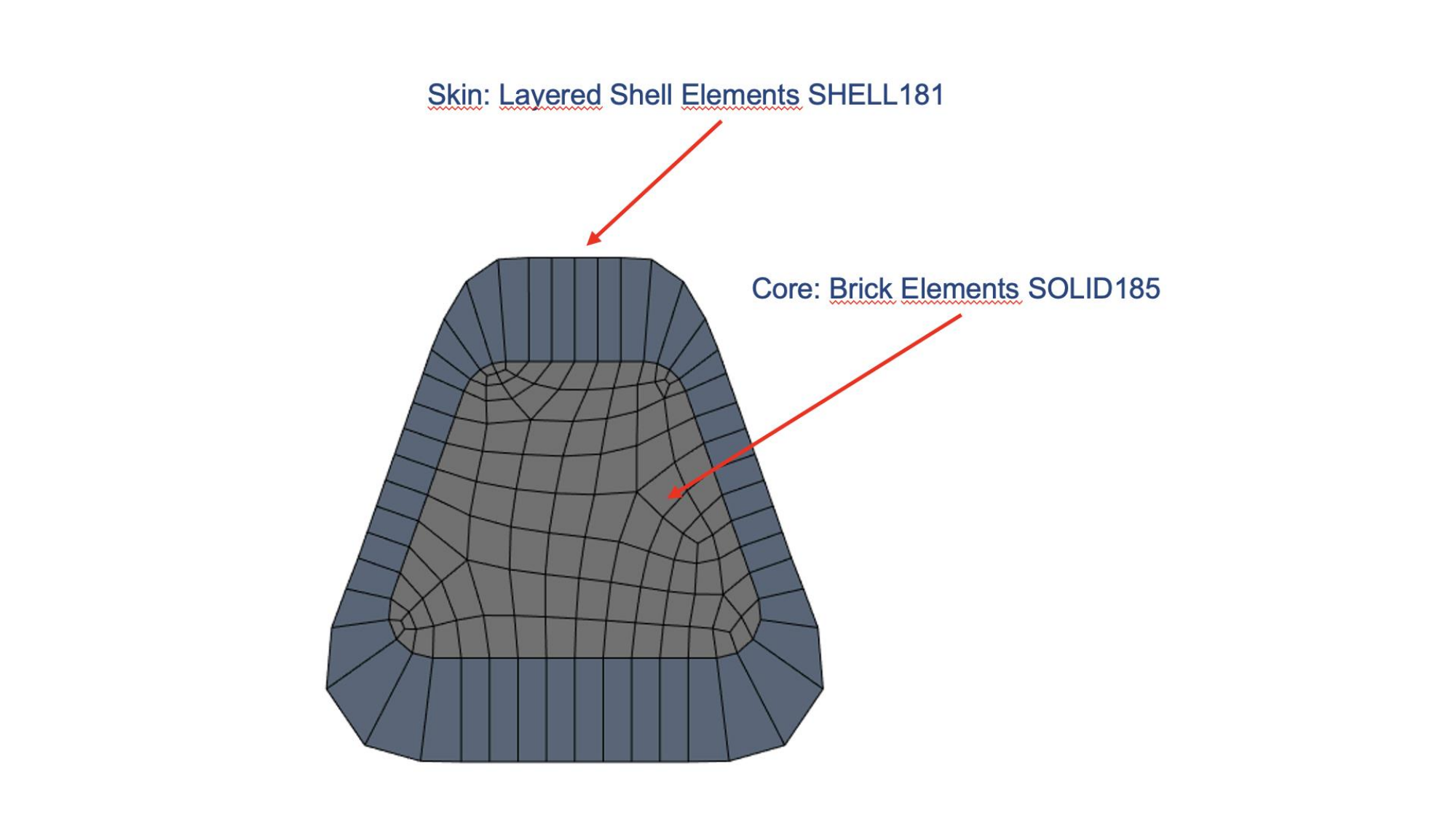}
\includegraphics[width=0.47\linewidth]{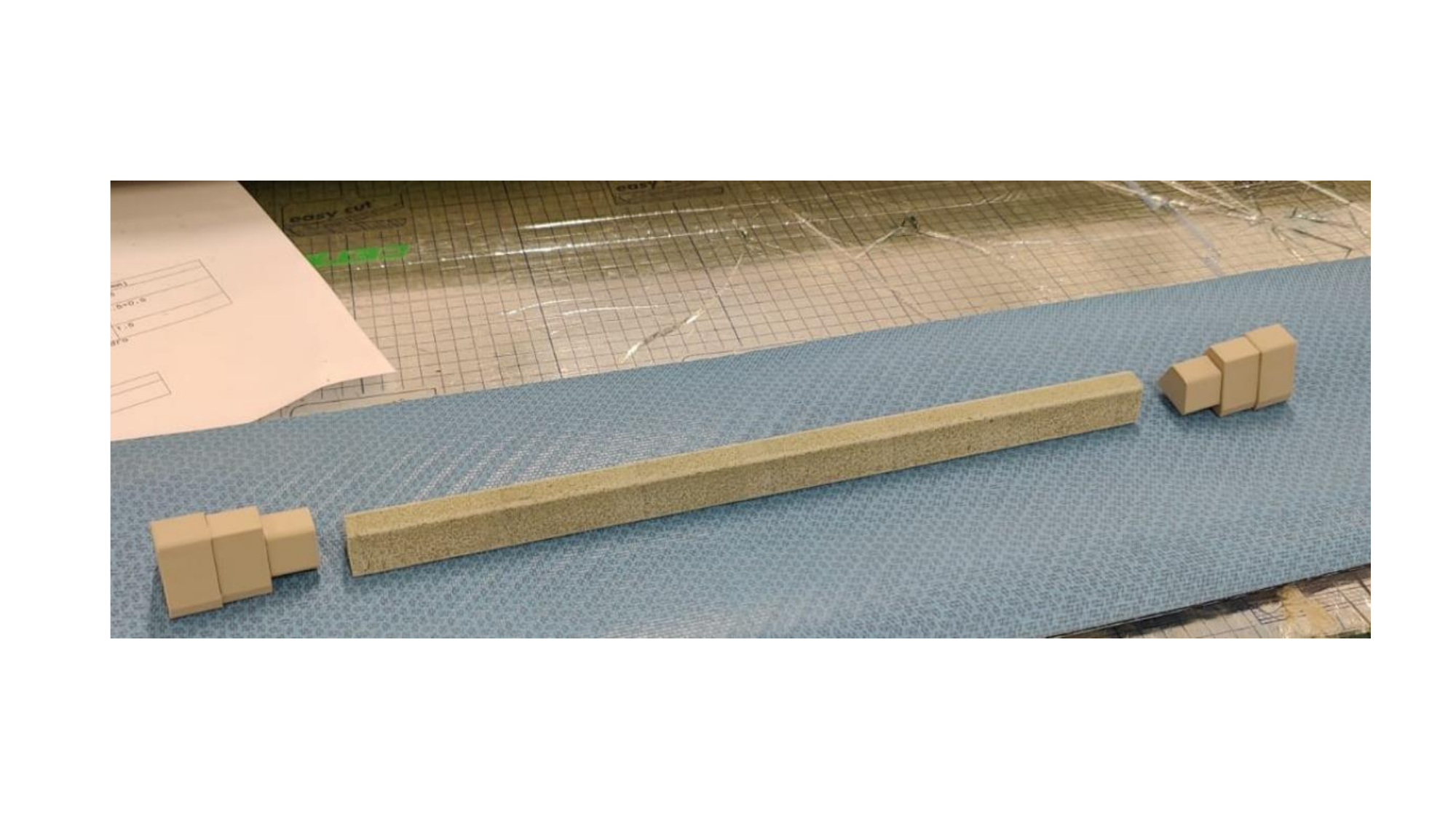}
    \caption{Almost-trapezoidal cross-section of a spoke, showing the core and the external skin in composite materials (left). Fabricated prototype of a spoke, 500\,mm long, together with the insertion ends in PEEK (right).}
    \label{fig:spokesxsect}
\end{figure}

Fig. \ref{fig:spokesxsect} (left) shows the almost-trapezoidal (with major base, minor base and height respectively of 34, 17 and 34\,mm) cross section of the spoke, while in Fig. \ref{fig:spokesxsect} (right) a 500,mm long fabricated prototype of a spoke is shown. The gas envelope surrounding the chamber is separated from the wire cage, and it can freely deform without affecting the internal wire position and tension.

\begin{figure}[ht]
    \centering
\includegraphics[width=0.9\linewidth]{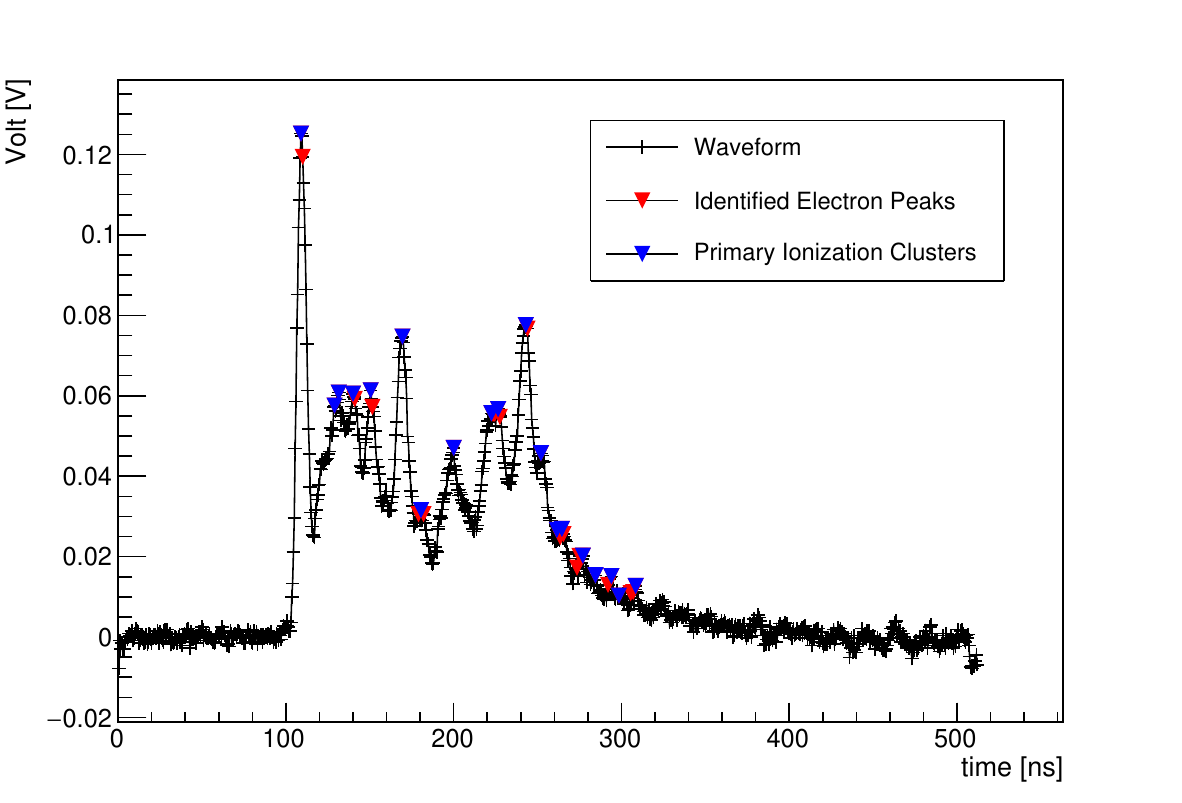}
    \caption{Typical signal waveform collected in Helium-Isobutane gas mixtures. Ionization clusters (marked in blue) can be counted after a clustering algorithm identifies individual ionization electrons (marked in red) belonging to the same cluster. The peaks of individual electrons are recognized by peak-finding algorithms. }
    \label{fig:sigwave}
\end{figure}

The mean number of ionization clusters per cm generated by a minimum ionizing particle in the drift chamber gas mixture is about $12.5$ at atmospheric pressure. By counting the number of ionization events per unit length (dN/dx)~\cite{Clucou} particles can be separated and identified with superior resolution compared to the conventional method based on ionization loss per unit length (dE/dx). Since in Helium-based gas mixtures the signals generated by ionization events and collected at the anode are separated in time from a few nanoseconds to a few tens of nanoseconds (depending on the drift distance from the anode), fast read-out electronics ($\sim$\,GHz sampling) and sophisticated peak-finding algorithms are required to efficiently identify the ionization clusters by acquiring and analyzing the chamber signal waveforms, an example of which is shown in Fig. \ref{fig:sigwave}.

Assuming an empirical parametrization of the dE/dx resolution obtained with the truncated mean technique~\cite{Reak} and a Poisson distribution for the number of the ionization clusters in a 90\% He–10\% iC$_{4}$H$_{10}$ gas mixture at atmospheric pressure, the resolution expected with the cluster counting method is about 2 times better than the usual dE/dx method. Analytical calculations assuming 2 m long tracks predict excellent K/$\pi$ separation for momenta up to $\sim$30 GeV/c except in the $0.85<p<1.05$\,GeV/c range, where separation could be recovered by a timing layer with a modest 100\,ps resolution.

\begin{figure}[ht]
    \centering
\includegraphics[width=0.90\linewidth]{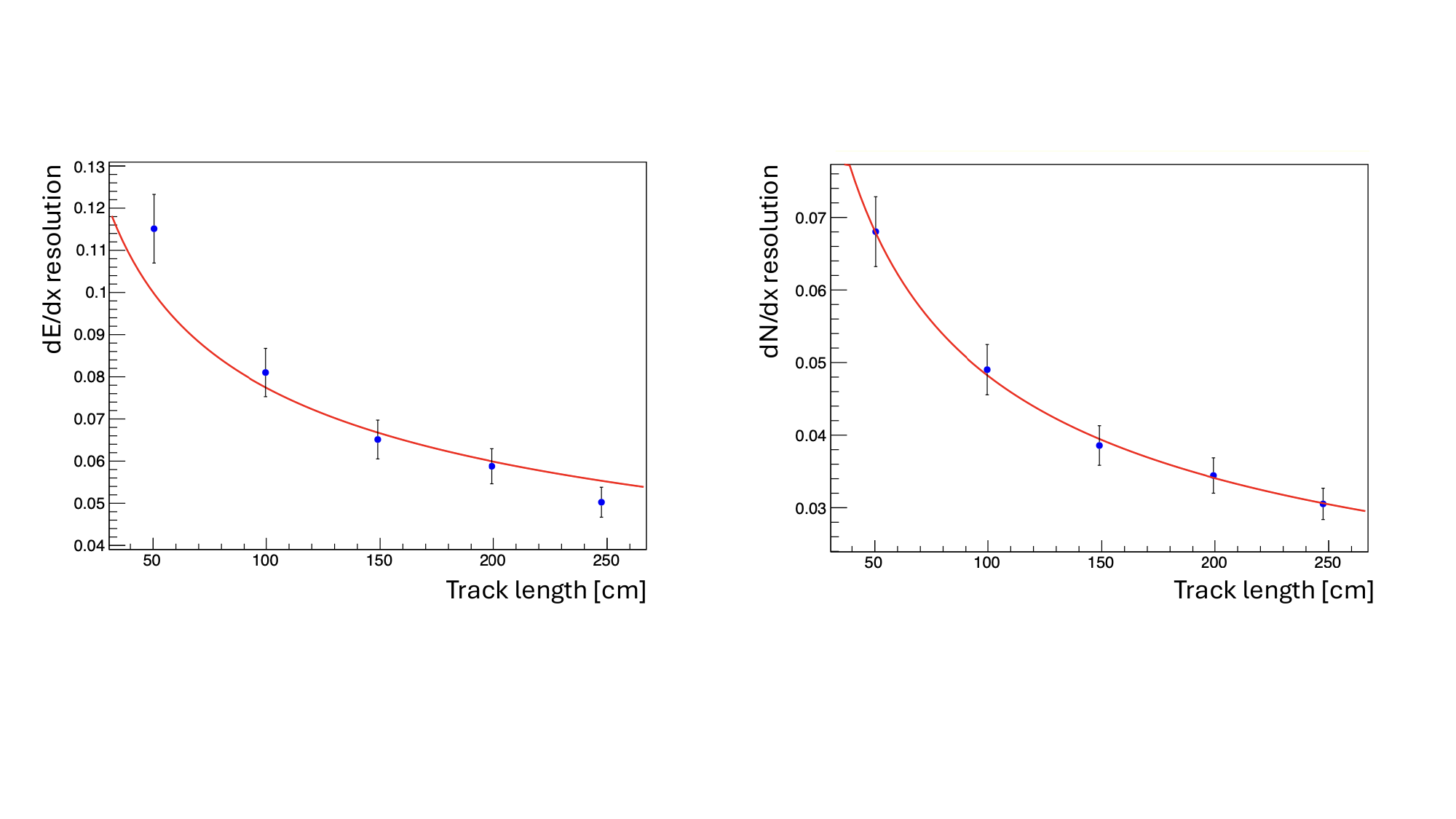}
 \vspace{-15mm}
    \caption{dE/dx resolution as a function of the track length (left). dN/dx resolution as a function of the track length (right). The superimposed red curves refer to a fit assuming a $L^{-0.37}$ (from an empirical parametrization) dependence and a $L^{-0.5}$ (Poissonian) dependence, respectively. An improvement of about a factor of two in the resolution is visible using the dN/dx technique. Details of the analysis are described in the text.}
    \label{fig:testresults}
\end{figure}

In order to verify these expectations, several muon beam tests have been performed at CERN. The performance of drift tubes of different sizes, sense wire diameters and operated with different He-based gas mixtures was studied using muons of different momenta and crossing angles. The collected signal waveforms have been acquired and analyzed with several algorithms, providing results compatible within the uncertainties. 
Fig.\ref{fig:testresults} shows preliminary results from a beam test performed at CERN with high momentum muons (40-180\,GeV/c). The particle separation power with dN/dx outperforms the traditional dE/dx truncated mean technique by about a factor of two. The analysis refers to the same set of tracks, made of the same hits, analyzed both with a 20\% truncated dE/dx mean (left) and with a still not optimized cluster reconstruction algorithm (right). Data analysis of two beam tests performed at CERN with muons in a 1-12\,GeV/c momentum range is ongoing, in order to confirm these results in the region of the relativistic rise of the most probable energy loss.

The cluster counting and timing techniques~\cite{Clucou} can also be exploited to improve the spatial resolution ($\sigma_{xy} < 100 \, \mu m$). A spatial resolution around $110\,\mu$m has been achieved in the 7\,mm cell size MEG2 drift chamber with the same gas mixture and very similar electrostatic configuration~\cite{PERFMEG}.
However, an improved spatial resolution is expected in the IDEA drift chamber, because of longer drift distances and the application of the cluster timing techniques. Indeed, for any given ``first cluster", the cluster timing technique could potentially exploit the drift time distribution of all successive clusters in determining the most probable impact parameter, thus reducing the bias and the average drift distance resolution with respect to those obtained by the method using only the first cluster.

%%%%%%%%%%%%%%%%%%%%%%%%%%%

    \subsection{Silicon wrapper \label{SiWrap}}
    The Silicon Wrapper is the last piece of the IDEA tracking system. It surrounds the drift chamber ($|z| \leq \SI{2.25}{m}$ and $r < \SI{2}{m}$), covering an area of more than $\SI{100}{m^2}$. The Silicon Wrapper provides full hermeticity (with the exception of sensor periphery regions) down to $\cos(\theta) \lesssim 0.989$. 

\begin{figure}[htbp]
    \centering
    \includegraphics[width=0.5\linewidth]{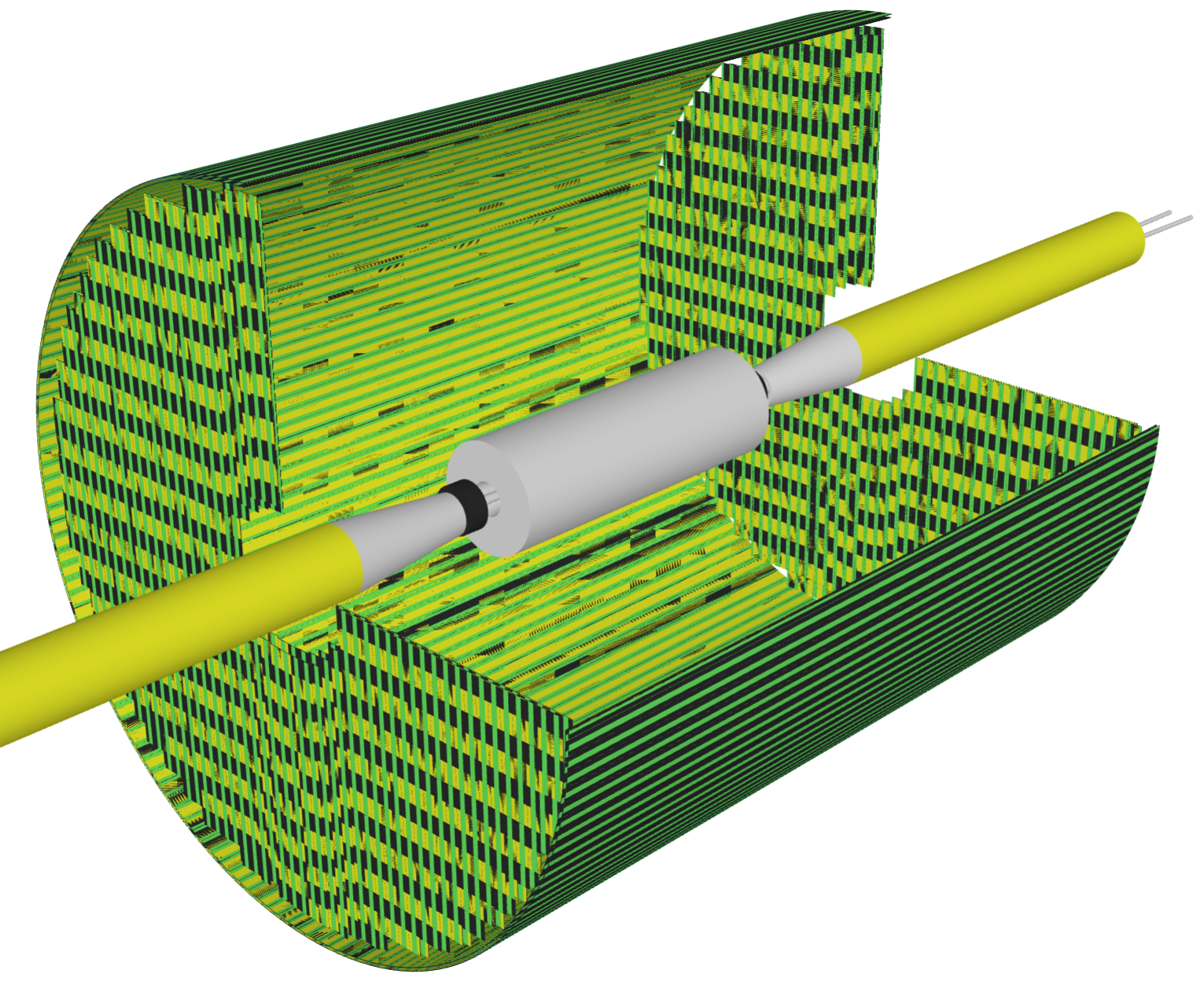}
	\caption{Visualisation of the Silicon Wrapper implementation in DD4hep, with the compensating solenoids, beam pipe and the vertex detector in the middle (drift chamber not shown).}
	\label{fig:siwr_vis}
\end{figure}

\begin{table}[ht] 
      \caption{Main parameters of the current IDEA Silicon Wrapper layout, assuming ATLASpix3-sized sensors ($42.2\times \SI{40.2}{\centi\meter\squared}$). The numbers for the disks are per side.}
    \label{tab:siwr}
    \centering
    \small
    \begin{tabular}{|c|c|c|c|c|c|}\hline
         \bf Layer & \bf Radius [m]  & \bf $z$ [m] & \bf Number of sensors & \bf Silicon area [\SI{}{\meter\squared}] & \bf Coverage \\ \hline\hline
         Barrel 1 & 2.04 & $\left|z\right| < 2.4$ & $\mathcal{O}(20,000)$ & $\approx 32$ & $\left| cos(\theta)\right| < 0.762$ \\ \hline
         Barrel 2 & 2.08 & $\left|z\right| < 2.4$ & $\mathcal{O}(20,000)$ & $\approx 32$ & $\left| cos(\theta)\right| < 0.756$ \\ \hline
         Disk 1 & 0.35 -- 2.04 & $\pm 2.30$ & $\mathcal{O}(7,500)$ & $\approx 12\times 2$ & $0.748 < \left| cos(\theta)\right| < 0.989$ \\ \hline
         Disk 2 & 0.35 -- 2.04 & $\pm 2.32$ & $\mathcal{O}(7,500)$ & $\approx 12\times 2$ & $0.751 < \left| cos(\theta)\right| < 0.989$ \\ \hline \hline
         \bf Total & & & $\mathcal{O}(70,000)$ & $\approx 112$ &  \\ \hline
    \end{tabular}
\end{table}

The sensor technology for the Silicon Wrapper has not been fixed yet. Candidates are, for example, two layers of silicon microstrip detectors or one layer of MAPS or low-gain avalanche diodes (LGADs). In case the LGAD option is chosen, the goal would be to reach a time resolution per track of $\sim \SI{100}{ps}$ for time-of-flight particle identification which, together with the cluster counting from the drift chamber, ensures a $\mathrm{K-\pi}$ separation power of more than three sigma up to momenta of $\sim$30 GeV/c. The time-of-flight measurement will also be crucial for long-lived particle searches at the FCC-ee~\cite{Verhaaren_2022}. The other options do not provide timing within the tracking system; in this case timing would be supplied by a LYSO layer in front of the EM calorimeter.

The requirement on the spatial resolution of the Silicon Wrapper is $\mathcal{O}(\SI{10}{\micro\meter})$. This enables enables a precise momentum measurement, especially for high-momenta particles and in the forward region where the drift chamber only provides a limited number of hits. The reasoning behind this small spatial resolution in $\theta$ is to allow the Silicon Wrapper to act as a precise and stable ruler for the detector acceptance definition ($<\mathcal{O}(\SI{10}{\micro\radian})$, which is critical for cross-section measurements. 

\begin{figure}[htbp]
	\centering
	\subfloat[Barrel]{
		\includegraphics[width=0.5\linewidth]{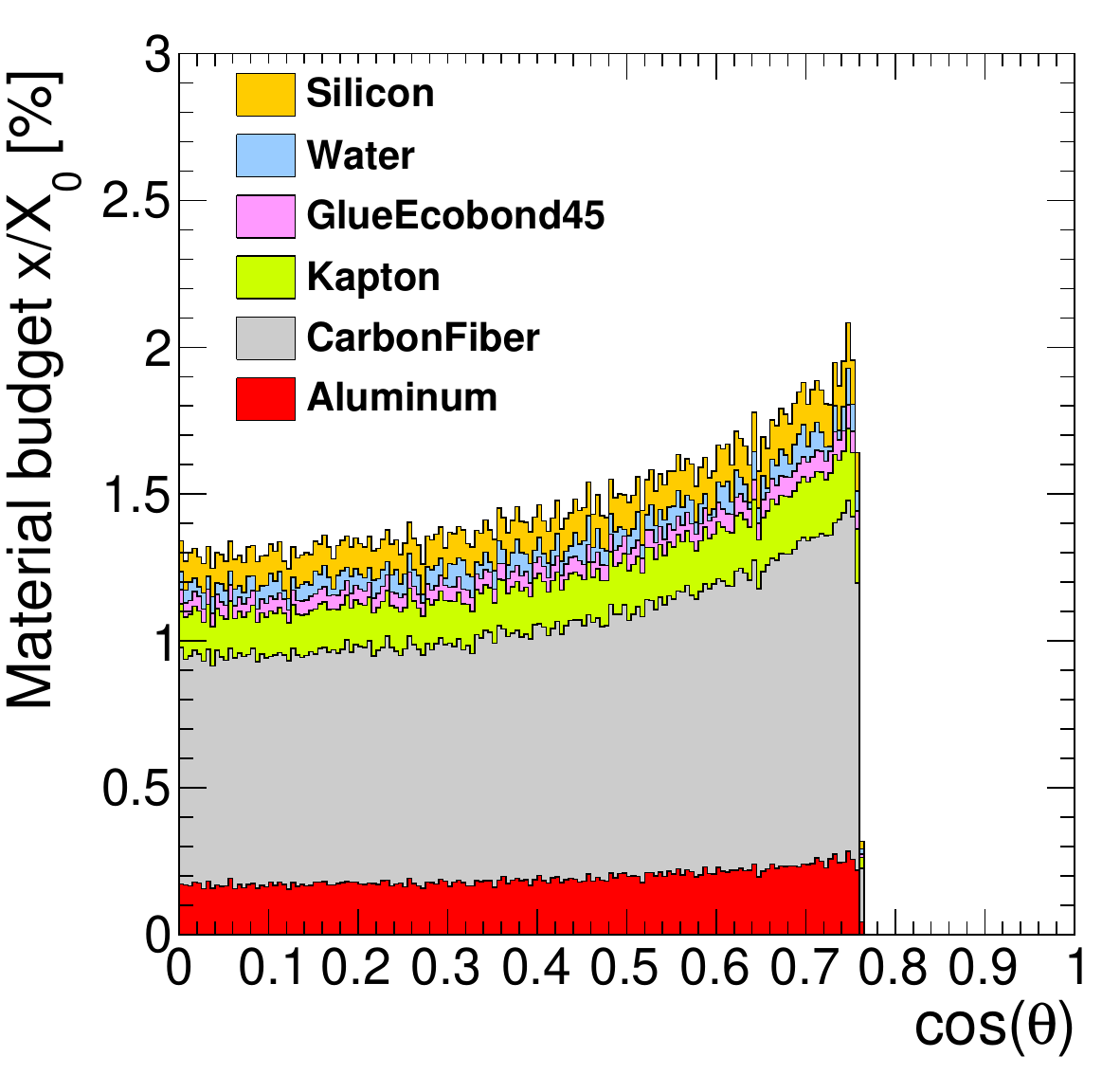}
	}
	\subfloat[Disks]{
		\includegraphics[width=0.5\linewidth]{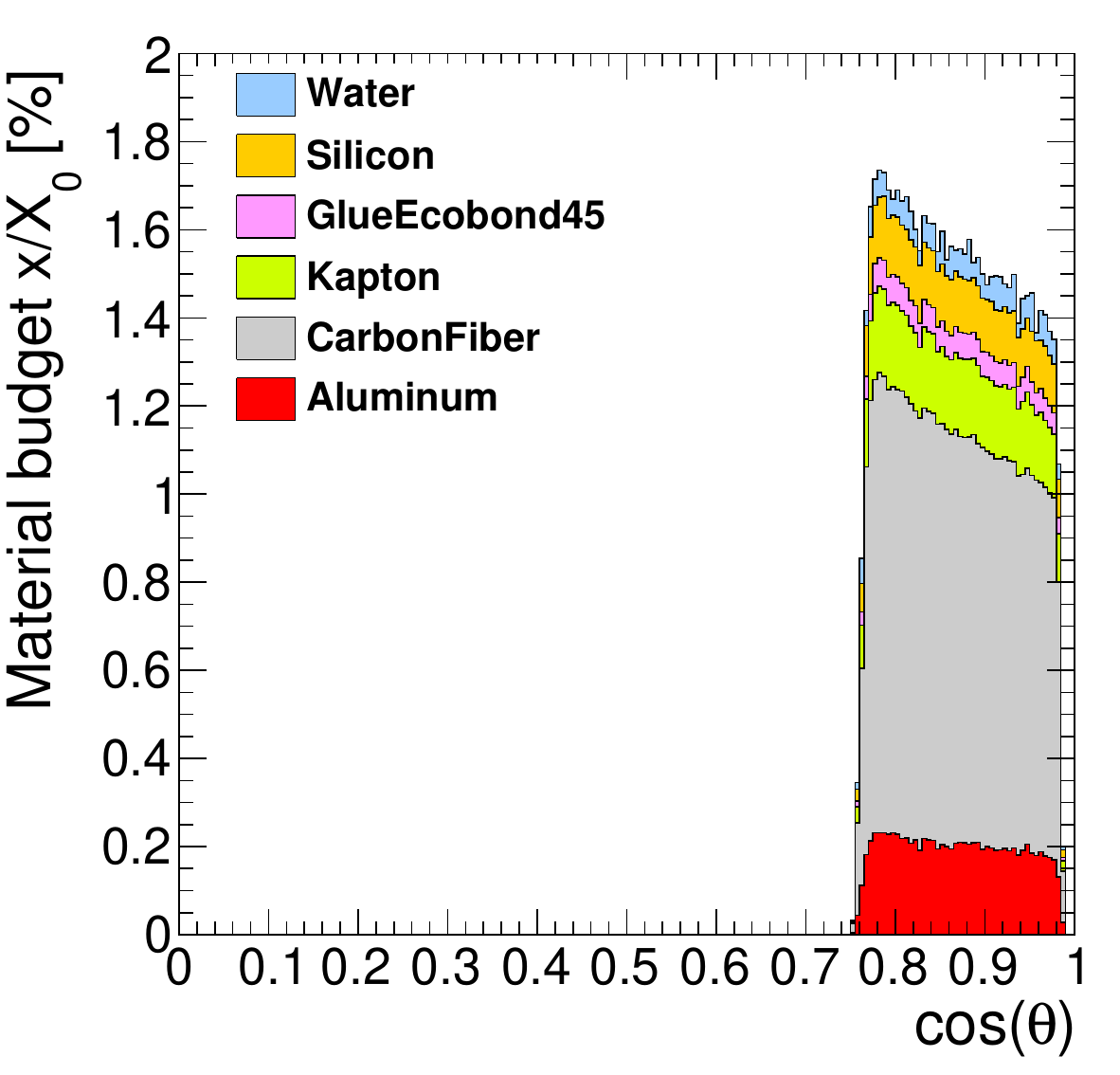}
	}
	\caption{Material budget in the DD4hep Silicon Wrapper implementation for the two barrel layers (a) and the two disks per side (b).}
	\label{fig:SiWr_material_budget}
\end{figure}

The current implementation of the Silicon Wrapper is based on MAPS. It is shown in Fig.~\ref{fig:siwr_vis} and consists of two barrel layers and two disks per side that overlap at their interface. Table~\ref{tab:siwr} lists the properties of the current layout. The barrel layers (disks) are made up of staves (modules) with sensors on both sides, flex circuits on top, and a support and cooling structure in the middle. The sensors have a thickness of $\SI{50}{\micro\meter}$. Each stave/module has a \SI{1.4}{mm} carbon fiber support, the same cooling pipes, and four readout flex circuits as for the outer vertex detector described in Sec.~\ref{Vtx}. The stave/module is inspired by the ATLASpix3 tile proposed for the CEPC tracker \cite{cepc_tracker1,cepc_tracker2}, and the global layout of the disks follows the one of the CMS endcap timing layer \cite{cms_endcap_timing_optimisation}. The current implementation of the Silicon Wrapper features a material budget of $\approx 0.7\%$ of $X_0$ per barrel layer/disk as shown in Fig.~\ref{fig:SiWr_material_budget}.

The Silicon Wrapper will feature a number of challenges that need to be addressed in the next years. The area to cover (\SI{112}{\meter\squared} in the current implementation) is very large and should be minimised to limit the number of sensors needed. An updated layout is being developed that targets only one barrel layer and disk per side. In the barrel, long longerons holding the sensors will be attached to support rings going around the drift chamber. For the disks, a design with two D-shaped half-endcaps per side will be pursued so that the Silicon Wrapper can be installed after insertion of the support tube holding the beam pipe, vertex detector, and luminosity calorimeters. These half-endcaps will be mounted onto the sides of the drift chamber, but it is necessary to ensure that all the drift chamber services can be guided out of the detector without creating a crack in the wrapper acceptance. 

Given the large area and high granularity, the aggregation of the signals, powering, and cooling will need to be addressed in an early design stage. 

%    \subsection{Tracker performance\label{TrkPerf}}

\clearpage\newpage
\section{Idea Calorimetry System \label{CalSys}}
To fully exploit the physics landscape of the FCC-ee, the performances of the calorimetry  must be improved in several aspects compared to the solutions adopted so far. Both energy and angular resolution must be optimized for different physics cases. For IDEA, we chose to aim for a detector with excellent energy resolution for both electromagnetic and hadronic showers, and high granularity, to enhance the final-state particle identification performance. We focused on dual-readout calorimetry, primarily to be able to resolve, in the 2-jet final states, events generated by W bosons, instead of Z bosons or Higgs bosons. The hadronic showers generated in the three cases are very hard to separate by current detectors. An energy resolution (really an invariant mass resolution) for hadronic showers of about 3\% at 100 GeV is required. The major problem is due to the presence, in hadronic showers, of two components, an electromagnetic one (electrons and photons) and a hadronic one, which produce different signals for the same energy loss. Each hadronic shower develops a different electromagnetic fraction, thus contributing to the energy shower measurement error.

Dual-readout (DR) calorimetry addresses this problem by sampling the showers with two processes that are differently sensitive to the passage of non-relativistic particles (essentially the hadronic component) and relativistic particles (essentially the electromagnetic component). A scintillating material produces a signal in both cases, while a material that emits Cherenkov light is activated only by relativistic particles. From the ratio between the two signals, it is possible to reconstruct the electromagnetic fraction of each shower, event by event, and cancel the effects of its fluctuation.

A simple way to implement a dual-readout calorimeter is to use a scintillating material (for example, scintillating fibers) alternating with a material that emits Cherenkov light (for example, clear fibers). Alternatively, one can try to isolate and measure the Cherenkov light that is emitted in a scintillating material.

Both approaches are proposed for IDEA calorimetry. A homogeneous electromagnetic crystal calorimeter, where the aim is to separate scintillating light from Cherenkov light, is followed by a fiber-sampling calorimeter, where scintillating fibers and clear fibers are placed on alternating lines and measured separately.

In order to assess the performance, several R\&D projects, for both crystal and fiber-sampling DR calorimeters, are being carried out within the Work Package 3 (on optical calorimeters) of the DRD-on-Calorimetry (DRD6) Collaboration~\cite{DRD6-proposal}.

\clearpage\newpage
\section{Crystal calorimeter \label{CryCal}}

The IDEA Dual-Readout Crystal calorimeter section is designed to achieve an electromagnetic energy resolution better than $3\%/\sqrt{E}\oplus1\%$ and includes the simultaneous readout of scintillation and Cherenkov photons which, integrated with the DR fiber hadronic calorimeter section, can provide a hadronic energy resolution of about $30\%/\sqrt{E}\oplus 3\%$ \cite{Lucchini_2020, Lucchini_2022}.

Another key feature compared to state-of-the-art homogeneous electromagnetic calorimeters is the increased transverse granularity and longitudinal segmentation which constitute powerful handles for particle identification and global event reconstruction based on the particle flow approach. The calorimeter can provide a time resolution at the level of 30~ps for electromagnetic showers with energy above 30 GeV. An additional layer consisting of a 6~mm thick LYSO crystal grid can be added in front of the calorimeter to provide cost-effective time resolution at the level of 20~ps for all charged particles, as described in \cite{Lucchini_2020}, and exploiting a technology similar to that of the CMS MTD barrel \cite{CMS_MTD_TDR,Addesa_2024}.
The baseline technological choices and calorimeter layout are presented below with a short overview of ongoing R\&D and prototyping efforts.

\subsection{Detector layout and technology}
\begin{figure}[!htbp]
\centering
\includegraphics[width=0.8\textwidth]{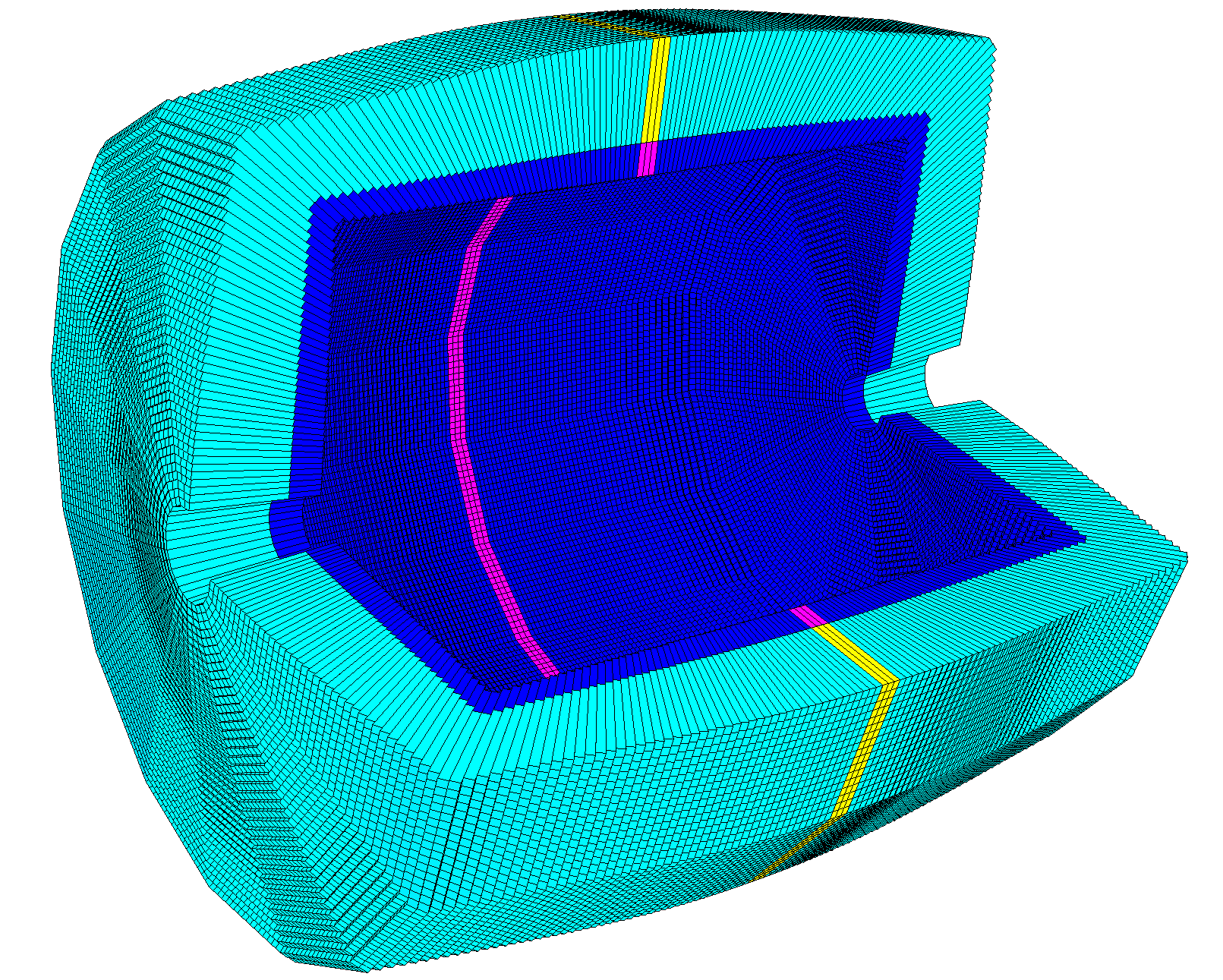}
\caption{3D cutout view of the dual-readout crystal calorimeter showing (dark blue) front crystal segments and (light blue) rear crystal segments. A nominal crystal front face size of $50\times50$~mm$^2$ is used for visibility. Three non-projective spacer rings, used to mitigate projective cracks by displacing either half of the detector in $z$, are shown in purple (front segments) and yellow (rear segments).}
\label{fig:crystal_geometry_3d}
\end{figure}

\begin{figure}[!htbp]
\centering
\includegraphics[width=0.99\textwidth]{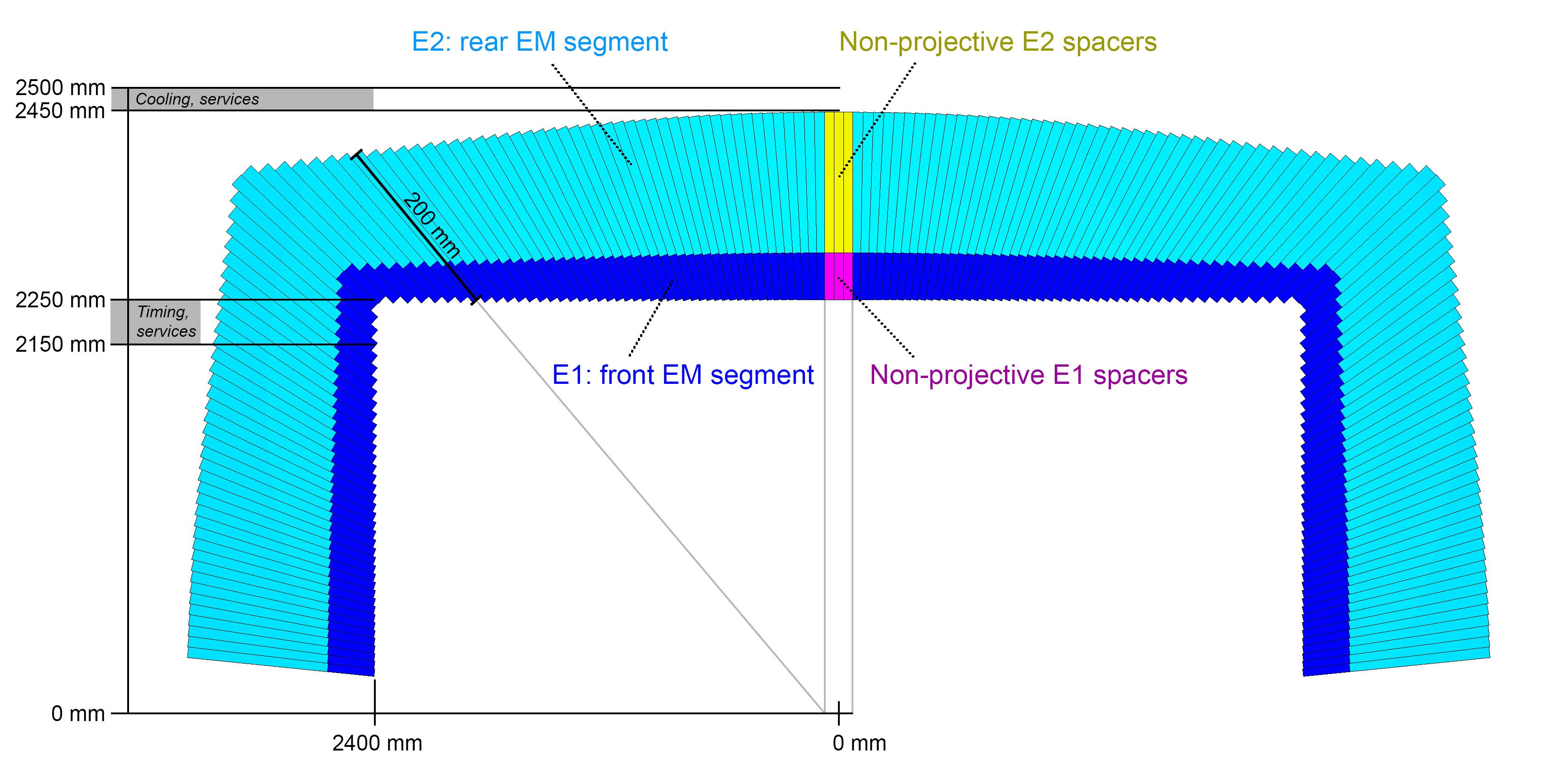}
\caption{View of a global $\phi$ slice of the crystal calorimeter showing projective front and rear electromagnetic segments (E1, E2) displaced in $z$ by non-projective segments (yellow, purple) to mitigate projective gaps. A nominal crystal front face size of $50\times50$~mm$^2$ is used in this image for visibility.}
\label{fig:crystal_geometry_sim_zoom}
\end{figure}

The crystal calorimeter is located between the tracking system and the solenoid with a radial envelope of 2150-2500~mm, and is divided into a central barrel section and two endcaps. The $\phi$ profile of the barrel is segmented as a regular polyhedron by a fixed number of global $\phi$ segments to allow for flat surfaces upon which a precision timing layer may later be instrumented. All crystal towers are initially oriented as projective trapezoids with respect to the interaction point; to mitigate projective cracks, the detector is then split into two halves and a displacement in $z$ is applied with the addition of non-projective spacer rings as shown in Fig.~\ref{fig:crystal_geometry_3d} and Fig.~\ref{fig:crystal_geometry_sim_zoom}. Each crystal tower has nominal front face dimensions of $10\times10$~mm$^2$ and is composed of two longitudinal segments of $6~X_0$ (front, E1) and $16~X_0$ (rear, E2) for a total depth of $22~X_0$.

\begin{figure}[!htbp]
\centering
\includegraphics[width=0.39\textwidth]{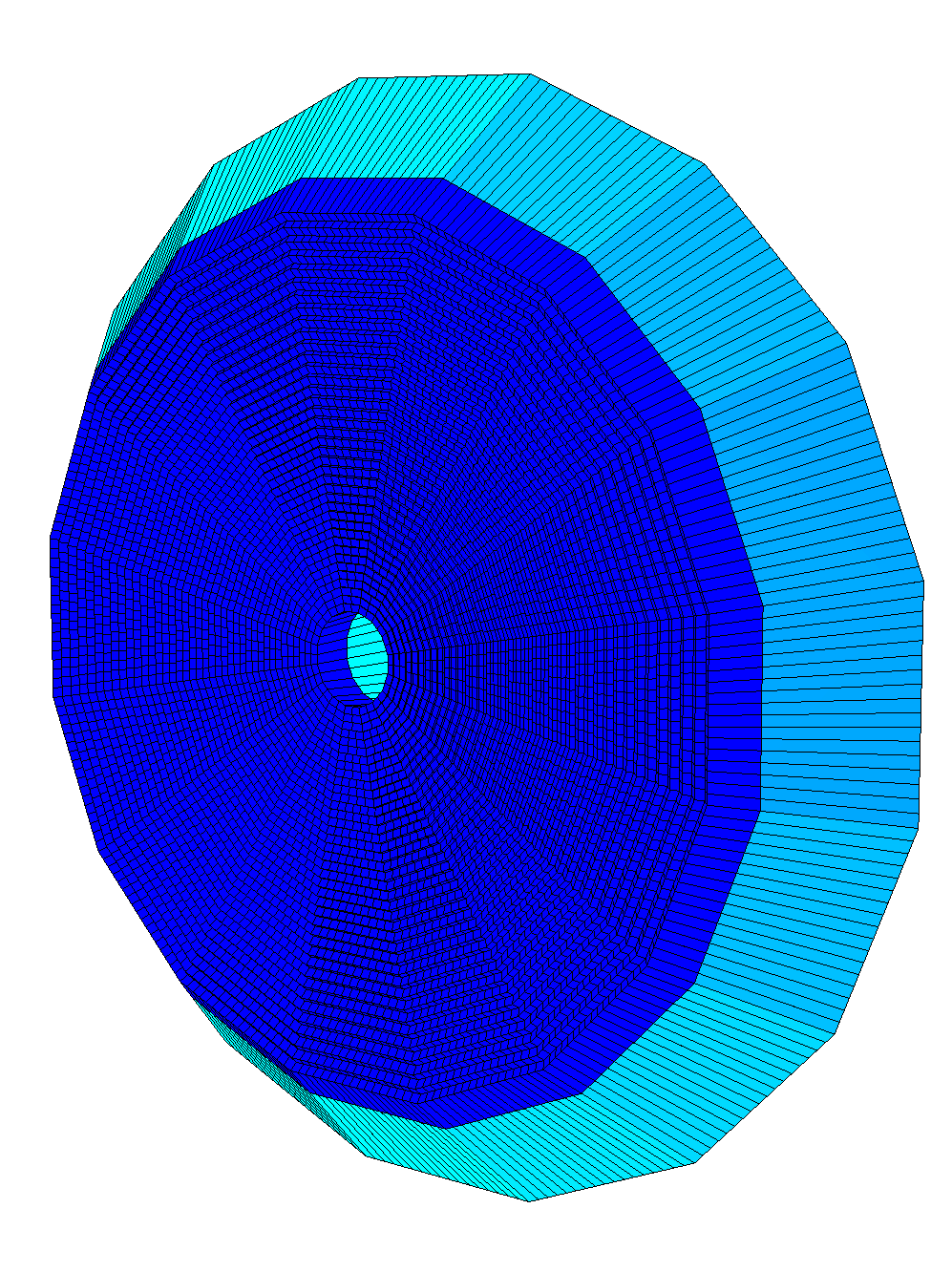}
\includegraphics[width=0.49\textwidth]{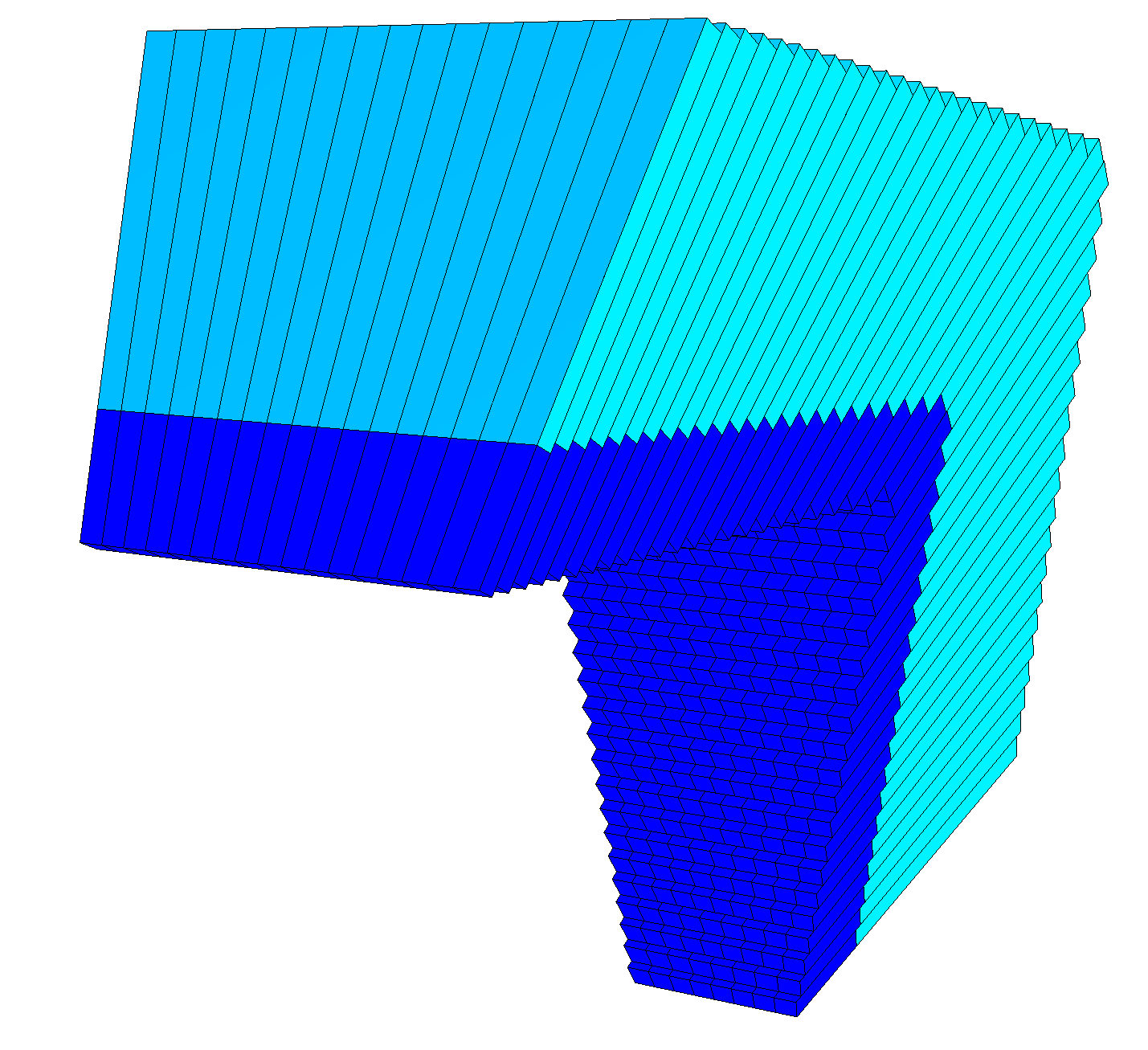}
\caption{(left) Endcap view of the crystal calorimeter showing the polyhedral structure. (right) Corner view of the barrel-endcap junction showing the varying number of crystals as a function of radius to maintain the nominal crystal face dimensions.}
\label{fig:scepcal_endcap}
\end{figure}

The barrel has a $z$-length of 4.8~m and the center point of the crystal front faces are located at a radius of 2250~mm. The endcaps connect continuously with the barrel and are therefore also segmented in $\phi$ with an outermost rear crystal radius of $\approx2386$\,mm and nominal pseudorapidity of 2.60 designed to coincide with the Silicon Wrapper endcap disks, corresponding to an inner radius of $\approx$ 360\,mm. The number of crystals per global $\phi$ segment in the endcap is adjusted with the radius as shown in Fig.~\ref{fig:scepcal_endcap} to maintain the nominal crystal face dimensions.

The geometry construction routine in the Key4hep framework \cite{key4hep} is configured to generate the individual crystal volumes dynamically from the input dimensions as shown in Tables~\ref{tab:scepcal_input_parameters} and~\ref{tab:scepcal_secondary_parameters}, therefore allowing the crystal dimensions to be perturbed easily for optimization studies. 
The total count of crystal towers and readout channels are reported in Table~\ref{tab:scepcal_final_parameters}. More details on the differentiable geometry and simulation implementation in the Key4hep framework are provided in \cite{chung2024}.

\begin{table*}[!b]
\centering
\caption{Input parameters for parameterized geometry construction.}
\label{tab:scepcal_input_parameters}
\begin{tabular}{lll}
\hline
Description                                         & Variable Name     & Value  \\\hline
Half Z-extent of the barrel                         & $Z_\text{B}$      & 2.40\,m \\
Inner radius of the barrel                          & $R_\text{inner}$  & 2.25\,m \\
Global number of phi segments                       & $N_\Phi$          & 128 \\
Nominal square face width of the front crystals     & $C_\text{fw}$     & 10\,mm \\
Front crystal length                                & $F_\text{dz}$     & 50\,mm \\
Rear crystal length                                 & $R_\text{dz}$     & 150\,mm \\
Number of projective spacer rings                   & $N_\text{Proj}$   & 3 \\
SiPM module thickness                               & $S_\text{th}$     & 0.5\,mm \\\hline
\end{tabular}
\end{table*}

\begin{table*}[t]
\centering
\caption{Secondary parameters calculated from input parameters.}
\label{tab:scepcal_secondary_parameters}
\begin{tabular}{lll}
\hline
Description                                                     & Variable Name  & Formula  \\\hline
Angular size of a single phi segment                            & $d\Phi$        & 2$\pi$/ $N_\Phi$ \\
Angular size of barrel region                                   & $\Theta_B$     & atan($Z_\text{B}/R_\text{inner}$) \\
Angular size of endcap region                                   & $\Theta_E$     & atan($R_\text{inner}/Z_\text{B}$) \\
Number of barrel segments in $\theta$                           & $N\theta_B$    & floor($2~Z_\text{B}/C_\text{fw}$) \\
Number of endcap segments in $\theta$                           & $N\theta_E$    & floor($R_\text{inner} / C_\text{fw}$) \\
Angular size of a single barrel segment in $\theta$             & $d\theta_B$    & $(\pi-2~\Theta_E) / N\theta_B$ \\
Angular size of a single endcap segment in $\theta$             & $d\theta_E$    & $\Theta_E/N\theta_E$ \\
Number of barrel segments in $\phi$ in a single phi segment     & $N\phi_B$      & floor($2\pi R_\text{inner}/(N_\Phi C_\text{fw})$) \\
Number of endcap segments in $\phi$ in a single phi segment$^*$ & $N\phi_E^*$    & floor($2\pi R_\text{inner}^*/(N_\Phi C_\text{fw})$) \\
Angular size of barrel segments in $\phi$                       & $d\phi_B$      & $d\Phi / N\phi_B$ \\
Angular size of endcap segments in $\phi$                       & $d\phi_E^*$    & $d\Phi / N\phi_E^*$ \\\hline
\end{tabular}
\end{table*}

\begin{table*}[t]
\centering
\caption{Final crystal/readout counts and endcap dimensions from the parameterized geometry construction using input values from Table~\ref{tab:scepcal_input_parameters}.}
\label{tab:scepcal_final_parameters}
\begin{tabular}{lr}
\hline
Quantity                                                               & Value  \\\hline
Number of total barrel crystals                                        & 1,360,128 \\
Number of barrel front crystal readout channels (1 SiPM per crystal)   & 680,064 \\
Number of barrel rear crystal readout channels (2 SiPMs per crystal)   & 1,360,128 \\
Number of total endcap crystals                                        & 251,136 \\
Number of endcap front crystal readout channels (1 SiPM per crystal)   & 125,568 \\
Number of endcap rear crystal readout channels (2 SiPMs per crystal)   & 251,136 \\
Endcap innermost radius (center point of innermost front crystal face) & 360.2\,mm \\
Endcap outermost radius (center point of outermost rear crystal face)  & 2386.8\,mm \\
Endcap maximum pseudorapidity                                          & 2.595 \\\hline
\end{tabular}
\end{table*}

The two-layer segmentation represents a tool for particle identification based on the longitudinal energy deposit pattern with the front segment measuring the e.m. shower before it reaches its maximum lateral development thus effectively reducing the Molière radius for separation of neighboring e.m. showers. At the same time, limiting the segmentation to two layers allows for no dead material inside the detector volume and maintains the calorimeter to be as homogeneous as possible in proximity to the shower maximum, while simplifying the signal readout and integration aspects. 

\begin{figure}[!tbp]        
    \includegraphics[width=0.99\linewidth]{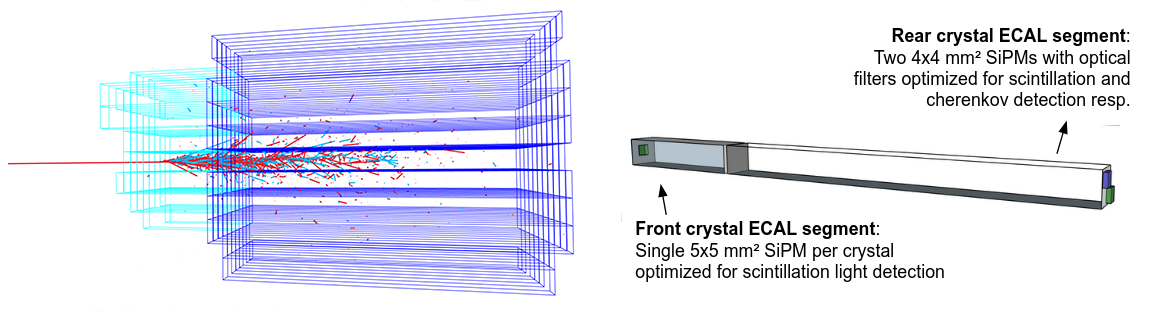}
    \caption{Simulated EM shower display from a 10\,GeV electron em shower (left) and schematic drawing of a segmented crystal calorimeter prototype cell and module with SiPM based dual-readout (right). }
    \label{fig:maxic_layout}
\end{figure}

The light is read out from the crystals using Silicon PhotoMultipliers (SiPMs) which are located on the front face of the front crystal segment (E1) and at the rear face of the rear segment (E2), where the respective front-end boards, cooling system and other services will also be located. One SiPM (S1) with active area of about $6\times 6$~mm$^2$ is glued on the front crystal segmented and two SiPMs (S2 and C2) of the same size are glued to the rear crystal as shown in Fig.~\ref{fig:maxic_layout}. The protective window of one of the two rear SiPMs is replaced with a thin optical filter (thickness $\sim0.1~\rm mm$) designed to filter out optical photons with wavelength around the scintillation maximum while letting pass other optical photons originating from Cherenkov radiation.
In this way a dual readout of the light, i.e. a scintillating (S) signal and a Cherenkov (C) signal, is performed on the rear crystal, which is then used in energy corrections for early showering hadrons.
It was demonstrated elsewhere \cite{Lucchini_2020} that the dual-readout in the front crystal segment is not required given the negligible probability for hadrons to start showering in the first $6~X_0$ (i.e. about $0.25\lambda_I$).

%crystal choice
The baseline crystal choice consists of lead tungstate crystals, PbWO$_4$, which have a long history in previous calorimeters at high energy physics colliders \cite{CMS_ECAL_TDR} due to their excellent calorimetric properties (mainly a small Moliere radius, $R_M$, of 2.0~cm and a short radiation length, $X_0$, of 0.89~cm). Other crystal candidates under study are BGO and BSO crystals which, given their higher light output, slower decay time and different emission peak of scintillation light can offer a complementary approach for the development of such a calorimeter. More details on the key parameters of these crystals are given in Table~\ref{tab:crystal_candidates}.

\begin{table}[!tbp]
\centering
\caption{Properties of scintillating crystal candidates.}
\vspace{0.2cm}
\begin{tabular}{l|c|c|c}
\hline
            	   	           &  PWO      & BGO          & BSO   \\
\hline\hline
Density [g/cm$^3$]            &  8.3      & 7.1          & 6.8  \\ \hline
$R_M$ [cm]                    & 2.00      & 2.23         & 2.33 \\ 
$X_0$ [cm]                    & 0.89      & 1.12         & 1.15 \\ \hline
$\lambda_I$ [cm]              & 20.7      & 22.7         & 23.4 \\ \hline
Refractive index at 420 nm    & 2.20      & 2.15         & 2.68\\ \hline
Light yield [ph/MeV]          & 130       & 7500         & 1500\\ \hline
Scintillation decay time [ns] & 10        & 300          & 100\\ \hline
Emission peak [nm]            & 420       & 480          & 470\\ \hline 
\end{tabular} 
\label{tab:crystal_candidates}
\end{table}

\begin{figure}[!tbp]
    \centering
        \includegraphics[width=0.495\linewidth]{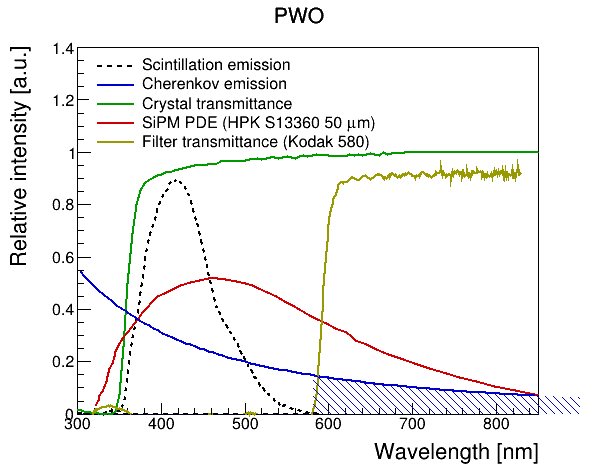}
        \includegraphics[width=0.495\linewidth]{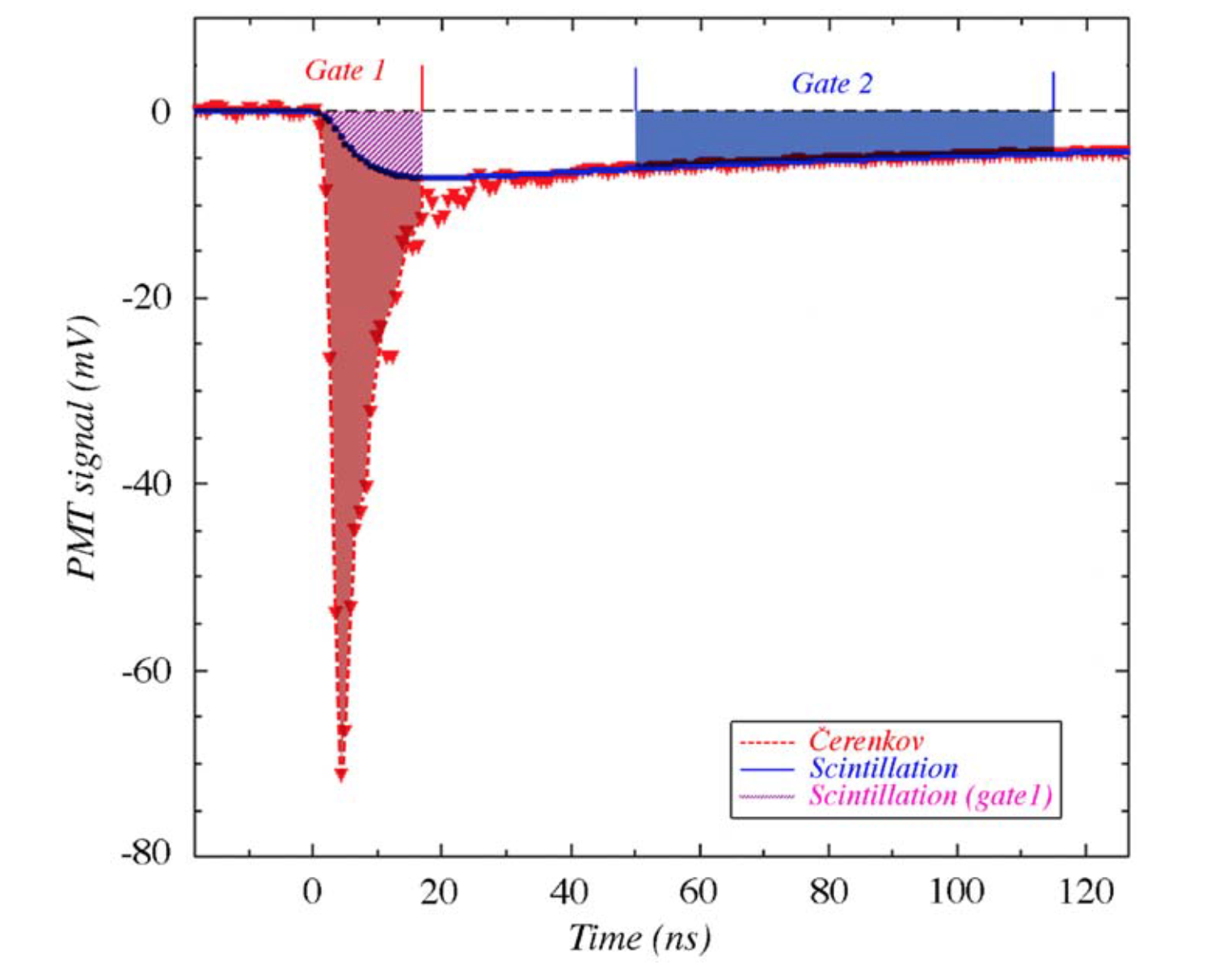}
        \caption{Strategy for implementation of dual-readout in PWO crystals (left), which consists of filtering out scintillation photons based on wavelength discrimination, and in BGO crystals (right)  which consists in isolating Cherenkov light based on the time distrubution of optical photons (prompt for Cherenkov and with a long decay for scintillation). }%\textcolor{red}{Update figure right with recent test beam plot.}}
        \label{fig:dr_implementation}
\end{figure}

Two complementary strategies for implementation of dual-readout are being explored:
\begin{itemize}
    \item The former which is optimal for PWO crystals is mostly wavelength-based, and consists in filtering out at least 99\% of the scintillation photons with a high-pass filter with cut-off wavelength around 580\,nm as shown in the left plot of Fig.\ref{fig:dr_implementation}. Based on preliminary calculations and crystal measurements it is expected that about 50 photoelectrons from Cherenkov radiation are detected per GeV of deposited energy with a contamination from scintillation photons smaller than 15\%  \cite{Lucchini_2020}. In this way two separate signals proportional to scintillation and Cherenkov light are measured integrating the charge of two independent SiPMs.
    \item The latter method also leverages the difference in the emission time profile between scintillation and Cherenkov photons and is more suited for BGO and BSO crystals. For these crystals, in fact, the number of scintillation photons is about 60 and 10 times higher than in PWO, respectively, and more difficult to filter efficiently given their emission being more shifted to 50~nm longer wavelengths. On the other hand, the scintillation decay time constants of BGO and BSO crystals are respectively of about 300 and 100\,ns (thus a factor 10 slower than PWO) making it possible to distinguish them from the Cherenkov photons which are instead emitted promptly with the e.m shower development. An example of measuring the S and C signal simultaneously from the same SiPM based on pulse shape analysis is shown in the right plot of Fig.~\ref{fig:dr_implementation}. Preliminary calculations and test beam results~\cite{Hirosky:2024anp} show that the target light yield $>50$\,phe/GeV from Cherenkov radiation can be achieved.
\end{itemize}

%SiPM choice
SiPM technology is in continuous evolution with a trend to develop devices with larger active area, smaller cell size and in particular with enhanced sensitivity in the near infrared region ($>780$~nm) driven by LIDAR applications \cite{Yang:2022xau}.
To achieve a light yield of 50 phe/GeV with Cherenkov photons with wavelength in the range 550-900~nm an effective PDE (weighed over this range) of 15\% is required and can be achieved with state-of-the-art SiPMs such as the S13360-6050 HPK model \cite{hamamatsu-sipms}.

For what concerns the detection of scintillation light the main challenge is the dynamic range intrinsically linked to the total number of cells of the SiPM. To be able to measure energy deposits from minimum ionizing particles (MIPs) up to 200\,GeV electrons (which can deposit about half of their energy in a single crystal rear segment), assuming a light output of 2000~phe/GeV it is required to detect without saturation up to $2\times10^5$ photoelectrons, thus requiring SiPMs with a number of cells at least two times larger. This is within reach using $6\times6$\,mm$^2$ active area SiPMs with 10~$\rm \mu m$ cell size that feature $3.6\times 10^5$ cells (e.g. model S14160-6010 from Hamamatsu \cite{hamamatsu-sipms}). 
%It should be noted that SiPMs with smaller cell size (5 and 7.5 um) were also produced by FBK \cite{} for R\&D on high energy experiments and dedicated developments in this direction could be pursued.

%Something on optical filters?
The filtering of scintillation light can be achieved for instance in PWO with high-pass filters with a cut-off wavelength around 580~nm as shown in Fig.~\ref{fig:dr_implementation} (left). Absorptive filters of this kind are available off-the-shelf (e.g. from Kodak or Everix companies) with a variety of transmittance spectra and can feature a thickness between 100 and 300$\,\rm \mu m$ which makes them particularly suitable for integration in front of the SiPM window or even as a replacement of the usual silicon/epoxy layer used for protection of the silicon active area. Filters of this kind which can filter more than 99\% of the scintillation light have been identified (e.g. Kodak 24 and Kodak 25 Wratten filters \cite{kodak-filters}).

Similarly in the case of BGO a notch-type filter can be used to select longer wavelengths beyond 650\,nm and also NUV light ($<400$\,nm), taking advantage of the fact that the emission spectrum of BGO is offset farther from the absorption cutoff. Filters with these characteristics, such as the Schott UG11 \cite{schott-filters} are readily available. For large scale production filters can be further optimized to enhance the overall light collection and S/C separation.

\clearpage\newpage
\section{Detector Solenoid \label{Sol}}
Both the CLD \cite{CLDReport} and IDEA particle detector concepts utilize a 2\,T solenoidal superconducting magnet, based on the reliable and well-established technology of aluminum stabilized NbTi Rutherford cables, satisfying all physics requirements for the collider operating points from the Z pole to the $t\Bar{t}$ center-of-mass collision energy. The electromagnetic design of the superconducting coil for the IDEA detector, originally positioned in front of a single-layer dual read-out fiber calorimeter, has been optimized to minimize the material along the particle trajectories. However, the recent conceptual design, implementing two separate dual readout calorimeters (a crystal calorimeter for the electromagnetic shower and a fiber calorimeter for the hadronic shower) increases the internal radius of the superconducting coil, now placed between the two calorimeters, see Fig. \ref{fig:idea_radenv}, reducing the constraint on the solenoid transparency. In this context, a new preliminary design for a high-temperature superconducting (HTS) solenoid developed by the INFN LASA \cite{LASA} superconducting magnet group is able to provide up to a 3\,T central magnetic field operating at T $>20$\,K. We summarize R\&D on the electromagnetic design, the mechanical stability, and magnet protection, while discussing the main challenges of the implementation of this technology compared to a conventional low-temperature superconductor (LTS) based approach.
% \begin{figure}[h]
%     \centering
%     \includegraphics[width=0.8\linewidth]{figs/solenoid/idea_detector_comparison.png}
%     \caption{comparison between the idea detector design reported in the fcc-ee conceptual design report (\textit{left}) and the innovative proposal under investigation (\textit{right}) of an hts solenoid (coil) in between a split dual readout calorimeter with enhanced performances}
%     \label{fig:detectorsketch}
% \end{figure}
\subsection{HTS Solenoid Design}
The use of high-temperature superconductors for next-generation particle detectors has the primary advantage of operating at a temperature of 20\,K, gaining a factor five in operational cost reduction of the cryogenic plant compared to the classical 4 K operation (and a further factor two, by operation at 40-50\,K, might be considered). Moreover, the use of HTS can reduce conductor volume by leveraging their higher operating current density compared to conventional low-temperature superconductors, which minimizes uninstrumented material and relaxes mechanical support requirements for the entire solenoid. High-temperature superconductors currently remain more expensive than low-temperature superconducting cables, however, in recent years the development and manufacture of many thousands of high-quality HTS conductors by various vendors all over the world, driven by fusion R$\&$D, has substantially decreased its cost.  At present, the “classical” aluminum-stabilized Nb-Ti Rutherford cable technology via co-extrusion, a widely exploited technology also by the INFN teams \cite{Acerbi1988,Fabbricatore1996,Rossi1999,Kircher2000,Fabbricatore2000,Acerbi2006}, is no longer commercially available with only limited efforts underway in research laboratories to revive this expertise. The use of Nb-Ti relies on soft soldering between the superconductor and a pure aluminum matrix, a technology that is also well-suited for HTS tapes. R$\&$D programs will be needed to implement either of these types of conductors for next-generation particle detectors, and we think that given the intrinsic advantage of HTS, it is worthwhile to make an effort in that direction. 
To dimension the superconducting magnet volume, an analytical formula for the maximum stress on thin cylinders under uniform pressure can be used. The relationship between the coil thickness and the average hoop stress on the conductor is a function of the central magnetic field value of the solenoid and the internal radius of the winding, see Equation \ref{eq:thickness}:
\begin{equation}
    t_{coil}=\left(\frac{R_{in}}{\sigma_h}\right)\left(\frac{B_{z}^{2}}{2\mu_0}\right)\hspace{3pt}\text{with}\hspace{3pt}B_z\left(z=0\right)=\frac{\mu_0}{2}\frac{NI}{\sqrt{R_{in}^2+\left(\frac{L}{2}\right)^2}}
    \label{eq:thickness}
\end{equation}
A maximum value of hoop stress $\sigma_h$ = 100\,MPa is considered in this analysis, with the target central field of 3\,T, resulting in a minimum coil thickness of 89.5\,mm. The electromagnetic design of the HTS solenoid is presently optimized to produce a 3 T central field leading to improved particle tracking resolution of the detector at particle collision energies above the Z resonance while being able to provide a 2 T solenoidal magnetic field within the requirements on field quality and superconducting coil stability.  

\begin{figure}[h]
    \centering
    \includegraphics[width=0.49\linewidth]{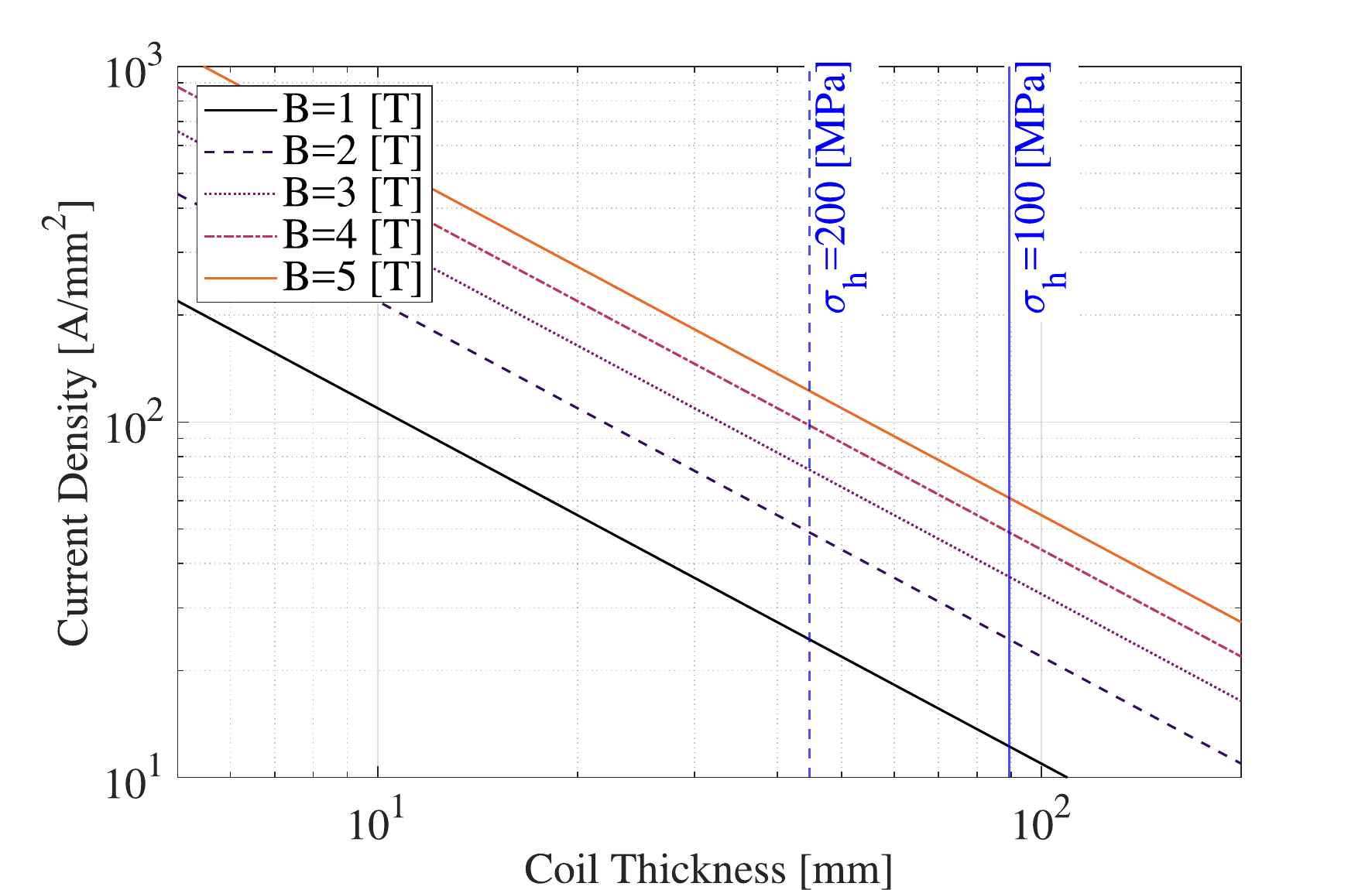}
    \includegraphics[width=0.49\linewidth]{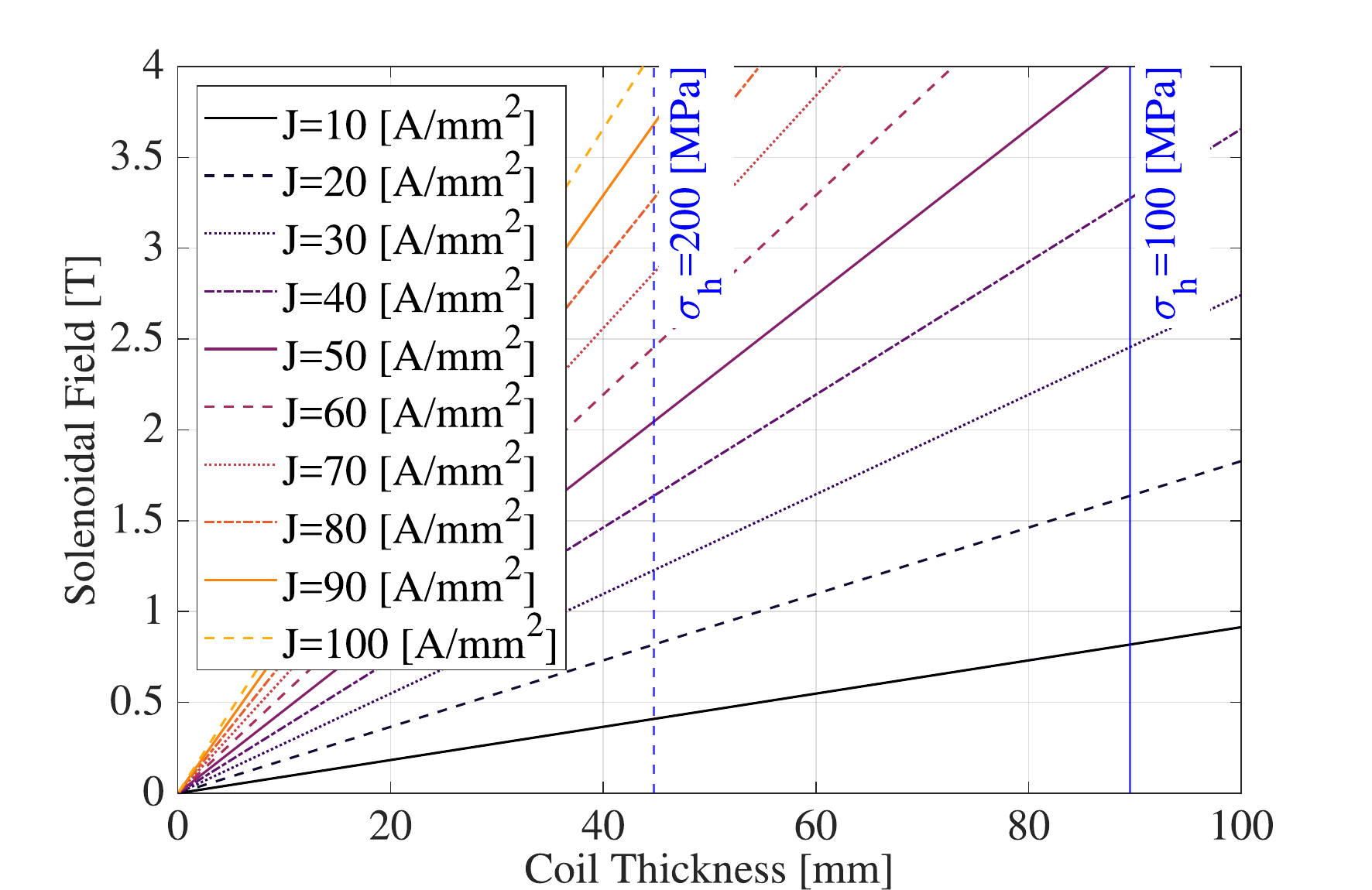}
    \caption{Scaling law of the solenoidal field (right) and current density (left) for a finite length solenoid assuming a thin solenoid approximation. Values of maximum hoop stress on the solenoid conductor under Lorentz forces are reported (110\,MPa being the present reference value).}
    \label{fig:analyticalevaluations}
\end{figure}

\begin{figure}[h]
    \centering
    \includegraphics[height=0.25\linewidth]{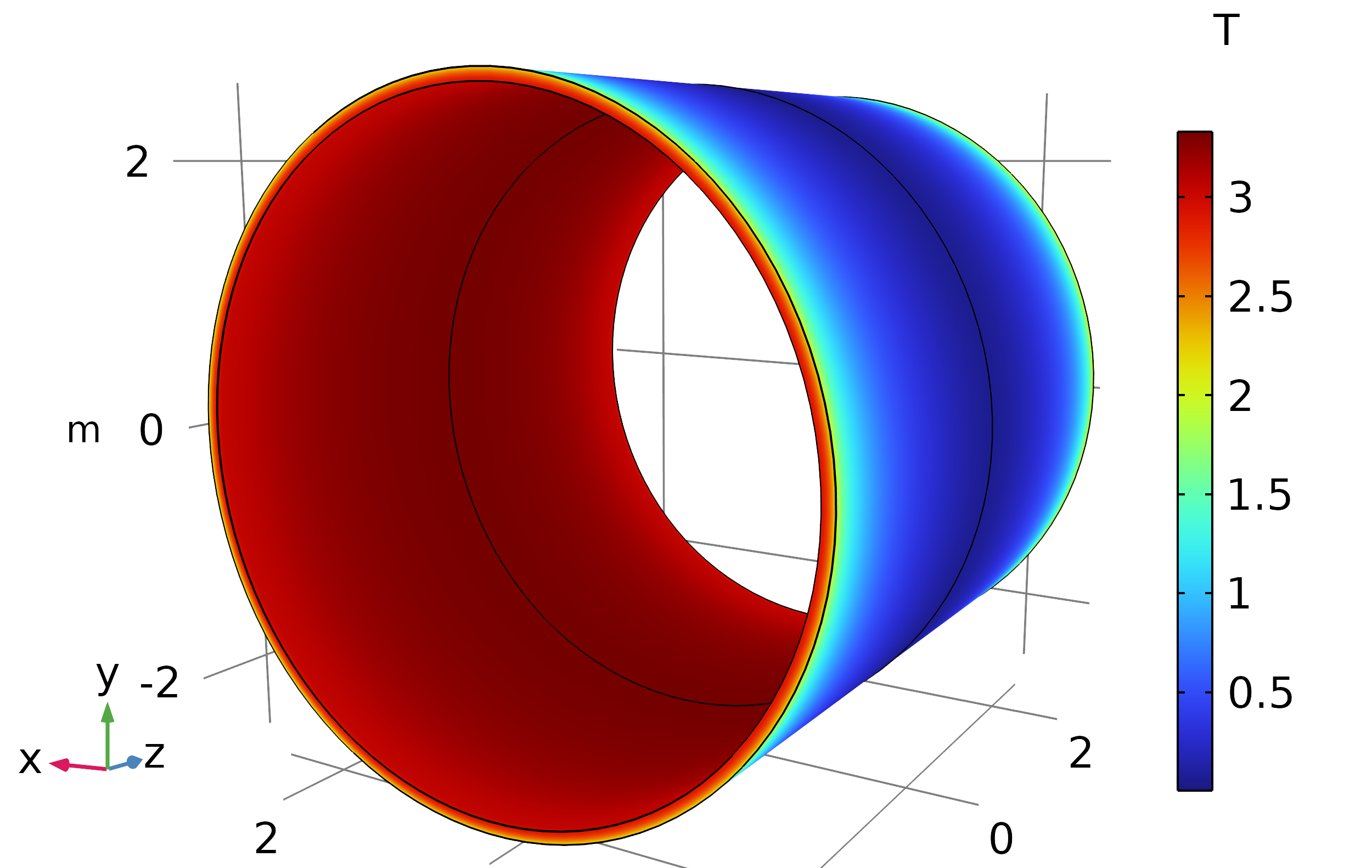}
    \includegraphics[height=0.25\linewidth]{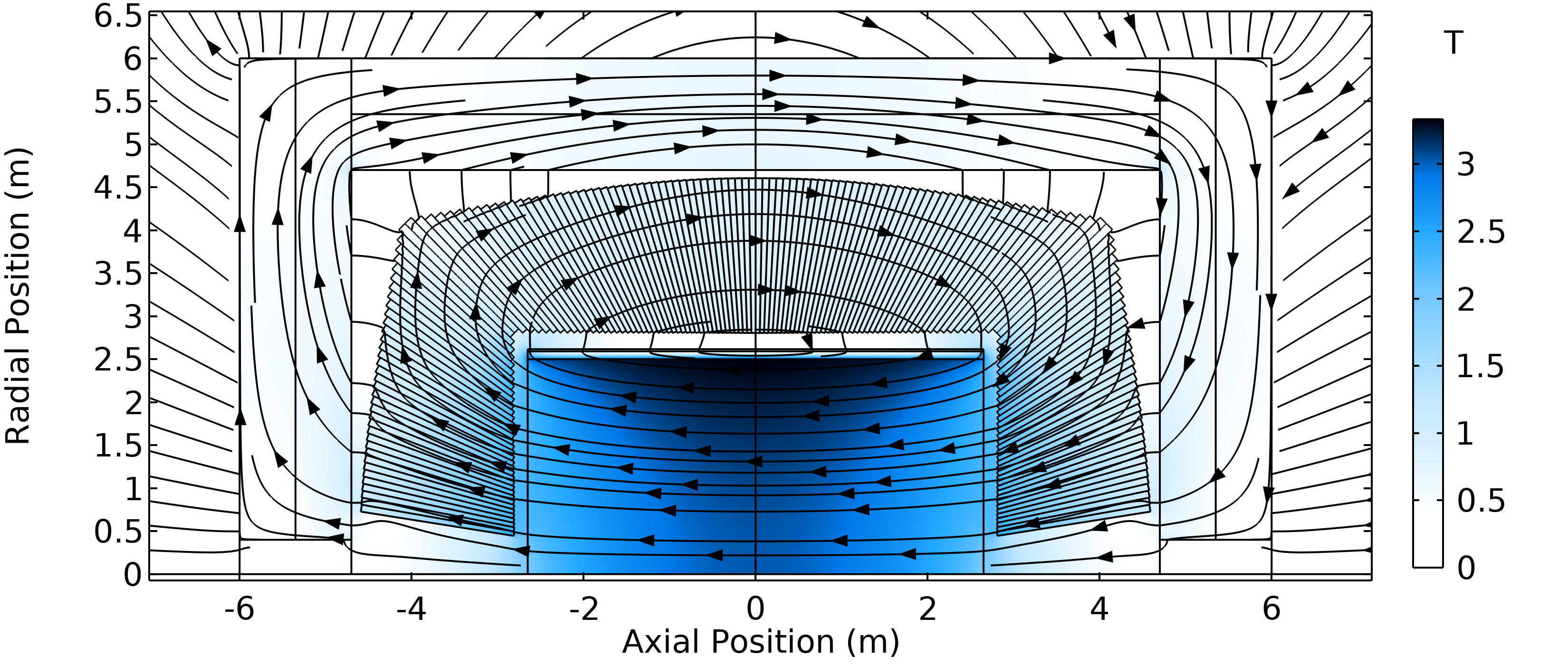}
    \caption{Magnetic Field maps produced by the 3\,T HTS superconducting solenoid. Peak magnetic field at nominal current on the superconducting coil (\textit{left}) and cross-section view of the particle detector (\textit{right}).}
    \label{fig:FieldMap}
\end{figure}

Using the equation for the magnetic field produced by a finite length solenoid with L = 5.3 m, R$_{in}$ = 2.5 m, to generate a magnetic field value of B$_z$ = 3 T, a total of NI = 17.4 MAturns or, equivalently, an average engineering current density of J$_{eng}$ = 37 A/mm$^2$ (J$_{eng}$ = 25 A/mm$^2$ in case of B$_z$ = 2\,T) is required, see Fig. \ref{fig:analyticalevaluations}.
Using these scaling laws, a 2D electromagnetic axi-symmetrical FEM model for a 3\,T solenoid has been reconstructed in Comsol multi-physics \cite{Comsol} considering also the effect of the iron detector yoke and the ferromagnetic (considering a 60\% filling factor of iron in the total volume, see Section \ref{HCal}) dual read-out fiber calorimeter. The field map of the superconducting coil and the shape of the produced magnetic field over the detector volume without considering the presence of the shielding solenoids and final focusing quadrupole of the interaction region (IR) are shown in Fig. \ref{fig:FieldMap}. The geometrical parameter of the cable considered for the electromagnetic design and the superconducting coil main features are reported in Table \ref{tab:CoilandCableParam}. 

\begin{table}[h!]
    \caption{Solenoid and Superconducting Cable Main Parameters}
    \centering
    \small
    \begin{tabular}{l|c|c|l|c|c}
    \hline
         Bore field & 3 & [T] & \# of tapes in a cable & 7 & \\
         Nominal current & 8168 & [A] & Cable thickness & 11.47 & [mm]\\
         Coil thickness & 91.7 & [mm] & Cable height & 24.0 & [mm]\\
         Number of turns & 220$\times$8 & & Stabilizer thickness & 9 & [mm]\\
         Inductance & 12.7 & [H] & Stabilizer edge height & 5 & [mm]\\
         Stored energy & 412 & [MJ] & Insulation thickness & 1 & [mm]\\
    \hline
    \end{tabular}
    \label{tab:CoilandCableParam}
\end{table}

\begin{figure}[h!]
    \centering
    \includegraphics[width=0.3\linewidth]{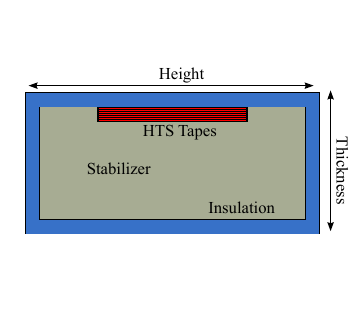}
    \includegraphics[width=0.65\linewidth]{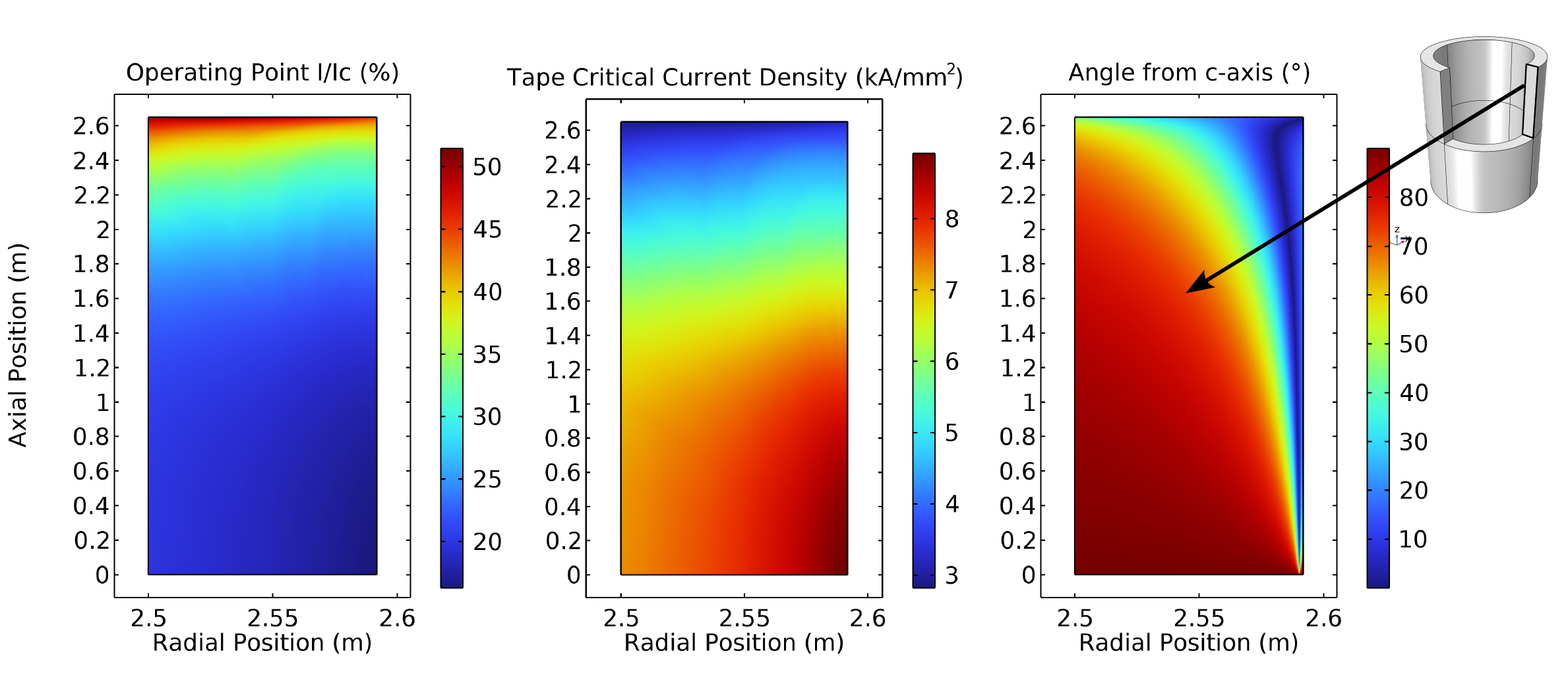}
    \caption{Cable design considered for the superconducting coil with the 7 HTS tapes highlighted in red on the upper part of the cable. On the right: map of the operating point of half solenoid compared to the cable critical current at T = 20\,K, tape current density as a function of the field orientation and angle of the field compared to the broad side of the tape.}
    \label{fig:CableCrossSection_MagnetPerformance}
\end{figure}

A classical approach using “U-shaped” channel, see Fig. \ref{fig:CableCrossSection_MagnetPerformance} (left), made of aluminum alloy Al-2.0$\%$Ni is presently considered for the cable stabilizer offering high mechanical strength (E = 70\,[GPa], yield strength = 170\,[MPa] ) compared to other aluminum alloys and high thermal stability during a quench event with a still acceptable value of RRR (Residual-Resistivity-Ratio) = 170. The number of superconducting 12\,mm tapes within the cable has been optimized to maximize the operating margin at the peak field on the coil volume. A maximum of I/I$_c$ = 51.5$\%$ at the edge of the solenoid has been obtained from the electromagnetic optimization of the solenoid dimension with most of the coil volume operating below 25$\%$ of the critical current ($I_c$) of the superconducting tape, see Fig. \ref{fig:CableCrossSection_MagnetPerformance} (right). The electromagnetic design can be further optimized achieving more uniform temperature margin operation, allowing for significant reductions in HTS conductor usage in the central region, leading to a substantial cost decrease in the total solenoid assembly. The calculated margin values account for the angular dependence of the tape's critical current, based on measured data from samples manufactured by Faraday Factory \cite{FaradayCompany}, considering the field's orientation relative to the tape's broad surface aligned to the magnet rotational axis. 

\subsection{Mechanical and Thermal Stability}
The cable geometry and the material fractions have been optimized to achieve high thermal stability of the superconducting solenoid in case of a quench event. To evaluate the enthalpy margin of the superconducting coil, the electromagnetic stored energy density of 19.12\,kJ/kg can be assumed to dissipate uniformly over the entire winding volume obtaining a safe maximum temperature value equal to 122\,K. Instead, conventional detection and discharge magnet protection schemes must be considered to protect the magnet during the quench development in the case of localized energy deposition. Depending on the aluminum alloy considered as stabilizer material, a maximum hot spot temperature of around 160 K is foreseen during a quench event assuming a conservative 10\,s quench detection time to activate the protection scheme and a maximum 2\,kV peak voltage to ground during the magnet discharge. An additional detailed quench propagation simulation at the cable geometry level will be performed to fully characterize the magnet protection technology and method required to avoid superconductor degradation and magnet safety. Even if the maximum hot spot temperature obtained is acceptable and the peak voltage to ground is within the admissible voltage range, an early detection scheme must be studied to avoid temperature rise in the winding. 
\begin{figure}[h!]
    \centering
    \includegraphics[width=0.55\linewidth]{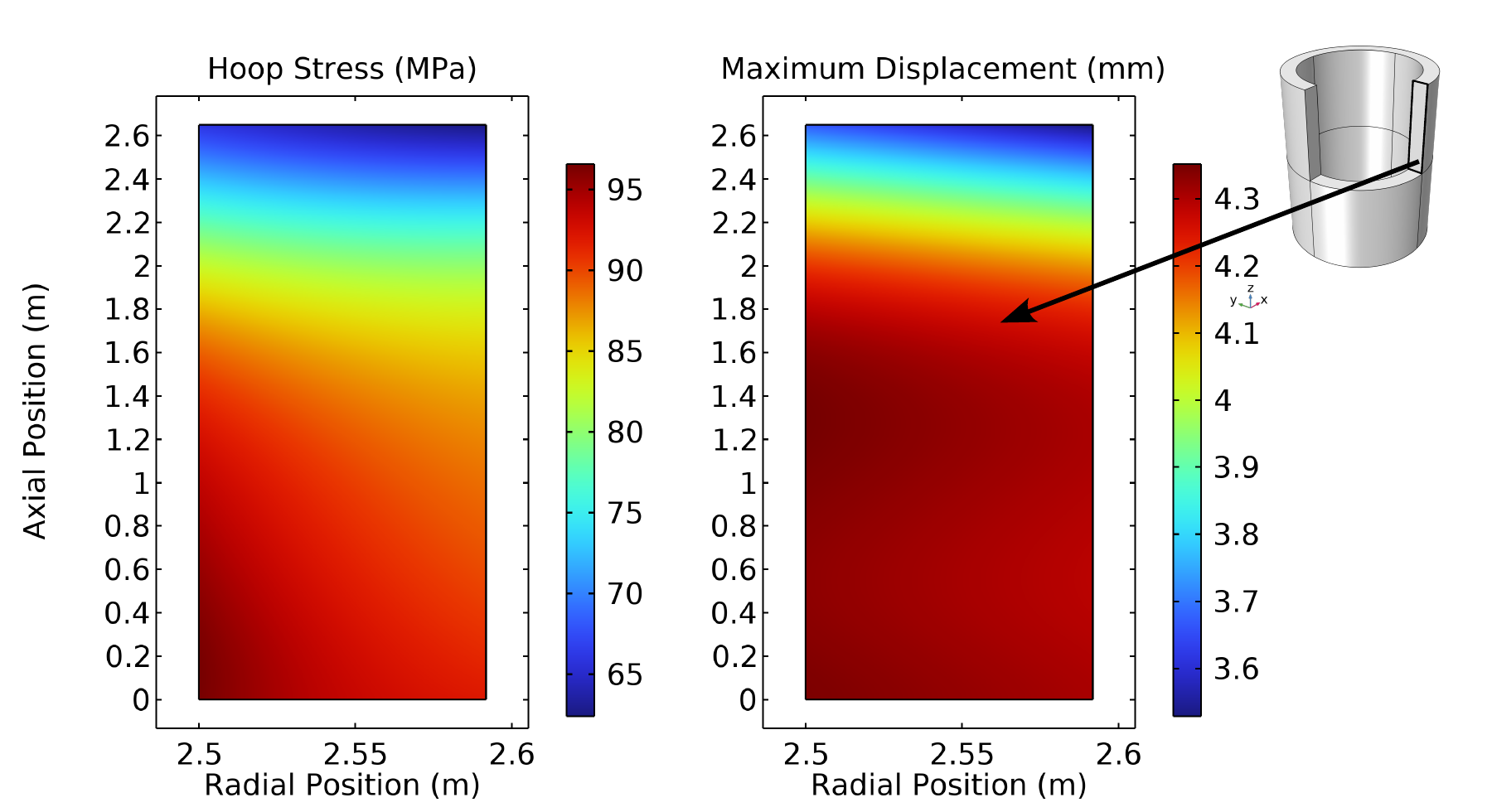}
    \includegraphics[width=0.3\linewidth]{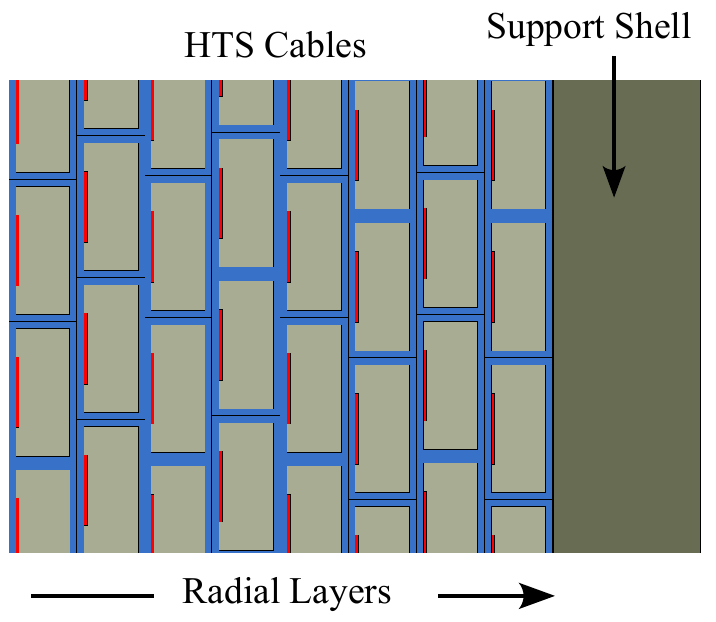}
    \caption{Maximum average hoop stress and displacement amplitude on the conductor volume under e.m. forces at nominal current of the magnet energization (\textit{left}). Sketch of conductor assembly with representation of the restraining cylinder outside of coil volume in the radial direction (\textit{right}).}
    \label{fig:MechanicalAssembly}
\end{figure}
In addition, to avoid degradation of the superconducting layers in the cable, 2D axisymmetric simulations considering, as a first step, only the contribution of the e.m. forces acting on the coil volume, have been performed evaluating the maximum stress on the conductor. Using equivalent anisotropic homogenized material properties (see \cite{Seungyoung_2024} for the HTS tape properties) and an additional 25\,mm thick aluminum 5083 alloy restraining cylinder (see Fig. \ref{fig:MechanicalAssembly}), already considered as supporting structure in existing solenoids for particle detectors \cite{CERN-LHCC-97-018,CERN-LHCC-97-010}, a maximum tensile hoop stress on the coil of 97\,MPa and longitudinal strain of 0.18$\%$ on the conductor are obtained. Even if these values are within the material performance limits (maximum yield stress of 170\,MPa at 4 K for the Al2.0$\%$wtNi alloy and 0.3$\%$ of maximum strain for the high-temperature superconducting layer) finer detailed model will be used to address the overall mechanical performance of the coil volume during all operating conditions (magnet assembly, cryogenic cool-down, and magnet energization). Considering the total thickness of the cold mass equals 116.7\,mm and the normalized radiation lengths for each material reported in Table \ref{tab:CoilTransp},%
\begin{table}[h]
    \centering
    \caption{Coil Transparency}
    \small
    \begin{tabular}{l|c|c}
    \hline
         Material & Thickness & X/X$_0$\\
    \hline
         HTS tape & 4.0 [mm] & 0.296\\
         Al2.0$\%$wtNi & 72 [mm] & 0.862\\
         Insulation & 16 [mm] &  0.056\\
         Shell & 25 [mm] & 0.278\\
    \hline
         TOTAL & 117.3 [mm] & 1.492\\
    \hline
    \end{tabular}
    \label{tab:CoilTransp}
\end{table}%
the thickness of the cold mass, equivalent to X/X$_0$= 1.492, is within the superconducting coil design constraints and does not affect the detector energy resolution.
Preliminary mechanical structure and cryostat dimensioning evaluations are under development at LASA where research and development activities will be carried out to enhance the technological readiness level (TRL) of HTS superconducting magnets supporting sustainable solutions for future collider projects.
%\section{Preshower \label{Pre}}
\clearpage\newpage
\section{Hadronic calorimeter \label{HCal}}
%% General introduction 
The proposed hadronic section for the IDEA calorimeter is based on a fiber-sampling dual-readout calorimeter~\cite{dualreadoutfibercalo}, consisting of alternate rows of scintillating and clear (Cherenkov) fibers inserted in a metal absorber. In the initial design of the IDEA detector, the full calorimetric system was built with a single, compact and unsegmented, fiber-sampling calorimeter. Only recently the electromagnetic crystal section has been added as baseline option and included in the full simulation. As a consequence, the present understanding of the hadronic calorimeter performance is mainly based on a standalone fiber-sampling solution. This standalone fiber-sampling design will be validated in detail in simulation and test beam studies since it will provide the reference benchmarks for hadronic shower measurements which should not be degraded by the presence of the electromagnetic crystal section.

\begin{figure}
\begin{center}
\includegraphics[width=0.75\textwidth]{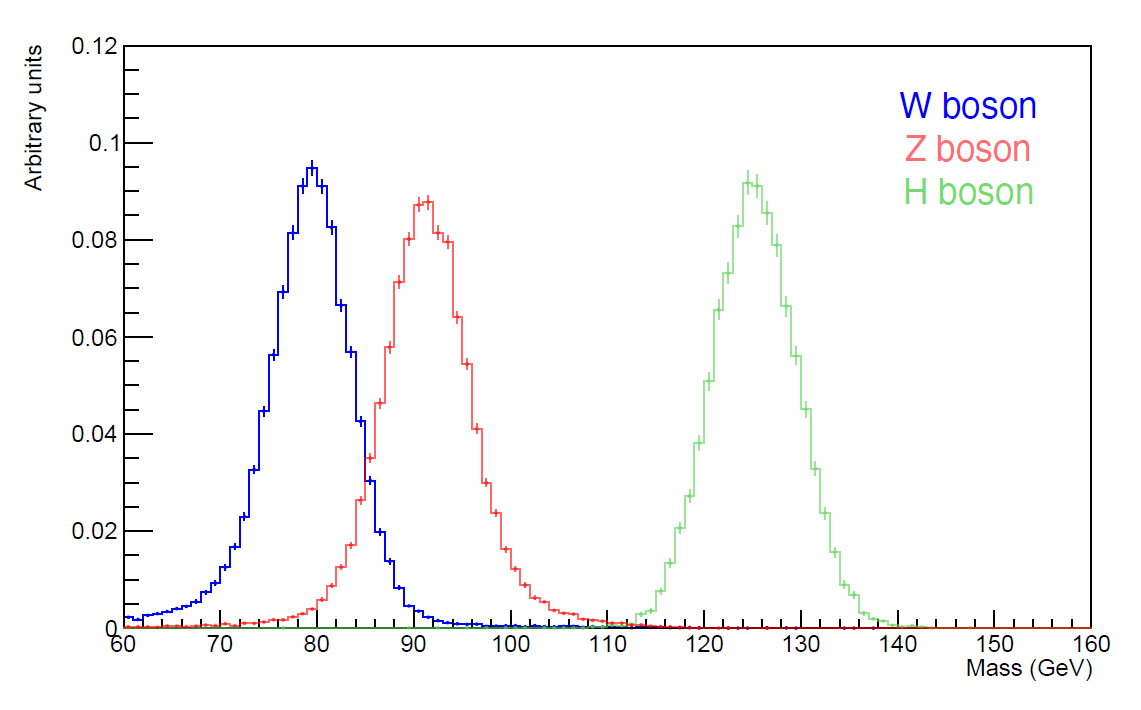}
\caption{Reconstructed invariant mass distributions for $W/Z/H \rightarrow jj$ fully-simulated events in a highly granular DR fiber-sampling calorimeter~\cite{Pezzotti_phdthesis}.}
\label{fig:DR-WZH2jj}
\end{center}
\end{figure}

The hadronic energy resolution was estimated, following a full simulation of the IDEA (standalone) fiber-sampling calorimeter~\cite{Pezzotti_phdthesis}, at about $30\%/\sqrt{E}$, with a small constant term. The expected achievable separation for the $W/Z/H \rightarrow jj$ peaks is shown in Fig. \ref{fig:DR-WZH2jj} for an ideal fiber-sampling calorimeter. Simulation studies show that this detector can also provide excellent standalone particle-identification capabilities~\cite{dualreadoutpid}.

The geometrical description of the present implementation is provided in Section~\ref{sec:hadcalo_design}. Here it is useful to underline that longitudinally unsegmented scintillating and clear fibers are inserted in metal capillary tubes oriented with a quasi-pointing geometry towards the interaction vertex. This way, they can provide a highly granular shower sampling in the transverse direction, i.e. an excellent 2-D imaging capability. Longitudinal shower information could only be accessible through signal time-of-arrival measurements as discussed in Section~\ref{sec:futurerandd}. The sampling granularity is presently designed for a standalone calorimeter (without crystals in front), i.e. it is driven by the requirement to allow the separation of the showers from the two $\gamma$ of a $\pi^0$ decay. This granularity needs to be differently tuned given the EM crystal section in front.  However, a highly granular sampling can provide very powerful input to deep neural networks and particle-flow algorithms.

Both iron and brass tubes are considered for the absorber material, where iron has the advantage of providing most flux return of the magnetic field and is therefore our current baseline. Small prototypes, capable of containing electromagnetic showers, were built With both materials and tested with beams.

%In the full-fiber solution (no crystals in front) the tubes are assembled with a small tilt angle of $\approx 2.5^\circ$ in both the $\theta$ and $\varphi$ coordinates to minimize the probability of high energy particles traveling a significant distance in a single fiber before starting showering. In the combined solution, the tilt angle requirement is less stringent. However, the tilt angle is very relevant for the calibration procedure, since the dual-readout works if the calorimeter is calibrated at the electromagnetic scale.\\

\subsection{Hadronic calorimeter design}
\label{sec:hadcalo_design}

\begin{figure}[h!]
    \centering
    \includegraphics[width=0.55\textwidth]{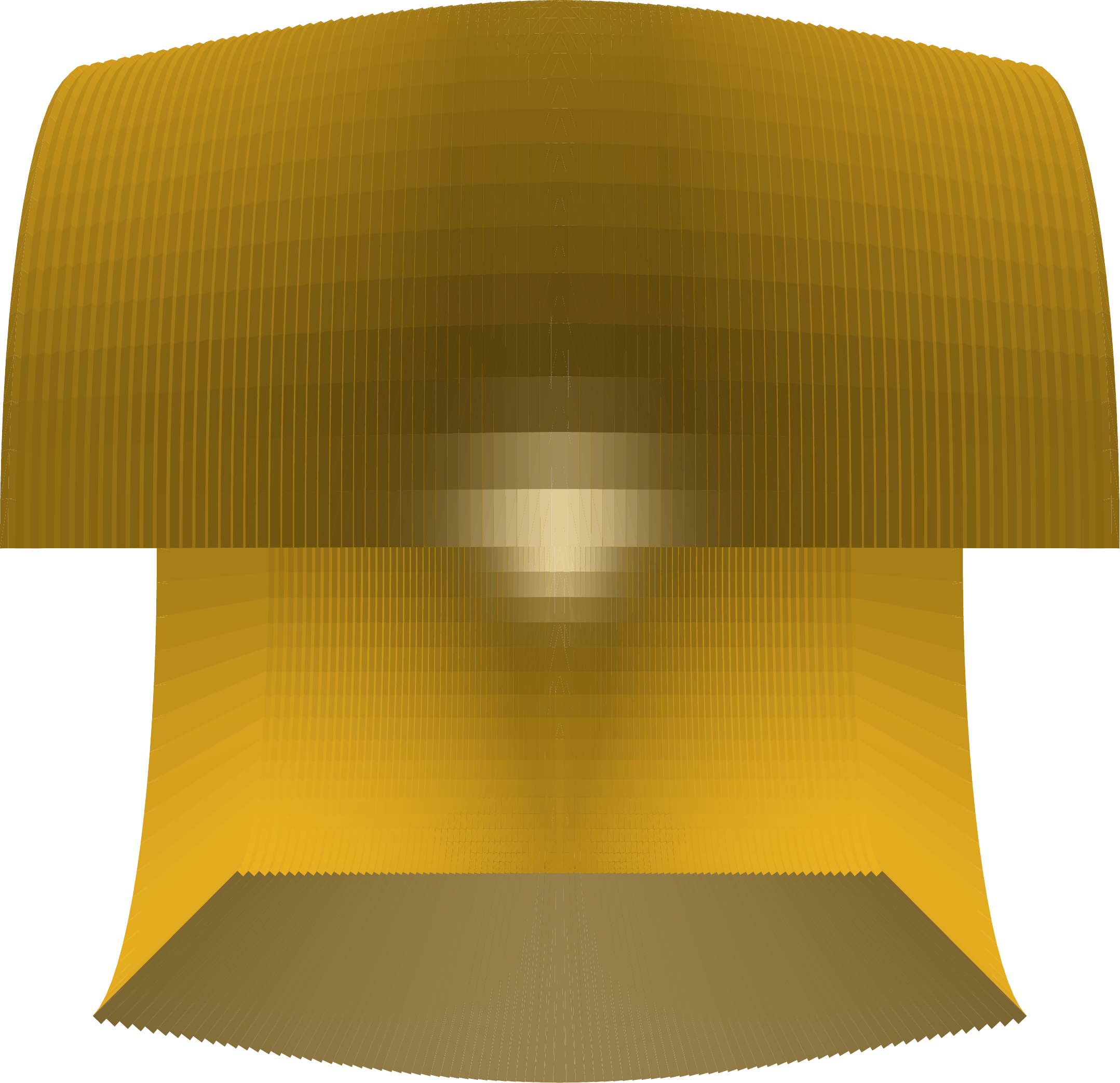}
    \includegraphics[width=0.33\textwidth]{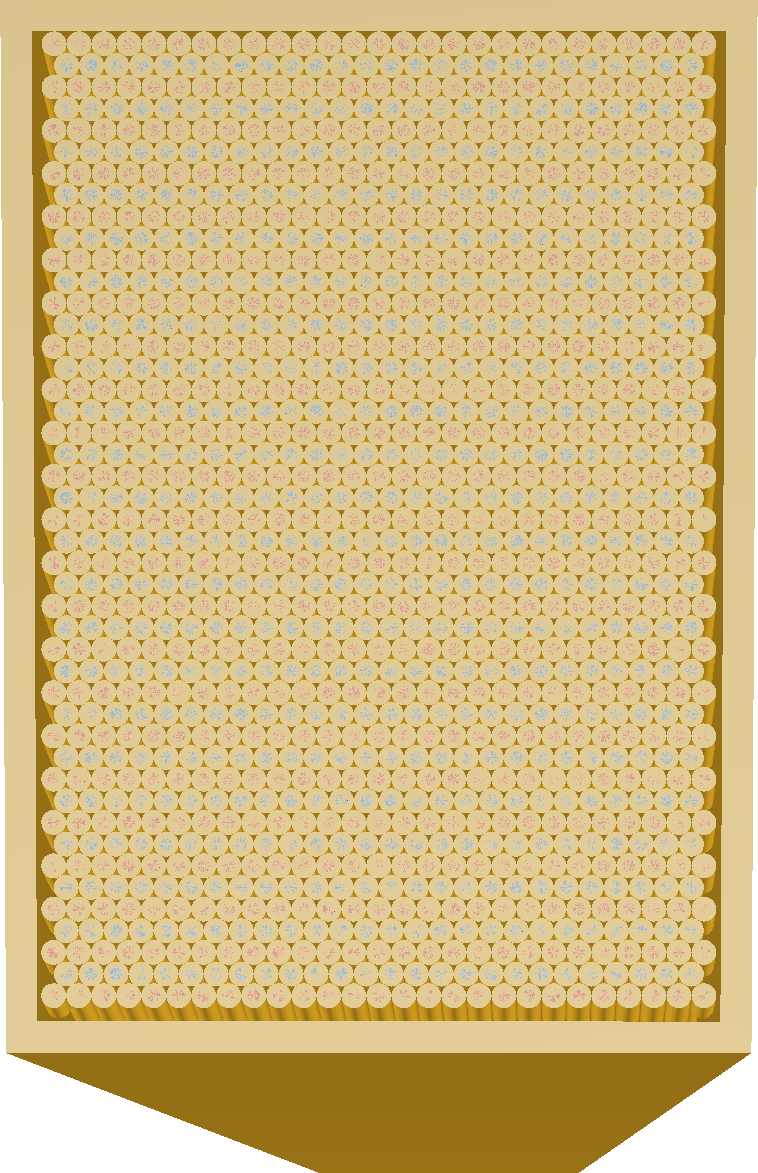}
    \caption{Barrel region of the dual-readout calorimeter, with a few towers taken out for visualization purposes (left). Backside of one tower with the assembly of tubes and the support structure around them, forming one tower (right).}
    \label{fig:tube_calo_geometry}
    \includegraphics[width=0.57\textwidth]{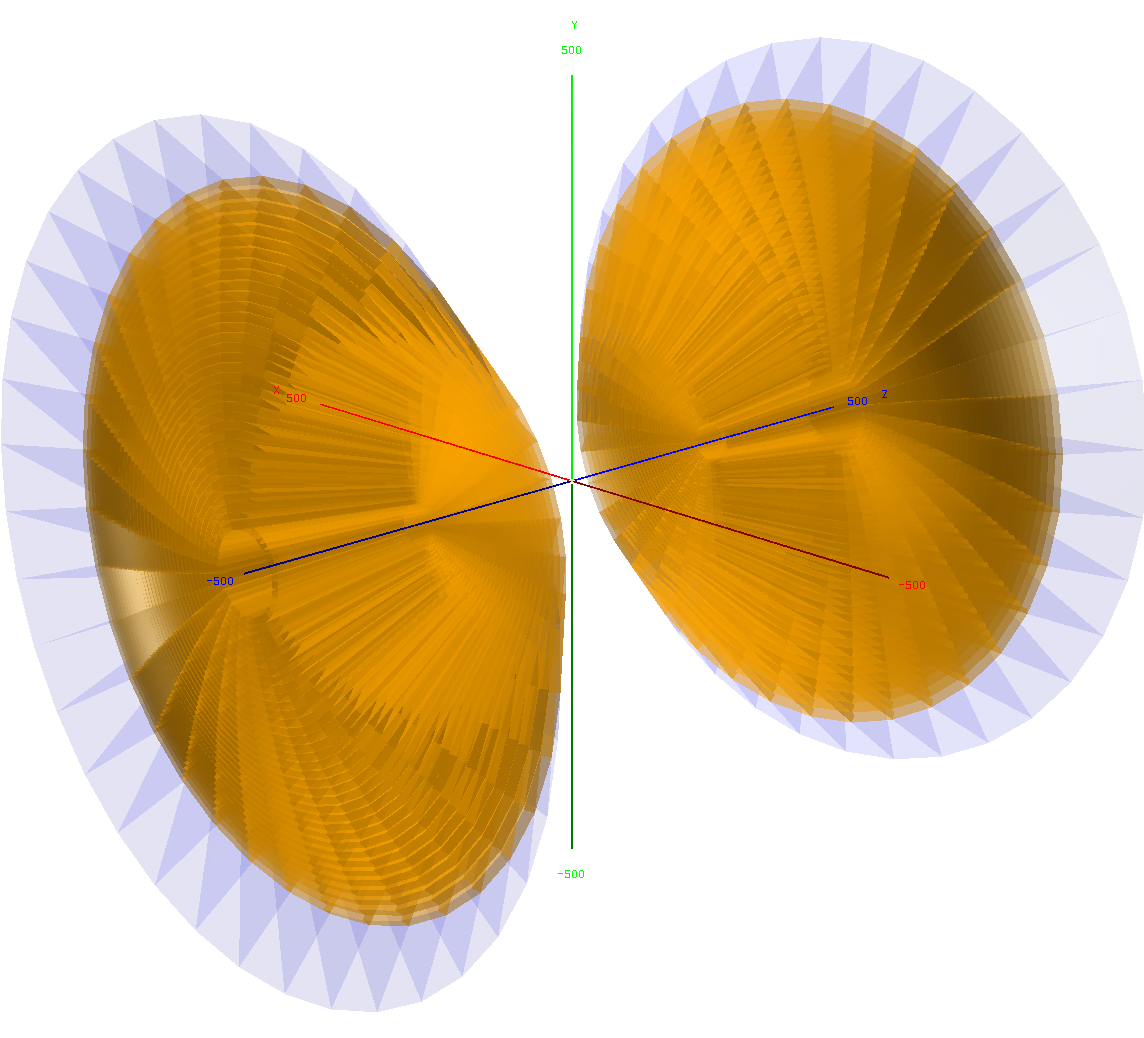}
    \includegraphics[width=0.33\textwidth]{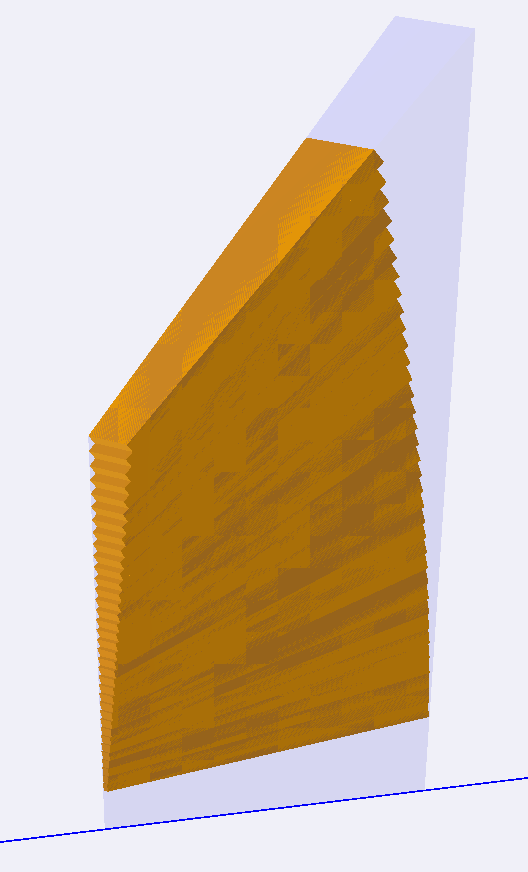}
    \caption{Endcap region of the dual-readout calorimeter, considering 35 $\eta$-towers and $36$-rotations around the beam pipe. The trapezoidal structure (a $\phi$-slice) containing the $\eta$-towers is displayed in light blue (left). A single $\phi$-slice populated with 35 $\eta$-towers (right).}
    \label{fig:tube_calo_geometry_endcap}
\end{figure}

The IDEA capillary tube dual-readout calorimeter consists of optical fibers inserted in metallic tubes arranged to projective towers. 
The capillary-tube technology was pioneered in recent years by a consortium of European Institutes led by INFN and represents the chosen solution to evolve from a conceptual design to a technical design. In this configuration the tube material acts as the absorber component; currently, two materials are considered: brass and iron. The tubes are stacked together in a hexagonal pattern to form towers, which are trapezoids pointing to the interaction point. 
This geometry was implemented recently in the new IDEA full simulation, for both the barrel and the endcap regions, implemented as DD4hep sub-detectors. An image of the barrel and endcap geometries can be seen in Figure~\ref{fig:tube_calo_geometry} and Figure~\ref{fig:tube_calo_geometry_endcap}, respectively.

The barrel and endcap geometries are implemented as two separate sub-detectors, which can be simulated independently from one another. In the baseline design, the transition from one to the other is assumed to occur at a \SI{45}{\degree} angle with respect to the beam axis. The inner radius (lowest z coordinate) of the barrel (endcap) is 2.8\,m, and the tower length is 1.8\,m. The tube outer radius is 1 mm and the fiber radius is 0.5\,mm. The current geometry implementation is modular and scales according to such parameters. The endcap calorimeter is designed to form a \SI{90}{\degree} angle with the barrel one. 
Due to the symmetry in $\phi$, the barrel and endcap are both created by multiple placings of one "stave". Each stave is populated with towers in the $\theta$ direction, where each tower is different, apart from a forward-backward symmetry. In total, 72 $\phi$-staves are considered and this defines the tower granularity in the $\phi$ direction. Inside a barrel stave, 45 towers are positioned along the $\theta$ direction for the barrel calorimeter. The endcap calorimeter uses 35 towers in the $\theta$ direction in each stave. In total, around 80 million tubes are placed in this geometry, roughly divided into 60 million for the barrel and 20 million for the two endcaps.

\begin{table}
    \centering
    \begin{tabular}{ll}
        \toprule
        \emph{Name} & \emph{Value} \\
        \midrule
        Inner Radius & $\SI{2.8}{\m}$ \\
        Outer Radius & $\SI{4.6}{\m}$ \\
        \midrule
        Tower $\Delta\theta$ (\textit{barrel}$/$\textit{endcap}) & $\SI{1}{\degree}/\SI{1.125}{\degree}$\\
        Tower $\Delta\phi$ & $\SI{5}{\degree}$ \\
        Barrel Support Thickness & \SI{1}{\mm} \\
        \midrule
        Tube Radius & $\SI{1}{\mm}$ \\
        Fibre Radius & $\SI{0.5}{\mm}$ \\
        \bottomrule
    \end{tabular}
    \caption{Geometry parameters for the dual-readout calorimeter.}
    \label{tab:geo:parameters}
\end{table}

\subsection{Mechanical construction }
%with a description of both EU and Korean options) -> Gabriella, Korean}
%
%  Edited a little bit on Feb 11 (NA)
%
Over the past two decades~\cite{dream-rd52}, DR prototypes have been developed by the DREAM/RD52 Collaboration using scintillating and clear fibers embedded in an absorber structure, oriented approximately parallel to the direction of incoming particles. These two fiber types were read out separately using either PMTs or SiPMs.

The fibers typically had a 1-mm diameter. For Cherenkov signal detection, the fiber cores were made of fused silica or PolyMethyl MethAcrylate (PMMA), with a fluorinated polymer single cladding. Several vendors manufacture such fibers, including Polymicro (FSHA-type fibers), Toray (Raytela PJR-FB750), Mitsubishi (ESKA SK40-1500), and Kuraray (Clear-PS). Scintillating fibers, on the other hand, featured a polystyrene core with PMMA single or double cladding, emitting in the blue region (wavelength ~400–450 nm, depending on the manufacturer), with examples including BCF-12 from Saint-Gobain, SCSF-81J and SCSF-78 from Kuraray. Key performance parameters include light yield, trapping efficiency, and optical attenuation. While scintillating fibers generally provide sufficient light yield,  lower yield of Cherenkov radiation presents a more challenging optimization. Emerging fiber and waveguide structures with larger numerical apertures and higher refractive indices  offer advantages for Cherenkov light production.  For example, a numerical aperture of 0.72 is achievable using a multi-clad structure around a polystyrene core, leading to a light trapping efficiency of 5.5\%.  Although, it is still in an R\&D stage, even higher trapping efficiencies are possible with air-clad fused-silica core fibers where the numerical aperture approaches 0.9. 

The absorber structure design has evolved alongside increasingly capable prototypes. The fiber positioning within the absorber affects the stochastic term via the sampling fraction and frequency, while the precision of fiber placement contributes to the constant term. The choice of absorber material significantly influences the production rate of secondary particles in hadronic showers, impacting signal generation and defining transverse and longitudinal containment dimensions. Additionally, the absorber must provide adequate structural integrity for the multi-ton-scale calorimeter in the IDEA detector.

Currently, two techniques are under investigation, both promising for full-scale experiment construction. The first, developed under the INFN-approved HiDRa R\&D project, employs capillary tubes. The second, advanced by a consortium of Korean institutes, utilizes finned plates produced by skiving, where fibers and copper rods are embedded to form a granular structure. At present the former looks able to provide better mechanical precision, while the latter provides a simpler and faster construction technique. 3D-printing was also investigated, but it is still not suitable due to cost and manufacturing time.

\subsubsection{Capillary Tubes Technique}
\label{sec:capillary_tubes}

In the past years, we built a prototype for a dual-readout calorimeter based on the capillary tube technique as the absorber structure.
The construction of a prototype, \mbox{10 $\times$ 10\,cm$^2$} in cross section and 1\,m long, is described in a technical paper \cite{Karadzhinova-Ferrer_2022}, and its performance when exposed to a beam of positrons is described in Ref.~\cite{Ampilogov:2023zxb}, and further detailed in Section~\ref{par:TB_fibreSamplDR}.
Based on the experience gained with this prototype, we decided to base the construction technique for the HiDRa calorimeter on the capillary tubes. 
Stainless steel capillary tubes for the full-hadronic containment prototypes are 2.5\,m long, with 2 $\pm$ 0.03\,mm outer diameter and 1.1 $^{+0.1}_{-0.0}$\,mm inner diameter. Mechanical tolerances on the diameters, length ($\pm 1$\,mm ), straightness\footnote{tube should be able to rotate smoothly on a flat surface} and quality of both ends were set and the producer was able to provide capillary tubes within the specification at an affordable price. 

\begin{figure}
   \begin{center}
      \begin{minipage}{0.49\textwidth}
         \centering
         \includegraphics[width=0.9\textwidth, angle=0]{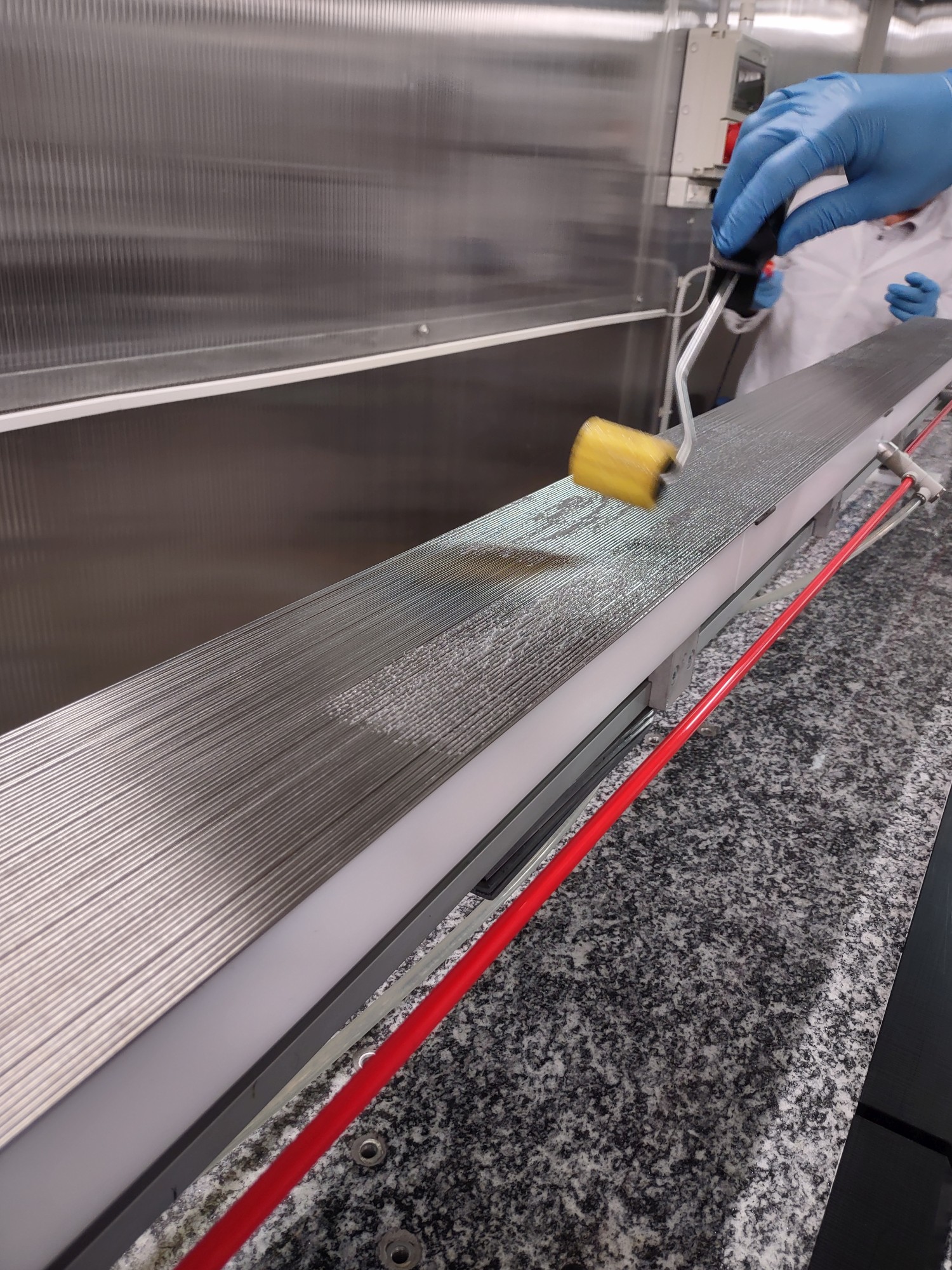}
      \end{minipage}
      \begin{minipage}{0.49\textwidth}
         \centering      
          \includegraphics[width=0.95\textwidth, angle=0]{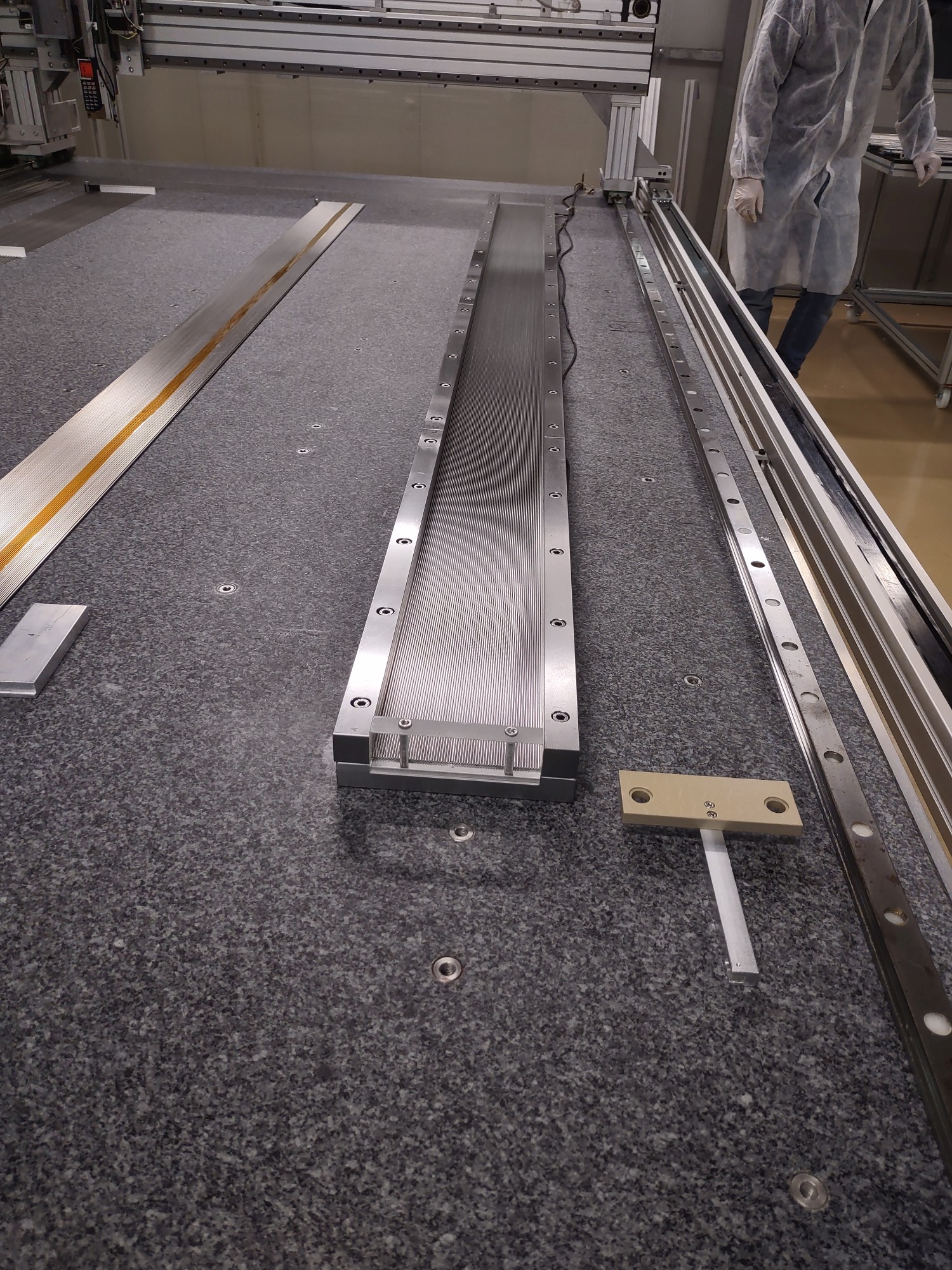}
      \end{minipage}         
\caption{Left: The handling tool with one layer of tubes already engaged and ready for glue distribution. Right: The high precision assembly tool. }
\label{fig:capillaryTool}
\end{center}
\end{figure}

Thanks to the high quality of the capillary tubes, the tooling for the module assembly was simplified, with respect to what was used for the EM-size prototype.
The module is built by gluing layers of capillary tubes in a reference tool machined at high precision, as shown in Fig.~\ref{fig:capillaryTool}.  
The tubes are prepared and aligned in the reference tool. A handling tool, suspended from a crane using chains attached to rotating pins, allows for engaging the tube layer through a vacuum system and moving it for glue distribution and piling up. This technique has proved to be fast and to guarantee both excellent precision and good reproducibility.

Once the module is built, the fibers are manually inserted in each tube, in alternate rows of scintillating and clear fibers. At present, the construction rate is time consuming, and  needs to be made more efficient for the construction of a full size calorimeter.

% added by HDYOO (begin) %
\subsubsection{Alternative mechanical options: Skived Fin Heat Sink, 3D metal printing, and Drilling Techniques}
\label{sec:alternative_cuform}
In the  RD52 project multiple methods were explored to form the metal absorber. After extensive mechanical R\&D, the initial solution was the machining of grooves in the cooper sheets. This technique was used to successfully build two modules. For the IDEA calorimetry system new fabrication options have been considered on top of the RD52 studies, in addition to the capillary tube technique described above. A consortium of Korean institutes has studied various copper forming options such as the skived fin heat sink (SFHS) manufacturing and 3D metal printing. Advanced drilling techniques are also being evaluated for the construction of the 4$\pi$ detector. 

The SFHS process is a precision copper-forming method to create fine structures for the DR calorimeter. Thin, closely spaced fins are sliced and bent from a solid copper block without any bonding and soldering. This technique ensures excellent mechanical stability and  mass-production manufacturing capability. The skived copper fins provide 
precise channels for the placement of optical fibers and are closed by adding rectangular-shaped copper plates. The Korean consortium has built 50 cm long prototype modules and a large (30x30x250 {cm}$^3$) prototype detector used in 2023 (CERN PS T9) and 2024 test-beams (CERN SPS H8), respectively~\cite{HDYoo_CEPC2024}.

3D metal printing is an advanced copper-forming technique that enables the fabrication of complex geometries with high precision. Using methods such as laser power bed fusion or binder jetting, pure copper can be printed layer by layer, allowing for intricate structures that would be difficult to achieve with traditional machining. A 50 cm long prototype module was built in 2022 and tested in 2023~\cite{SWKim_CALOR2024}.

\begin{figure} [htp]
\begin{center}
\includegraphics[width=0.65\textwidth]{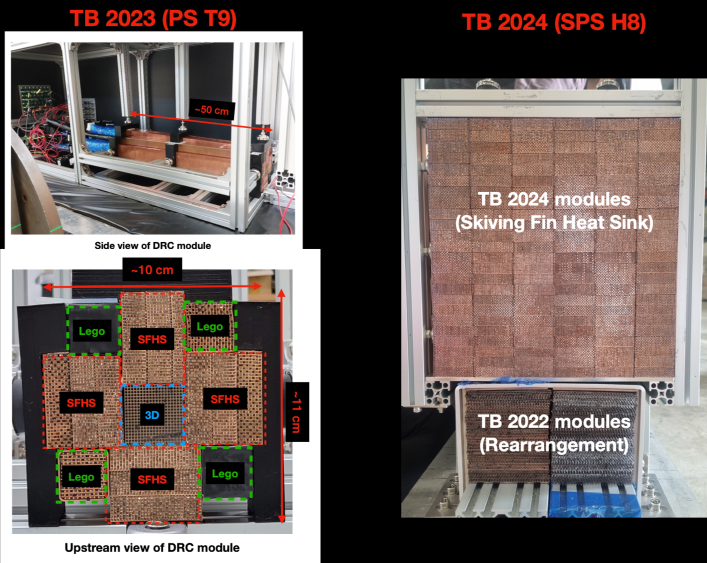}
\caption{DR prototypes using alternative mechanical options: 3D metal printing (left) and skiving fin heat sink (right). The two bottom modules on the right figure are based on the groove machining technique.}
\label{fig:figure_alternativeCuForm1}
\end{center}
\end{figure}

Each manufacturing option (shown in Fig.~\ref{fig:figure_alternativeCuForm1}) was also evaluated by measuring quantities such as air volume, radiation/nuclear interaction lengths, and sampling fraction. In particular the 3D metal printing provides excellent performance achieving smaller air volume and  $\lambda_I$. However it is the most expensive solution at the moment, therefore more investigation into manufacturing options is warranted. Drilling is also being considered as an alternative and more cost effective option to replace the 3D metal printing. The drilling technique was also observed to reduce the air volume up to 3.6$\%$, reducing $\lambda_I$ to 22.26 cm. 

\begin{table}[htp]
\centering
\caption{Parameters of alternative mechanical options}
\begin{tabular}{|l|c|c|c|c|c|c|}
\hline
Method & Cu ($\%$)& fibers ($\%$)& Air ($\%$)& $X_0$ (cm) & $\lambda_I$ (cm) & Sampling fraction ($\%$) \\
\hline
Ideal & 65.1 & 34.9 & 0 & 2.16 & 21.16 & S(4.02) C(4.54) \\
Machining & 54.1 & 32.7 & 13.2 & 2.59 & 25.10 & S(4.48) C(5.07) \\
SFHS & 48.1 & 33.6 & 18.3 & 2.91 & 27.77 & S(5.10) C(5.77) \\
3D printing / drilling & 57.8 & 34.9 & 7.3 & 2.43 & 23.54 & S(4.48) C(5.06) \\
\hline
\end{tabular}
\label{tableMechOptions}
\end{table}%

% added by HDYOO (end) %

\subsection{Signal readout and feature extraction}
%sensors (SiPMs and MCP-PMTs, dSiPMs ? ), frontend electronics (charge integrating ASIC and waveform samplers),  exploitation of DNNs as a function of readout architecture  -> Romualdo, Korean, Nural}
The light produced inside the calorimeter is transported via optical fibers to light sensors located on the back side of the absorber material, simplifying the services required for the operation of the whole detector. A series of prototypes previously equipped with PMTs were replaced by MicroChannel Plate PhotoMultiplier Tubes (MCP-PMTs)  or solid-state light detectors (SiPM). The latter are of great interest because they have single-photon sensitivity, high Photon Detection Efficiency (PDE), excellent timing resolution with magnetic field immunity and benefit from the fast evolution of silicon technology and investments by multiple companies to achieve high-quality mass production.

The use of SiPMs in fiber DR calorimetry also offers the possibility to sample the lateral shower profile with high granularity, improving particle ID performance and the identification of complex final states containing non-isolated objects (each SiPM is coupled to a single fiber). R\&D is ongoing in order to exploit the time profile of developing showers as additional information, providing an effective 3D segmentation as described in Section~\ref{sec:futurerandd}. 

An extensive R\&D program is being carried out within the collaboration with the aim of qualifying the different strategies in beam tests. In this phase, different readout architectures and light sensors are used to balance complexity with the possibility to study the calorimetric performance and calibration techniques. The results are then used to fine-tune Monte Carlo simulations and better constrain requirements for the light sensors and the readout scheme. The main R\&D programs are briefly discussed in next subsection, along with more prospective strategies that are now at an early stage but are  in line with the timeline of the physics programs.

\subsubsection{Front-end and readout electronics for hadronic prototypes}
\label{par:front-endAndReadout}

\begin{figure}
\begin{center}
\includegraphics[width=0.85\textwidth]{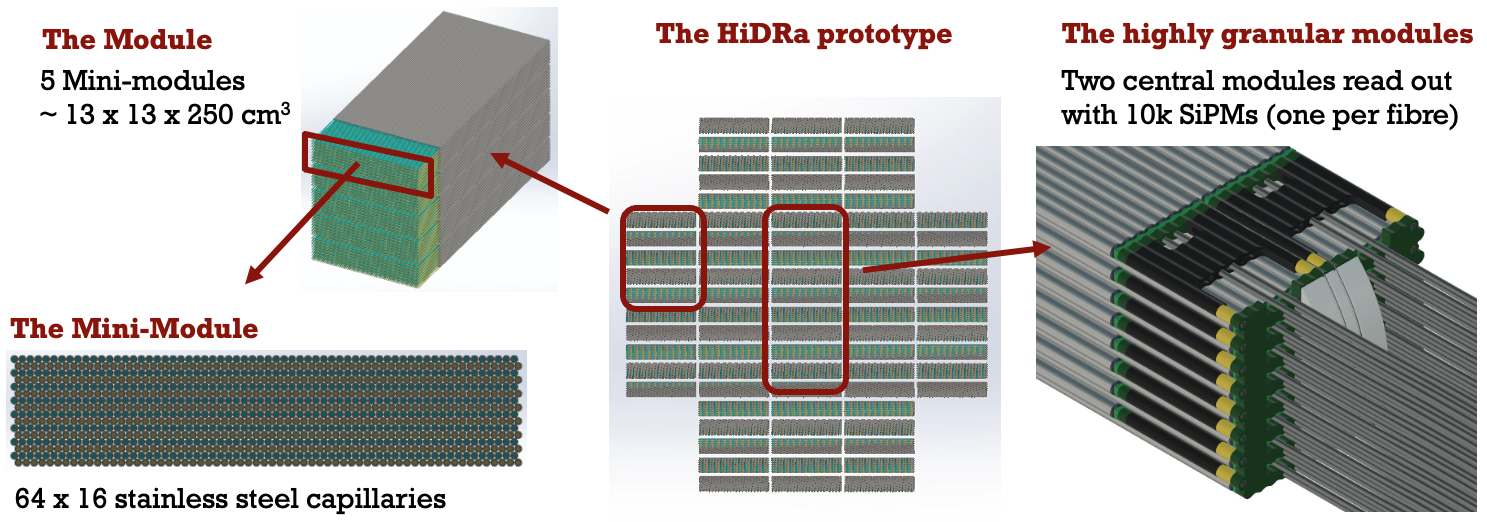}
\caption{The schematic view of HiDRa is shown in the center, the segmentation into modules and minimodules on the left and details of the mechanical integration of part of a highly granular minimodule on the right. The output signals from each group of 8 SiPMs are connected in parallel to feed a single readout channel.}
\label{fig:HidraLayout}
\end{center}
\end{figure}

A recent prototype was qualified on beam in 2021 and 2023 (see Section~\ref{par:TB_fibreSamplDR}). It was divided into 9 modules; the central one (highly granular) was equipped with 320 SiPMs (one per fiber). Among other results (see subsection~\ref{par:TB_fibreSamplDR}), this prototype was used to define a calibration procedure and to measure the number of photoelectrons (p.e.) per GeV produced by the Cherenkov and scintillating light generated in the shower( $\approx 60$\,p.e./GeV for Cherenkov and $\approx 280$\,p.e./GeV for scintillating light (\cite{calor2022})). 
These numbers were used as references for the SiPM and readout electronics specification.

A new generation of fiber DR prototype (HiDRa) is under construction to achieve hadronic containment and identify a true-scalable design for the central module equipped with SiPMs. The sensors used to equip the previous prototype were not mechanically compliant with the detector since the package ($2.63 \times 2.1$\,mm$^{2}$) did not fit in the space available in the rear part of the absorber (the tube outer diameter was 2\,mm). This requirement is even more strict because we need to avoid contamination between the scintillating and Cherenkov light collected by neighboring SiPMs. For this reason, we are using SiPMs with a sensitive area of $1 \times 1$\,mm$^{2}$ and an overall packaging of $1.1 \times 1.1$\,mm$^{2}$, that reduces the extra space to only 100\,$\mu$m. A custom PCB board will be equipped with 8 SiPMs with closely selected breakdown voltages (within 100\,mV) mounted to match the spacing of the fibers (2\,mm) and all SiPMs in the PCB will feed the same readout channel (see Fig. \ref{fig:HidraLayout}).

\begin{table}[htp]
\centering
\caption{Main parameters of the SiPM used for the EM-size prototype compared with the SiPMs considered for the HiDRa prototype. The values are extracted from the vendor's specifications and are relative to an operating temperature $\text{T} = 25^{\circ}$\,C. }
\begin{tabular}{|l|c|c|c|}
\hline
 &  &  &  \\
Parameter & S14160-1315PS & S16676-15(ES1) & S16676-10(ES1) \\
 &  &  &  \\
\hline
Effective photosensitive area (mm$^{2}$) & 1.3 $\times$ 1.3 & 1 $\times$ 1 & 1 $\times$ 1 \\
Pixel pitch ($\mu m$) & 15 & 15 & 10 \\
Number of pixels & 7284 & 3443 & 7772 \\
Recommended operating voltage (Vop) & +4 V & +4 V & +5 V \\
PDE at the Vop ($\%$) & 32 & 32 & 18 \\
Direct cross-talk at the Vop ($\%$) & $<$ 1 & $<$ 1 & $<$ 1 \\
Dark count rate (kHz) & 120 (360 max) & 60 (200 max) & 60 (200 max) \\
Gain (10$^{5}$) & 3.6 & 3.6 & 1.8 \\
\hline
\end{tabular}
\label{tableSiPM}
\end{table}%

The identified SiPM is customized by Hamamatsu to precisely fit the fiber spacing geometry. In Table \ref{tableSiPM}, we compare the main characteristics of the S14160-1315PS SiPM (sensor used in the EM-size prototype) with the two devices considered for the HAD-size prototype.
The S16676-15(ES1) SiPM will be used for the Cherenkov signals (which have a much lower light yield and need a high PDE) and the S16676-10(ES1) for the scintillation signals (that cover a wider dynamic range and need a lower gain to avoid saturation). This choice, based on light yield measurements with beam, will allow us to operate the sensor(s) in a linear regime over a large energy range. As shown in the table, by reducing the photosensitive area we have a reduction in the number of pixels which limits the linearity range by a factor of 2. This is not expected to be a problem for the Cherenkov light (more than 5 times less intense than the scintillating light) but it could be problematic for the scintillating light. This brought us to the conclusion to use SiPMs with a different microcell pitch.

Readout electronics can use different strategies i.e. charge integration or waveform sampling. The implementation of the charge integration with a precise time stamp (O(10-100) ps) is probably easier and could satisfy most of the requirements set by the dual-readout fiber calorimetry although additional information on timing and signal shape may provide a powerful input for event reconstruction, likely improving the performance of Particle Flow Algorithms (PFAs) applied to dual-readout calorimetry. 
The highly granular module of the HiDRa prototype will be read out with the CITIROC 1A by Weeroc \cite{citiroc:weeroc} integrated on front-end readout boards (FERS - A5202) produced by CAEN \cite{FERS:CAEN}.  A highly granular DR prototype (HG-DREAM) developed by TTU employs the DRS waveform digitizer (CAEN V1742) as well a 10-14 GS/s waveform digitizer (AARDVARC)  by NALU Scientific. Alternative solutions continue to be considered (\ref{par:alternativesDR_readout}).

% added by HDYOO (begin) %
\subsubsection{Alternative front-end and readout electronics based on MCP-PMTs}
\label{sec:alternativeMCPPMT}
In the current design, one photosensor is connected to each fiber, exploiting the maximum possible granularity of the DR calorimeter and the SiPM is considered the best solution. The MCP-PMT is an advanced photon detection device that combines the high sensitivity of a traditional PMT with fast timing and spatial resolution. The SiPMs are compact, solid-state devices with high photon detection efficiency, low operating voltage, and immunity to magnetic fields, however, they suffer from higher dark noise and limited timing resolution. In contrast, MCP-PMTs offer superior time resolution at picosecond scale, excellent single-photon detection capability, and lower noise. 

\begin{figure} [tph]
\begin{center}
\includegraphics[width=0.85\textwidth]{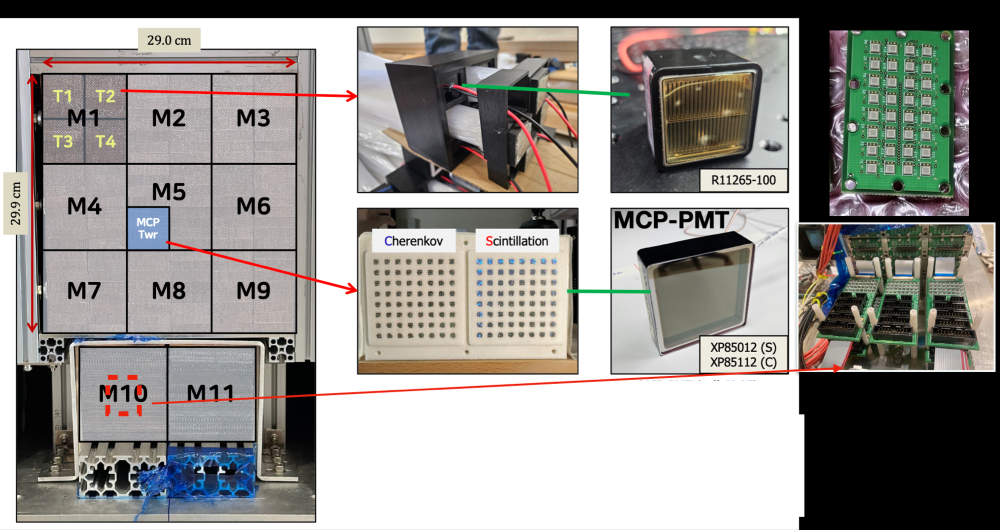}
\caption{High granularity system installed in the DR prototype that has been tested at CERN SPS H8 in 2024, built by Korean consortium.}
\label{fig:figure_mcppmt}
\end{center}
\end{figure}

The Korean consortium has explored not only SiPM-based (400 channels) but also MCP-PMT based (128 channels) signal readouts for high granularity detector systems in the DR prototype used in the 2024 test-beam experiment. Figure ~\ref{fig:figure_mcppmt} shows the installed MCP-PMT and SiPM readout systems in the prototype~\cite{HDYoo_CEPC2024}. 

The weaknesses of the MCP-PMTs are that they are more fragile, expensive, and susceptible to magnetic fields. Such problems can be addressed by recent R\&D performed by the Argonne National Laboratory (ANL) group~\cite{ANL_MCPPMT} in collaboration with the  Korean consortium. New MCP-PMT fabrication methods  led to a $10\times 10$\,cm$^2$ large area picosecond photosensor device. This R\&D program uses 3D printed MCP technology to provide low-cost, highly-pixelated MCP-PMTs working in high radiation and magnetic field environments. The customized design will be produced and tested in the near future. 

A  data acquisition (DAQ) readout electronics system with fast timing resolution is essential to reconstruct the longitudinal profile of a hadronic shower in the unsegmented DR calorimeter. A customized DAQ system developed with NOTICE Korea digitizes and records analog signals from the readout system using 20 DAQ boards and a Trigger Clock Board (TCB). Each DAQ board contains 32 channels, 640 channels in total, and utilizes the Domino Ring Sampler 4 (DRS4) chip, developed at PSI~\cite{drs4-manual},  to digitize signals with 12-bit ADC resolution in 1024 time bins. The TCB board synchronizes all DAQ boards, allowing simultaneous signal digitization upon receiving a trigger signal. It has archived under 100\,ps of timing resolution on the data processing with multiple channels. Further studies and upgrades will be performed in the future~\cite{Korea_TB2022}.
% added by HDYOO (end) %

\subsubsection{Perspectives for readout architectures}
\label{par:alternativesDR_readout}

Single Photon Avalanche Diodes (SPADs) are highly sensitive photodiodes capable of detecting single photons by operating beyond breakdown voltage, and SPAD arrays (SiPMs) offer a valid alternative to PMTs. While the optimal performance is typically achieved through the independent optimization of sensors and readout electronics, CMOS technologies have recently gained ground for SPAD implementations. This trend is driven by commercial applications, such as laser ranging, which have prompted semiconductor companies to optimize CMOS technologies for SPADs. 
As a result, more foundries are offering SPAD-compatible processes, which allow for improved detector performance: peak photon detection probability (PDP) nearing 50$\%$, dark count rates below 1\,MHz/mm$^2$, timing jitter of a few tens of picoseconds (\cite{gramuglia22}). 
Several manufacturers are developing 3-D integrated SPADs designed for consumer applications. These technologies further improve the trade-off between fill factor and embedded sensor functionality, although the costs can be significant for the field of high-energy physics. 

ASPIDES is a two-year INFN project, approved in 2025, which aims to develop a planar detector using standard CMOS technology, naturally suited for a monolithic sensor in which the front-end electronics and sensing element share the same substrate.  The 2-D monolithic sensors will feature:
\begin{itemize}
\item an all-digital output via a digital processing chain or analogue-to-digital conversion;
\item asynchronous counting over a dynamic range of three decades of the fired micro-cell;
\item time-of-arrival and duration measurements with a resolution better than 100 ps;
\item adjustable thresholds for noise rejection.
\end{itemize}

%(1) an all-digital output via a digital processing chain or analogue-to-digital conversion, (2) asynchronous counting over a dynamic range of three decades of the fired micro-cell, (3) time-of-arrival and duration measurements with a resolution better than 100 ps, and (4) adjustable thresholds for noise rejection. 
These capabilities meet the requirements of various applications, including dual-readout calorimetry.

\begin{figure}
\begin{center}
\includegraphics[width=0.85\textwidth]{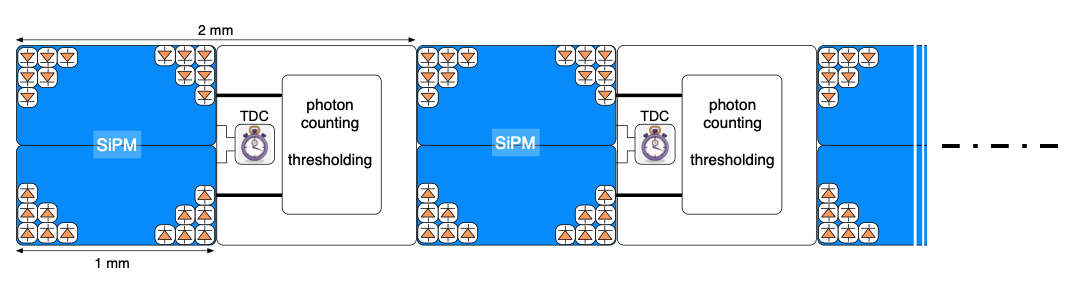}
\caption{Modular structure for a dSiPM in dual-readout calorimetry.}
\label{fig:dSiPM_layout}
\end{center}
\end{figure}

The fabrication process selected for the sensor implementation is based on 110\,nm CMOS image sensor technology. The technology offers SPADs with a Dark Count Rate (DCR) of few hundreds of kHz/mm$^2$, a Photon Detection Probability (PDP) close to 50$\%$ at 450\,nm \cite{garcia18} and a sufficient integration density to encapsulate all necessary functions in the available area without unacceptably degrading the PDE of the sensor. A simplified block diagram of the dSiPM to be developed by ASPIDES illustrates how the sensing area can be subdivided into several macrocells to optimize electronic processing. In applications where detector geometry allows this, the processing electronics will be positioned outside the sensitive area to maximize the fill factor and PDE. 
For dual-readout calorimetry, ASPIDES aims for a linear array of 8 SiPMs, 
$1 \times 1$\,mm$^2$ each and a pitch of 2\,mm (see Fig. \ref{fig:dSiPM_layout}). This layout allows inter-SiPM regions to accommodate processing circuits, enhancing integration without impacting detection efficiency. 

\subsection{Test beam results} 
\label{par:TB_fibreSamplDR}
%-> Iacopo, Giacomo, Korean

A dual-readout calorimeter prototype using the capillary tube  technique described in Section~\ref{sec:capillary_tubes} was tested with beam in 2021 and  2023. Nine identical modules with size  $3.3 \times 3.3 \times 100\,\mathrm{cm^3}$ were glued together, yielding a prototype with size of about $10 \times 10 \times 100\,\mathrm{cm}^3$. Each module was assembled by gluing together 320 100-cm long brass capillary tubes, each enclosing a 100-cm long optical fiber, in alternate rows of scintillating and clear (Cherenkov) fibers. The effective radiation length was estimated to be 22.7\,mm, while the Moli\`ere radius was 23.8\,mm. 

Each of the external eight modules was instrumented with two Hamamatsu R8900 PMTs~\cite{R8900}, one optically connected with the bundle of Cherenkov fibers, the other with that of scintillation fibers. In the central module, each fiber was instead read out by an individual SiPM, as described in Section \ref{par:front-endAndReadout}, connected to the commercial FERS system produced by CAEN~\cite{FERS:CAEN}.

Based on the experience with the RD52 dual-readout prototypes, a set of auxiliary detectors placed on the H8 beam line at the CERN SPS, were used to assess the quality of the beam and to identify positrons among the beam particles. Particularly relevant for the results shown in this section are (moving in the direction of the beam): a set of three gas Cherenkov counters, where the $H_2$ pressure was set to optimize the separation between $\pi/\mu$ and $e$ for beam energies up to about 40\,GeV; a set of three thin plastic scintillators used to build a particle trigger signal; a pair of delay wire chambers, used to select particles from the center of the beam; a preshower detector, formed by a 1 $X_0$ of lead glued to a scintillator, which aided in the selection of high-energy electrons. All these detectors were upstream along the beam with respect to the prototype. Downstream with respect to the prototype, and behind a thick concrete shielding, was a plastic scintillator dedicated to the identification of muons.  

After a first gain-equalization step based on the SiPM multiphoton spectrum, the response from the nine modules was equalized by exposing them to a 20\, GeV beam aimed at their geometrical center: after the application of a positron selection, the response of each module was set to be equal to that of the central module. After the equalization step, the overall calorimeter scale was set by aiming a 20\, GeV positron beam to the geometrical center of the prototype front face, and imposing that the calorimeter response was equal to the beam energy corrected for the detector shower containment. While doing this, a set of fine energy corrections for the external towers was also estimated. 

\begin{figure}[htbp]
    \centering
    \subcaptionbox{Linearity of the response.}
      [.49\linewidth]{\includegraphics[trim={0 0 0 0},clip,width=\linewidth]{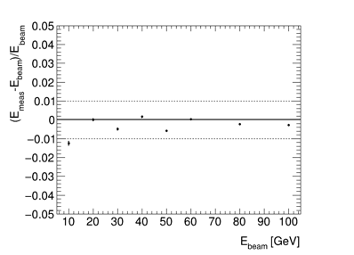}}
      %}
    %Sx Bot Dx Top
    \hfill
    \subcaptionbox{Resolution of the response.}
      [.49\linewidth]{\includegraphics[trim={0 0 0 0},clip,width=\linewidth]{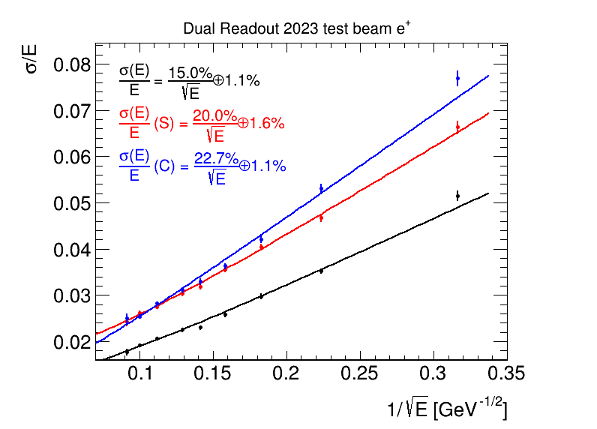}}
      %}
    %Sx Bot Dx Top
    \caption{Performance of the energy measurement when exposing the fibre calorimeter prototype to beams of positrons of various energies.} \label{fig:calo_fibre_response}
\end{figure}

The calorimeter response obtained with the 2023 test beam campaign is illustrated in Fig.~\ref{fig:calo_fibre_response}. The calorimeter response to positrons is linear in the range $10-100\ \mathrm{GeV}$. The combined energy resolution of the Cherenkov and scintillation channels displays a stochastic term of 15\%. The constant term (1.1\%) is affected by an intrinsic energy spread of the H8 beam line, which was estimated by the beam physicists to be of the order of 1\%.   

\begin{figure}[htb]
\begin{center}
	\includegraphics[width=0.5\textwidth]{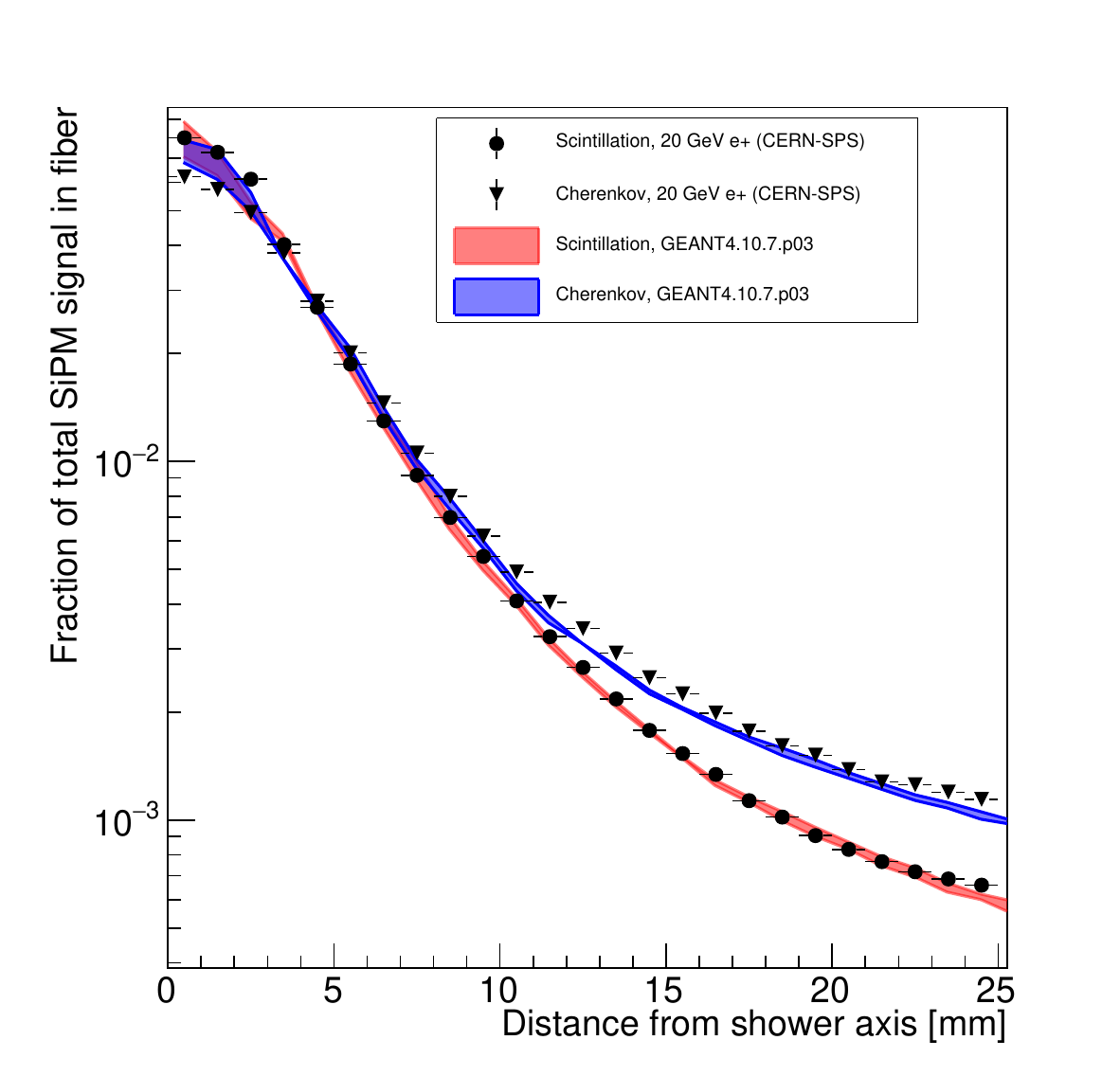}
\caption{Lateral shower profile of positrons with an energy of 20\,GeV. The black circles (triangles) refer to the scintillation (Cherenkov) signal in the central prototype module. The red and blue bands correspond to the simulation predictions for the scintillation and Cherenkov signal,  respectively.}
\label{fig:lateral_profile}
\end{center}
\end{figure}

Thanks to the single fiber readout of the central module and the small distance of 2~mm between the individual fibers, a highly granular measurement of the lateral development of electromagnetic showers could be performed and compared to the predictions of Geant4 test-beam simulations. The details of the measurements are discussed in Ref.~\cite{Ampilogov:2023zxb}. The lateral shower profile was defined as the fraction of energy deposited in a single fiber at a distance $r$ from the shower axis, identified as the the $x-y$ barycenter of the shower. The measurement is compared to the simulation predictions in Fig.~\ref{fig:lateral_profile}. 

The shower profile of the Cherenkov signal is wider than the one measured with the scintillation light. This observation confirms those of Refs.~\cite{Akchurin:2005rs} (for a different calorimeter setup) and \cite{Antonello:2018sna}. This is understood to be due to the fact that the early components of electromagnetic showers are collimated with the incoming positron direction and the Cherenkov emitted light falls outside the numerical aperture of the fiber. Overall, the result is a good display of the possibilities opened by a highly granular readout of the prototype, and a proof of the excellent description of electromagnetic showers in Geant4 simulations.

\begin{figure} [htbp]
\begin{center}
\includegraphics[width=0.85\textwidth]{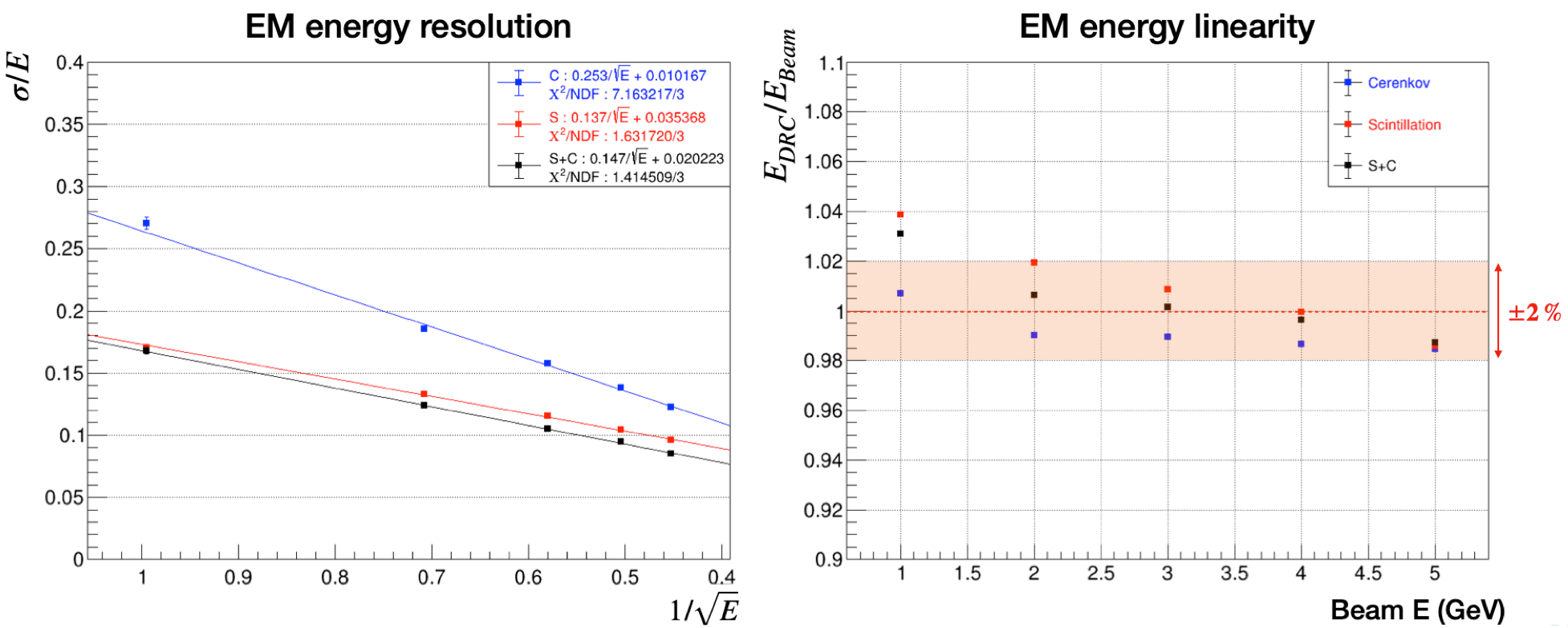}
\caption{Performance of the energy measurement using alternative prototypes with low energy positron beams: (left) resolution  and (right) linearity.}
\label{fig:figure_alternativeCuForm}
\end{center}
\end{figure}

% added by HDYOO (begin) %
DR prototype detectors built by alternative mechanical options to form fine copper structures have been tested in beam experiments in 2023 at CERN PS T9 with low-energy positron beams from 1 to 5~$\mathrm{GeV}$, which has never been explored in previous tests, including those reported by RD52. The modules and DAQ systems are described in section~\ref{sec:alternative_cuform} and~\ref{sec:alternativeMCPPMT} and experimental setups are similar to the descriptions above. The calorimeter response obtained with the low-energy test beam is shown in Fig.~\ref{fig:figure_alternativeCuForm}. The calorimeter response to positrons is mostly linear in the range $1-5\ \mathrm{GeV}$. The combined energy resolution of the Cherenkov and scintillation channels displays a stochastic term of 15\% and a constant term of 2.0\%, which is similar to the higher energy  tests~\cite{SWKim_CALOR2024}. 

% added by HDYOO (end) %

%subsection{Detector geometry and technologies}

%\subsection{Hadron and jet energy resolution with hybrid dual-readout method}\label{sec:hybrid_calo_performance}

%\textcolor{red}{The idea is to discuss here the resolution to hadrons and jets when combining the crystal (ecal) and fiber (hcal) calorimeter sections.
%Here could be briefly discussed the dual-readout method in a hybrid calorimeter and possibly the performance for jets.}

%\begin{figure}[htbp]
%    \centering
    %\includegraphics[width=0.495\linewidth]{figs/crystal/em_resolution.png}   
 %       \caption{Single hadron energy resolution and linearity as a function of the hadron energy.}
        %\label{fig:single_had_hybrid_performance}
%\end{figure}

%\begin{figure}[htbp]
%    \centering
        %\includegraphics[width=0.495\linewidth]{figs/crystal/cCompareRes_DRO_PFA_B2T__TRKSMEAR.pdf}
%        \caption{Jet energy resolution and linearity as a function of the jet energy for with and without the use of a custom dual-readout particle flow algorithm \cite{Lucchini_2022}.}
        %\label{fig:jet_hybrid_performance}
%\end{figure}

\subsection{Future R\&D}
\label{sec:futurerandd}
\begin{figure}[!htbp]
\centering
\includegraphics[width=10.3cm,clip]{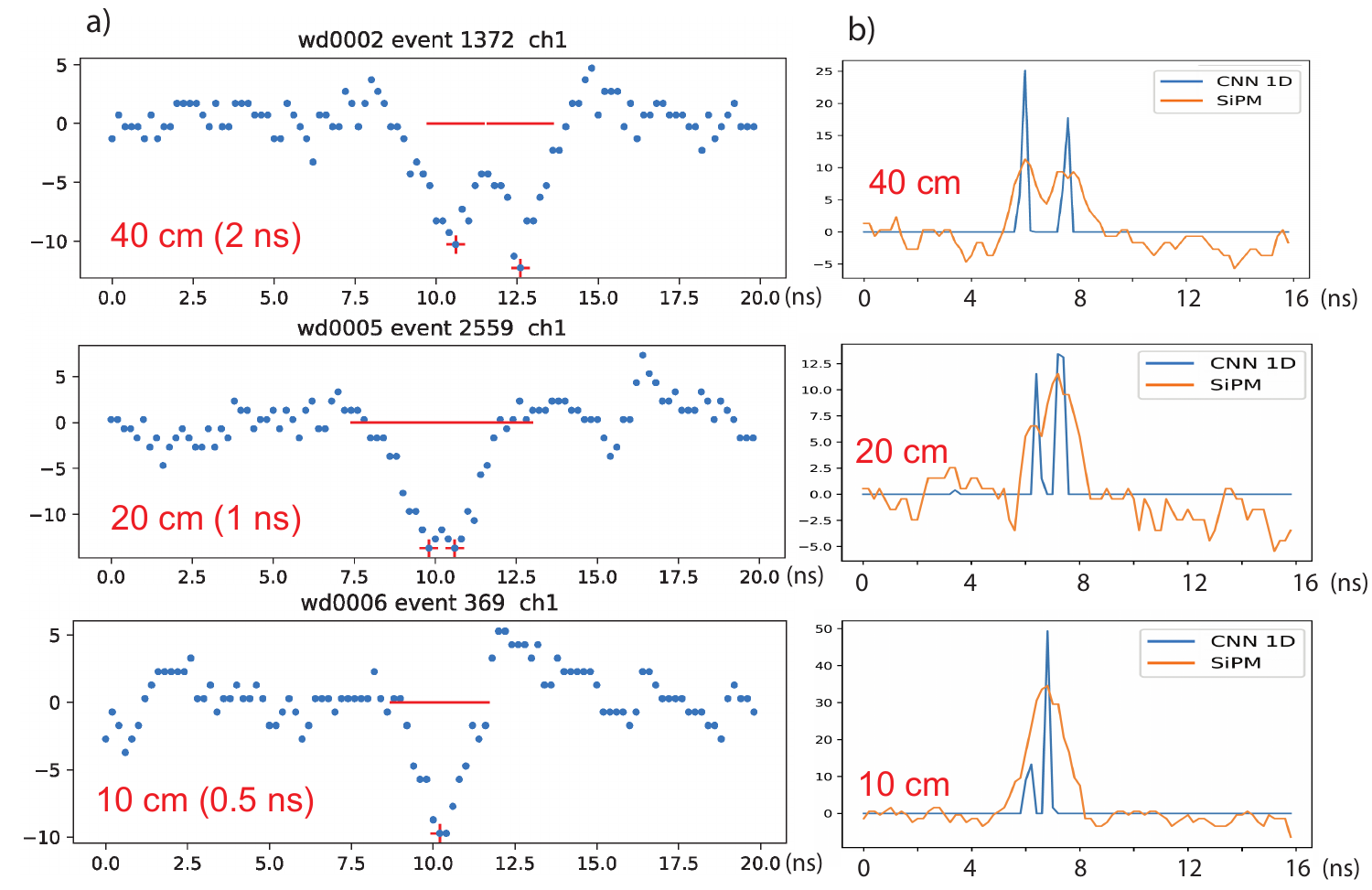} 
\caption{\small (a) The pulse doublets plotted in the left column were generated by an electron beam on fused-silica fibers that differed by lengths of 40, 20, and 10\,cm, but read out by the same SiPM (MicroFC 3$\times$3 mm$^2$ with Mini-Circuit TB-409-S66+ amplifier), and digitized by CAEN V1742.  The speed of light in fuse-silica fibers is approximately 20\,cm/ns.  It is clear that a 20-40 cm separation between energy deposits could be identified.   (b) When the energy deposition was separated by 10\,cm or less, however, the identification of the individual pulses in a train becomes possible when a convolutional neural network (CNN) algorithm is applied as illustrated by the bottom plot on the right column of pulse trains.  The vertical axes are in units of ADC counts where one photon approximately corresponds to 10 counts.
}
\label{fig:pulses}       
\end{figure}

The fiber calorimeter described here is longitudinally unsegmented. However, longitudinal segmentation can be achieved by timing the trains of Cherenkov photon pulses detected by the SiPMs at the fiber ends. This approach requires fast SiPMs, high-rate sampling waveform digitizers, and potentially on-detector intelligence to process and reduce data volume. For instance, a time resolution of 200\,ps corresponds to an effective longitudinal segmentation of approximately 4\,cm. Figure ~\ref{fig:pulses} shows the results of a feasibility test.

Longitudinal segmentation combined with fine transverse segmentation enables the calorimeter to ``image’’ showers, facilitating particle identification in jets through correlation with tracker information and providing access to jet substructure for identifying boosted $W,Z,H$, and top quarks \cite{Akchurin_CALOR2024}.
Incorporating time ($t$) as an additional measurement dimension for hadronic showers, along with neural network regression techniques such as Graph Neural Networks (GNNs) \cite{Bronstein_2017}, enhances energy measurement accuracy \cite{Akchurin_2021, instruments6040043}. The development of digital SiPMs (dSiPMs), as discussed in Section~\ref{par:alternativesDR_readout}, offers significantly improved timing resolution over analog SiPMs, underscoring the need for continued R\&D in this area.

Further improvements to the DR technique may be achieved by selecting clear fibers with different refractive indices. By using a higher index material ({\it e.g.}, 1.77 for sapphire instead of 1.46 for fused silica), the Cherenkov threshold for protons can be reduced from 350\,MeV to 200\,MeV, thereby capturing more abundant spallation protons and improving energy resolution. Additionally, the time delay between signals from fibers with different refractive indices provides a method for determining the energy deposit location within the calorimeter without requiring a reference timing signal from the accelerator \cite{Kunori_CALOR2024}.\\

% added by HDYOO (begin) %
Advancements in 3-D metal printing and drilling indicate that the intricate calorimeter absorber structure could be manufactured in the near future. Several prototypes have already been produced and tested~\cite{SWKim_CALOR2024} and new prototype with the drilling technique will be followed up. 
% added by HDYOO (end) %

\clearpage\newpage
\section{Muon system \label{Mu}}
%\subsection{Muon system requirements}

\begin{figure}[!htbp]
    %\captionsetup{justification=raggedright}  
    \begin{minipage}[b]{.45\textwidth}  
        \centering  
        \includegraphics[trim={0 0 0 0},clip,width=\linewidth]{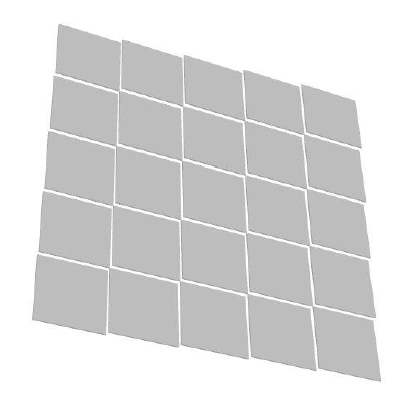}
        %Sx Bot Dx Top
        %\vspace{1cm} % per l'allineamento fine
    \end{minipage}\hfill
    \begin{minipage}[b]{.45\textwidth}  
        \centering  
        \includegraphics[trim={0 0 0 0},clip,width=\linewidth]{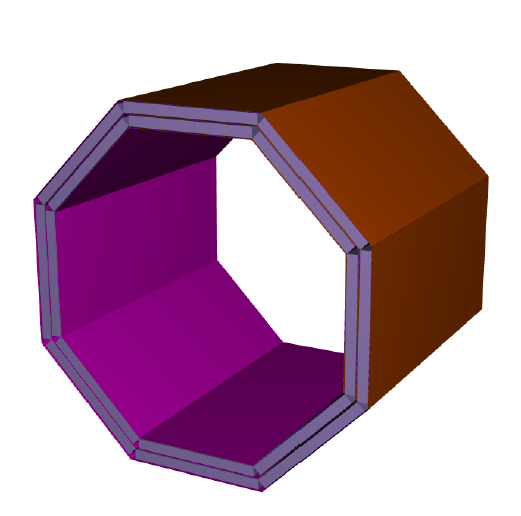}
        %Sx Bot Dx Top
        %\vspace{0.1cm} % per l'allineamento fine
    \end{minipage}
     \par
    \begin{minipage}[t]{.35\textwidth}
        \caption{Geant4 visualization of 50$\times$50 cm$^2$ $\upmu$-RWELL tiles in the muon apparatus.}
        \label{fig:muon}
    \end{minipage}\hfill
    \begin{minipage}[t]{.35\textwidth}  
        \caption{The barrel muon detection system for the IDEA detector.}
        \label{fig:muon_pre}
    \end{minipage}  
\end{figure}

The Muon detection system will follow the IDEA geometry, featuring a central cylindrical barrel region closed at both ends by two endcaps to ensure hermeticity (Figs.~\ref{fig:muon} and~\ref{fig:muon_pre}). This apparatus will consist of three or more layers of detectors covering the barrel and endcap regions, housed within the iron yoke that encloses the solenoidal magnetic field. 
Preliminary simulation studies indicate that the multiple scattering of muons originating from Z${^0}$ boson decays introduces a loss of positional accuracy of a few millimeters upon reaching the first muon detection layer.  Conversely, muons decaying from long-lived particles within the calorimeter apparatus exhibit a significantly smaller loss of accuracy, of the order of hundreds of microns, at the first detection layer. In addition to the effects of multiple scattering, momentum measurement performance is also considered. A spatial resolution of a hundred microns is required to achieve the precision necessary for accurate momentum reconstruction of long-lived particles.

To fulfill the required spatial resolution, the muon apparatus will utilize $\upmu$-RWELL detectors~\cite{MUON:micro-RWELL}, innovative, single-stage, and compact gaseous detectors belonging to the Micro-Pattern Gaseous Detector (MPGD) family.
To take advantage of the industrial production capabilities of this technology, a modular design has been adopted for the muon detection layers. Each basic $\upmu$-RWELL tile features an active area of 50$\times$50\,cm${^2}$ with a two dimensional strip readout.
A strip pitch of approximately 0.4$\div$1.5\,mm provides a typical spatial resolution in the range of 100$\div$500 $\mu$m, which corresponds in 2,500$\div$640 readout channels per tile.

The choice of detector tile size, strip pitch, and strip width represents a compromise among several factors: the largest $\upmu$-RWELL detector that can be industrially mass-produced, the maximum input detector capacitance tolerable by the Front-End Electronics (FEE) to maintain an adequate signal-to-noise ratio, the spatial resolution required by the IDEA experiment, and the  costs of electronics for the readout channels, which need to remain within reasonable budget constraints.

%The Muon detection system will follow the IDEA geometry, featuring a central cylindrical barrel region closed at both ends by two endcaps to ensure hermeticity (fig.~\ref{fig:muon},~\ref{fig:muon_pre}). 
%This apparatus will consist of three layers of detectors covering the barrel and endcap regions, housed within the iron yoke that encloses the solenoidal magnetic field.
%The chosen technology for the muon apparatus is the $\upmu$-RWELL detectors~\cite{MUON:micro-RWELL}, an innovative, single-stage, and compact gaseous detector belonging to the Micro-Pattern Gaseous Detector (MPGD) family.
%To take advantage of the industrial production capabilities of this technology, a modular design has been adopted for the three detection layers. Each basic $\upmu$-RWELL tile features an active area of 50$\times$50 cm${^2}$ with a two dimensional strip readout. A strip pitch of approximately 1.5\,mm provides a sufficient spatial resolution ($\mathcal{O}$ 500 $\mu$m) while minimizing the number of readout channels. This results in 670 readout channels per tile for the muon detector.\\
%The choice of the detector tile size, the strip pitch and strip width represents  a compromise between the largest $\upmu$-RWELL detector that can be industrially mass-produced and the maximum input detector capacitance tolerable by Front-End Electronics (FEE) in terms of S/N ratio.\\

%Table~\ref{tab:barrel-muon} summarizes the dimensions, the number of basic $\upmu$-RWELL tiles and the readout channels for three muon stations of the IDEA experiment.
%

\begin{figure}[htbp]
    %\captionsetup{justification=raggedright}  
    \begin{minipage}[b]{.49\textwidth}  
        \centering  
        \includegraphics[trim={0 0 0 0},clip,width=0.95\linewidth]{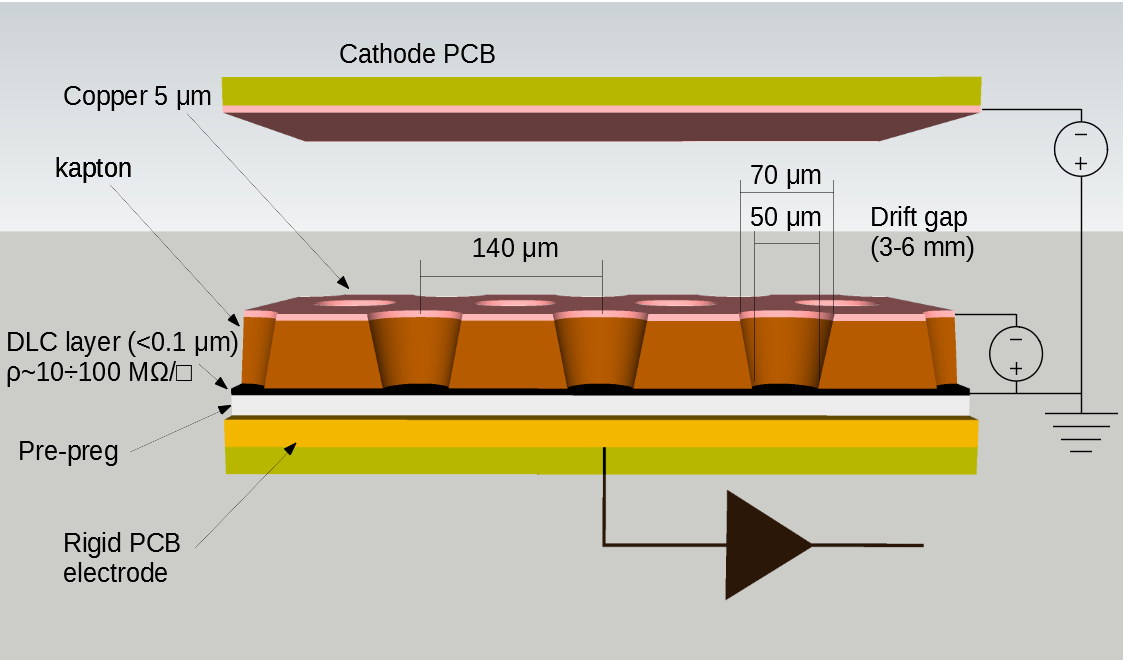}
        \caption{Layout of the \rwell.}
        \label{rwell-substrate}
        %Sx Bot Dx Top
        %\vspace{1cm} % per l'allineamento fine
    \end{minipage}
    %\hfill
    \begin{minipage}[b]{.49\textwidth}  
        \centering  
        \includegraphics[trim={0 20 0 20},clip,width=0.95\linewidth]{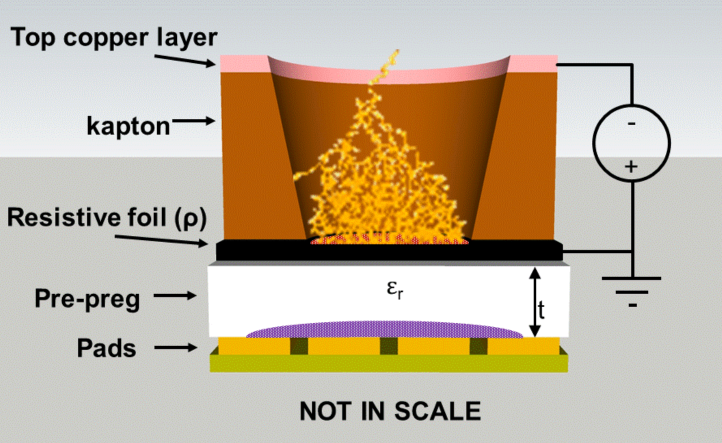}
        \caption{Principle of operation of the \rwell.}
        \label{principle-operation}
        %Sx Bot Dx Top
        %\vspace{1cm} % per l'allineamento fine
    \end{minipage}
    \par
\end{figure}

\begin{figure}[htbp]
    \begin{minipage}[b]{.49\textwidth}  
        \centering  
        \includegraphics[trim={0 20 0 20},clip,width=0.95\linewidth]{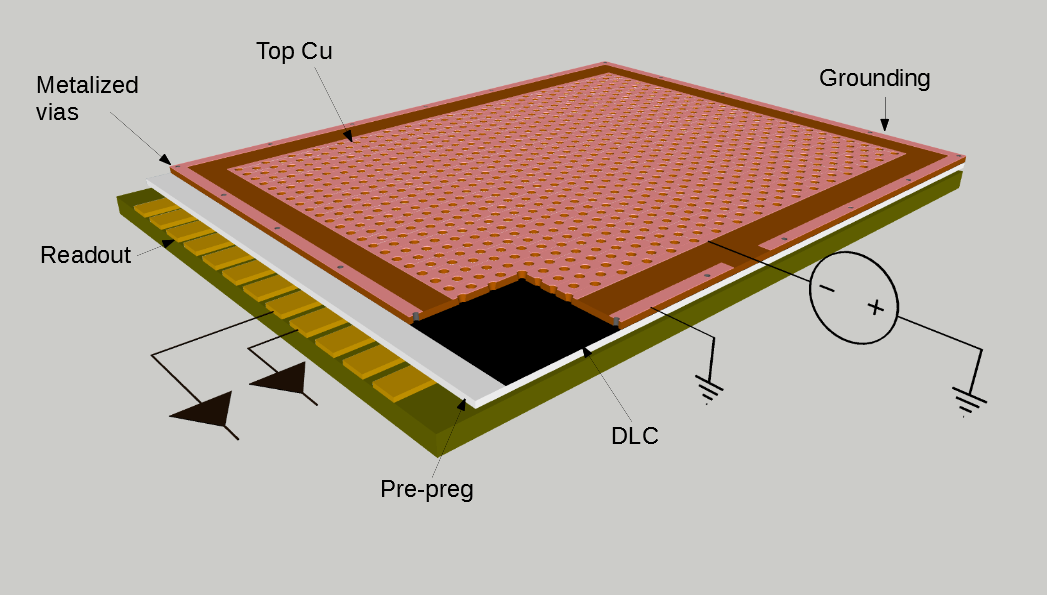}
         \caption{Sketch of the Single-Resistive layout.}
        \label{resistive-stage1}
        %Sx Bot Dx Top
        %\vspace{1cm} % per l'allineamento fine
    \end{minipage}
    %\captionsetup{justification=raggedright}  
    \begin{minipage}[b]{.49\textwidth}  
        \centering  
        \includegraphics[trim={0 0 0 0},clip,width=0.97\linewidth]{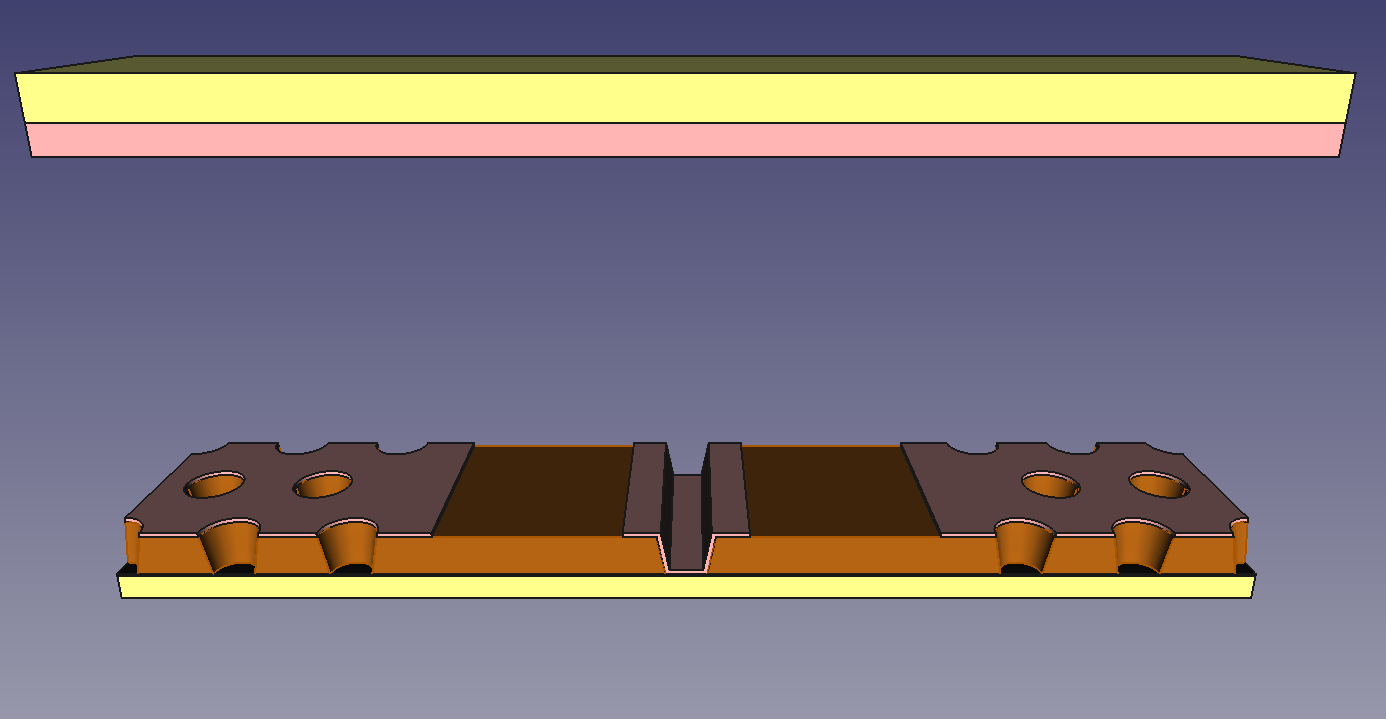}
         \caption{Sketch of the PEP layout.}
         \label{fig:sketch-PEP}
         %Sx Bot Dx Top
        %\vspace{1cm} % per l'allineamento fine
    \end{minipage}
    %\hfill
    \par
\end{figure}

%\begin{table}[hptb]
%\centering
%\caption{Dimensions of the three IDEA barrel muon stations, together with the number of detector tiles and the corresponding number of readout channels.}
%\label{tab:barrel-muon}
%\begin{tabular}{@{}cccccccc@{}}
%\toprule
%Station &
%  \begin{tabular}[c]{@{}c@{}}Radius\\ {[}m{]}\end{tabular} &
%  \begin{tabular}[c]{@{}c@{}}Lenght\\ {[}m{]}\end{tabular} &
%  \begin{tabular}[c]{@{}c@{}}Strip Pitch\\ {[}mm{]}\end{tabular} &
%  \begin{tabular}[c]{@{}c@{}}Strip Lenght\\ {[}mm{]}\end{tabular} &
%  \begin{tabular}[c]{@{}c@{}}Area\\ {[}m$^2${]}\end{tabular} &
%  \begin{tabular}[c]{@{}c@{}}\# of\\ tiles\end{tabular} &
%  Channels \\ \midrule
%1 & 4.52 & 9.0   & 1.5 & 500 & 260 & 1040 & 0.7$\times$10$^6$ \\
%2 & 4.88 & 9.0   & 1.5 & 500 & 280 & 1120 & 0.7$\times$10$^6$ \\
%3 & 5.24 & 10.52 & 1.5 & 500 & 350 & 1400 & 1.$\times$10$^6$ \\ \bottomrule
%\end{tabular}
%\end{table}

%\noindent The geometries and material of both barrel and end-cap have been implemented in the FCCSW Geant4 full simulation for the muon system. The algorithms for digitization and clustering will be implemented in 2025-2026.\\
Since the $\upmu$-RWELL technology has not yet been used to realize a full detector system, a rigorous R\&D program will be undertaken in the coming years to address integration issues. 
%The detailed layout of the detector, together with all its services, will need to be carefully developed. 
R\&D on the $\upmu$-RWELL is performed in synergy with Working Package 1 (WP1) of the Detector R\&D Collaboration for Gaseous Detectors (DRD1)~\cite{MUON:DRD1}. Another key aspect of the R\&D program will be the design and development of a dedicated FEE system based on a custom-made ASIC.

\subsection{Technology choice and detector design}\label{secTech}

% \textcolor{red}{Including R\&D studies - about 2.5 pages\\
% Editors: Gianni, Marco, Gianfranco

% \textcolor{red}{\it Ho inserito le varie citazioni nella sezione ma i riferimenti bibliografici vanno inseriti nel file bibliogragraphy.bib}

The  \rwell, Fig.~\ref{rwell-substrate}, is a single-amplification stage resistive 
MPGD~\cite{MUON:micro-tpc, MUON:urwell2019, MUON:urwell2018, MUON:micro-RWELL}, uniquely combining the advancements and innovations achieved in the MPGD field over recent years. 
The R\&D on \rwell aims to enhance stability under high irradiation while simplifying the construction procedures in view of an easy technology transfer to industry: an essential milestone for large-scale applications in fundamental research at the future colliders.

The detector is composed of two main components: the cathode, a simple Printed Circuit Board (PCB) with a thin copper layer on one side, and the $\upmu$-RWELL\_PCB, the core of the detector. The $\upmu$-RWELL\_PCB is a multi-layer circuit realized using standard photo-lithography techniques. It is composed of a WELL-patterned single copper-clad polyimide (Apical{\textsuperscript\textregistered}) foil\footnote{50 $\upmu$m thick polyimide covered on one side with 5 $\upmu$m thick copper, similar to the GEM base material.} that acts as the detector amplification element; a resistive layer, realized with a Diamond-Like-Carbon (DLC) film sputtered onto the bottom side of the polyimide foil, that acts as discharge limitation stage, and a standard PCB for readout purposes, segmented into strip, pixel or pad electrodes.
%
\iffalse
\begin{figure}[htbp]
    %\captionsetup{justification=raggedright}  
    \begin{minipage}[b]{.49\textwidth}  
        \centering  
        \includegraphics[trim={0 0 0 0},clip,width=0.95\linewidth]{figs/muon_figs/X-section_label.png}
        \caption{Layout of the \rwell.}
        \label{rwell-substrate}
        %Sx Bot Dx Top
        %\vspace{1cm} % per l'allineamento fine
    \end{minipage}
    %\hfill
    \begin{minipage}[b]{.49\textwidth}  
        \centering  
        \includegraphics[trim={0 20 0 20},clip,width=0.95\linewidth]{figs/muon_figs/avalanche.png}
        \caption{Principle of operation of the \rwell.}
        \label{principle-operation}
        %Sx Bot Dx Top
        %\vspace{1cm} % per l'allineamento fine
    \end{minipage}
    \par
\end{figure}
\fi

By applying a suitable voltage between the copper layer and the DLC, the WELL acts as a multiplication channel for the ionization produced in the drift gas gap (Fig.\ref{principle-operation}). 
The charge induced on the resistive stage spreads with a time constant~\cite{MUON:dixit} determined by the DLC surface resistivity and the distance between the resistive layer and the readout plane.

The spark suppression mechanism, similar to that of Resistive Plate Counters (RPCs)~\cite{MUON:RPC3, MUON:RPC1}, relies on a localized voltage drop caused by the streamer-induced current in the resistive layer, quenching the  multiplication processes. This allows the detector to operate at high gains ($\geq$10$^4$) with a single amplification stage.
A drawback, correlated with the Ohmic behavior of the detector, is the reduced capability to stand high particle fluxes. This effect is correlated to the average resistance faced by the charge produced in the avalanche which depends on the distance between the particle incidence position and the detector grounding line.
In a \rwell, the Single-Resistive Layout (SRL) (Fig.~\ref{resistive-stage1}) uses a single resistive layer with a grounding line around the active area. However, for large devices, the current path to the ground can be long and depends on the particle's incidence point. This issue is mitigated by introducing a denser grounding network as shown in the new high rate PEP layout\footnote{The name PEP is the acronym of the three main processes involved in the production of this layout: Patterning-Etching-Plating.}  (Fig.~\ref{fig:sketch-PEP}), recently introduced for the Phase II Upgrade of the Muon apparatus in the LHCb experiment~\cite{MUON:LHCbUpg2TDR}. The PEP layout involves creating conductive grooves etched through the Cu/Kapton foil to the DLC, forming a 2D high-density current evacuation grid. For the IDEA muon system, with rates up to 1\,kHz/cm$^2$, PEP grooves are planned every 10\,cm.

As an alternative to the baseline solution for the muon system based on the uRWELL technology, we are also considering ongoing developments based on resistive Micromegas~\cite{Alviggi:2024lir}~\cite{DellaPietra:2025mpgd}, which offer potential synergies and shared advancements.

\subsection{Layouts description and results}
% Editors: Riccardo, Marco 
%
In recent years, the R\&D program has focused on two main objectives: optimizing the DLC resistivity and strip pitch to minimize the number of electronics channels while achieving the required spatial resolution for the muon apparatus; and developing a 2-D layout capable of efficient and stable operation.
The study on DLC resistivity demonstrated stable and consistent performance within the resistivity range of 40$\div$80\,$\un{M\Omega/\Box}$. This finding relaxes any strict homogeneity requirement for DLC resistivity over a large-area \rwell and enhances the reliability of performance uniformity across such an area. At lower DLC resistivity ($\leq$ 10$\un{M\Omega/\Box}$), the efficiency plateau is reached at higher HV due to a slightly increased charge spread and threshold-related effects. 

The study on strip pitch from 0.4 to 1.6 mm shows that as the pitch increases, the collected signal charge decreases due to geometric and threshold effects, requiring higher gain for full efficiency. Additionally, the number of fired strips decreases, bringing spatial resolution closer to the $\mathrm{pitch}/\sqrt{12}$ limit. 

\begin{figure}[htbp]
    \centering
    \subcaptionbox{Tracking efficiency for different HV.\label{fig:rdfcc2022-efficiency}}%
      [.49\linewidth]{\includegraphics[trim={0 0 0 0},clip,width=\linewidth]{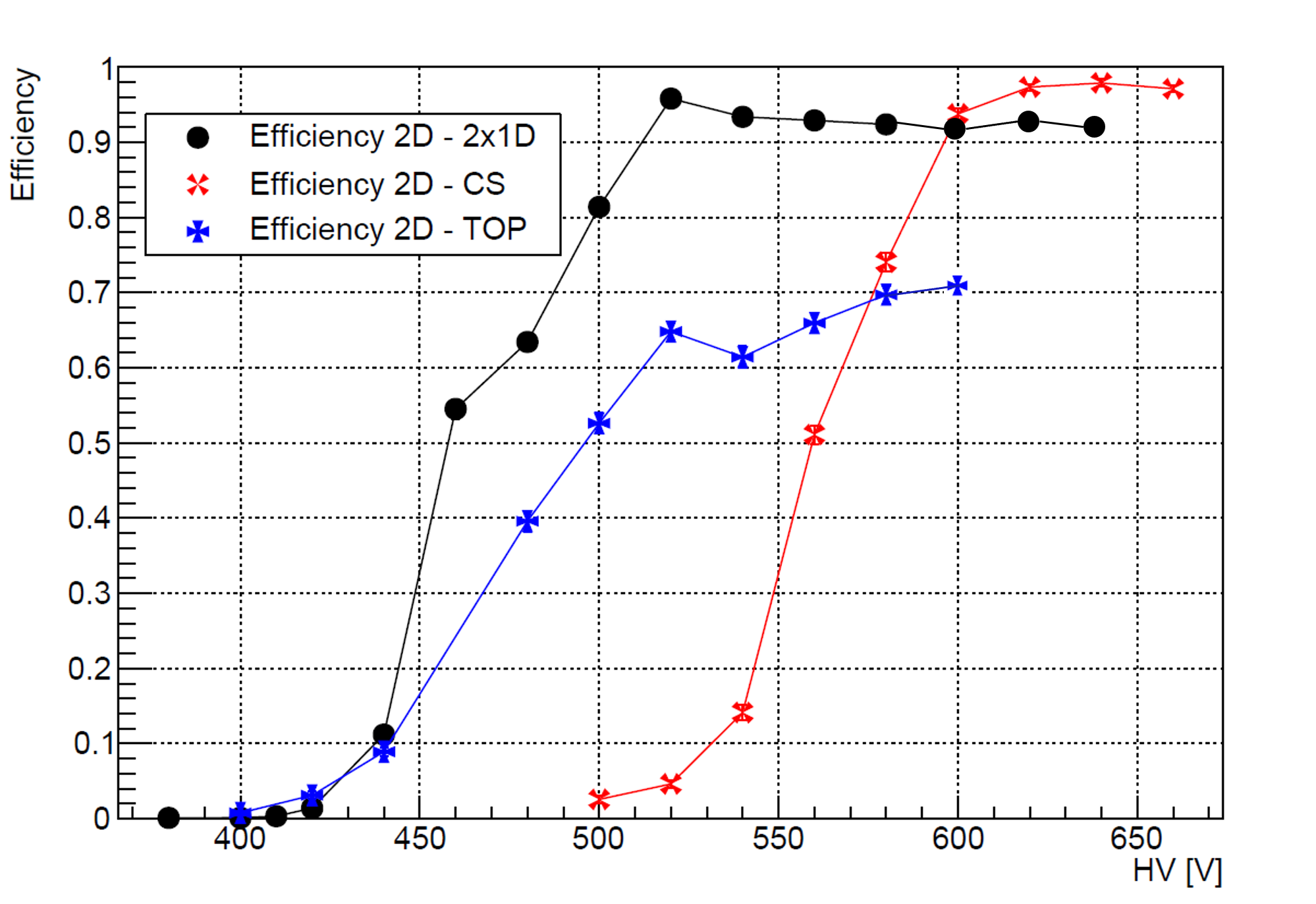}}
      %rdfcc_2D_eff_hv.pdf}}
      %}
    %Sx Bot Dx Top
    \hfill
    \subcaptionbox{Residuals width for different HV.\label{fig:rdfcc2022-residuals}}
      [.49\linewidth]{\includegraphics[trim={0 0 0 0},clip,width=\linewidth]{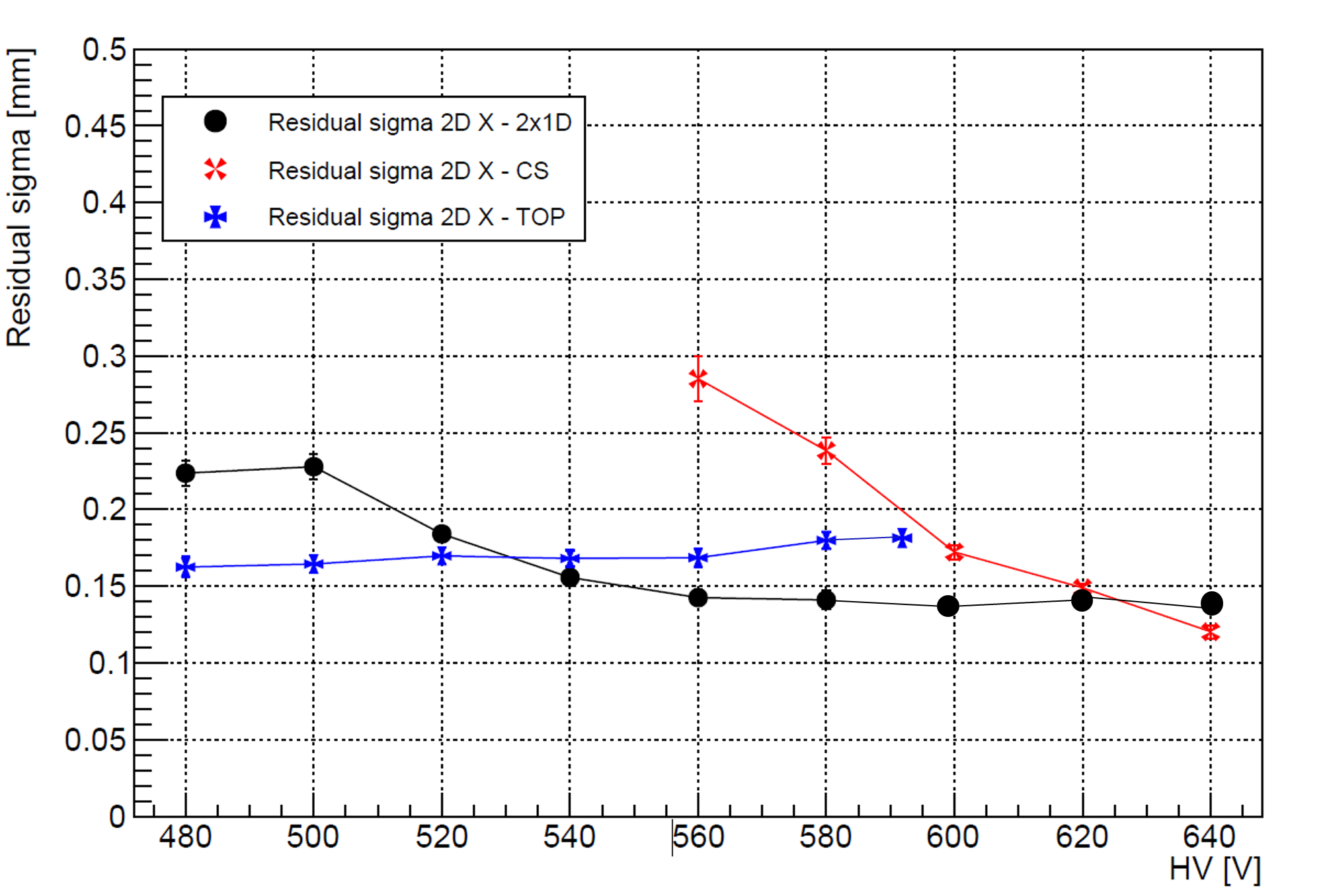}}
      %rdfcc_2D_res_hv.pdf}}
      %}
    %Sx Bot Dx Top
    \caption{Results of the 2D layouts test campaign.} \label{fig:rdfcc_2D-total}
\end{figure}

Besides the tuning of the parameters (resistivity, strip pitch and strip width) still in progress, the first ideas for the two-dimensional readout of a \rwell have been designed. % and are under construction.
A commonly used 2-D layout involves embedding two parallel layers of strips in the readout plane at a defined angle (e.g., XY or XV), as implemented in COMPASS triple-GEM detectors~\cite{MUON:altunbas2002}. 
However, this approach, which equally shares the charge between the two views, requires a high detector gain and is therefore not optimal for a single-stage amplification detector such as the the \rwell, as discussed in Ref~\cite{MUON:LevShekhtman}. Alternative 2-D readout designs are being explored, with three layouts currently under study:
%One of the most common 2-D layout techniques is to embed in the readout plane two parallel layers of strips, with a defined angle between them (XY, XV), as done for example for COMPASS triple GEM detectors~\cite{MUON:altunbas2002}. In this layout (\emph{à la Compass}) the charge is equally shared between the two readout views, thus requiring a high detector gain. Of course, for a single amplification stage detector, such as the \rwell, this is not be the optimal choice as discussed in \cite{MUON:LevShekhtman}. Alternative approaches are investigated and here are reported three layout under study:
\begin{compactitem}
\item two one-dimensional detectors, coupled through a common cathode. This is the simplest solution as it consists of using two standard detectors. They can operate at the usual gas gain with the readout strips completely separated. This layout is a feasible option thanks to the \rwell overall small thickness, due to the compactness of the \rwellPCB.
\item a single two-dimensional \rwell, with the standard 1-D readout on the PCB plus a strip patterned top electrode for the second coordinate. 
The main advantage of this layout 
%with respect to the previous readout scheme 
is that it does not require an increase of the detector gas gain. 
%In this option, the top electrode and the PCB readout strips are grounded while the DLC  layer is kept at a positive voltage.
\item a single two-dimensional \rwell based on the capacitive-sharing anode readout as described in Ref~\cite{MUON:gnavo2023}.
%The concept of capacitive-sharing is based on a spatial arrangement of metallic pads between dielectric layers in a vertical stack of these layers that allows the transfer of the induced charge signal from the \rwell amplification through capacitive coupling between the pad layers to the coarsely segmented strip electrodes of the anode readout PCB of the detector.  
This layout offers high spatial performance with a significant reduction of electronics channels required to read out a very large area apparatus.
\end{compactitem}
The first versions of these designs were produced between 2022 and 2023. A dedicated test beam campaign evaluated the performance of the three layouts using a muon beam at the SPS-H8 beamline at CERN. The results, shown in Fig.~\ref{fig:rdfcc_2D-total}, demonstrate good spatial resolution for all three layouts. However, the efficiency results reveal an efficiency plateau of approximately 70\% for the second layout (blue lines), due to dead areas on the amplification electrode required for segmentation. The third layout (red lines) delivers very good performance despite requiring a higher high voltage (HV) on the amplification stage. The R\&D for this activity will continue to optimize the layout configurations, ensuring 2-D readout performance while maintaining detector operational stability.

\subsection{Electronics}\label{secElec}
%Editors: Gigi, Riccardo
The results shown in the previous section are evaluated with APV-25 electronics and SRS readout system \cite{MUON:SRS, MUON:APV}. This readout system is widely used in the gas detector R\&D community but it can not be used in experiments such as LHC or FCC due to a maximum trigger rate of about 1\,kHz. Alternatives are needed to profit from the performance of the ASIC and a design optimized for the \rwell and its final layout. A possible solution is identified in the TIGER/GEMROC system which is a compact, modular, scalable, and highly customizable~\cite{MUON:TIGER}. 
%The chip TIGER (Torino Integrated GEM Electronics for Readout) was developed in recent years for the readout of the Cylindrical Gas Electron Multiplier (CGEM) of the BESIII experiment.
% It is a mixed-signal chip with 64 channels on an area of 5$\times$5\,mm$^2$. 
% Each channel has a full readout chain with amplification, signal conditioning, and discrimination, as well as a data payload containing the timing and charge information for each event. 
TIGER is versatile for the readout of radiation sensors up to 50\,fC and for high rates up to 60\,kHz. Its output is passed to two shapers optimized for time and charge measurement. The peak time of the shaper for the time branch was set to the expected charge collection time (60\,ns) to enable measurements with low jitter. The shaper for the energy branch has a slower peak time (170\,ns) for better charge resolution and ENC optimization. 
%A dedicated off-detector electronics was developed at INFN-Ferrara using FPGA-based modules. These readout cards (GEMROC) provide interfaces to up to eight TIGERs in terms of power supply and data processing. 
% At the heart of each GEMROC is a development kit based on an Intel/ ALTERA ARRIA V GX family FPGA connected to an interface card designed for the CGEM detector. The GEMROC modules can process the received data in two different ways. In triggerless mode, the data received from the activated TIGERs is merged and transmitted via the Ethernet output port using the UDP protocol. In trigger-matched mode, a finite state machine selects the hits to be transmitted as follows. The incoming data from each TIGER pair is stored in a latency buffer circular memory, which is divided into pages of 32 memory locations each. This memory is intended to buffer the incoming TIGER data until a selection trigger occurs. 
A first integration test between \rwell and TIGER was performed on a 10$\times$10 cm$^2$ prototype. 
%A noise scan from a FEB containing two TIGER chips is reported in fig.~\ref{fig:rdfcc2024-tiger_noise}. 
The average noise is 0.3-0.4\,fC allowing the detector to operate at a threshold of about 1\,fC. %Further characterization of the TIGER/GEMROC performance as \rwell readout are ongoing cosmic ray and testbeam using the same set of \rwell with 40\,cm long strips and a pitch of [0.4,1.2,1.6]\,mm. Due to the higher length of the strips, the noise level increased to 2-3\,fC. 
% Nevertheless, the reconstruction of information as time and charge had no large impact as shown in fig. \ref{fig:rdfcc2024-tiger_tb}.
Further studies between TIGER and \rwell integration are ongoing to confirm the expected performance achieved with APV/SRS system. In the future, an optimization of the TIGER parameters is planned to match the needs of the \rwell layout chosen to optimize the performance of the 2-D readout.

\iffalse
\begin{figure}[htbp]
    \centering
    \subcaptionbox{Noise measurement of the 128 channels of two TIGER chips connected to a \rwell.\label{fig:rdfcc2024-tiger_noise}}%
      [.49\linewidth]{\includegraphics[trim={0 0 0 0},clip,width=\linewidth]{figs/muon_figs/noise_tiger_urwell_10cm.png}}
      %}
    %Sx Bot Dx Top
    \hfill
    \subcaptionbox{Time and charge distribution of hits recorded with TIGER electronics.\label{fig:rdfcc2024-tiger_tb}}
      [.49\linewidth]{\includegraphics[trim={0 0 0 0},clip,width=\linewidth]{figs/muon_figs/tiger_charge_time.png}}
      %}
    %Sx Bot Dx Top
    \caption{Integration of \rwell and TIGER chip.} \label{fig:TIGER}
\end{figure}
\fi

\vspace{1cm}

\clearpage\newpage

\section{Simulation and Performance \label{CombPerf}}

The specific FCC-ee experimental environment and its rich electroweak, QCD, and flavour physics programme offered by the very large event samples anticipated at the Z boson resonance 
(the so-called Tera-Z run), bring entirely new challenges to the detector requirements previously driven by the needs of a Higgs factory at a linear collider. 
The statistical uncertainties expected on key electroweak measurements, both at the Z resonance and at the $WW$ threshold, call for a superb control of the systematic uncertainties,  and put commensurate demands on the acceptance, construction quality and stability of the detectors.
The specific discovery potential of feebly coupled particles in the huge FCC-ee event samples 
should also be kept in mind when designing the detectors. 
The IDEA detector concept has been used as a baseline for several physics benchmark analyses to understand the needs of the detector performance for the FCC-ee physics program. 

The FCC software fully adopts Key4hep\cite{key4hep}, a comprehensive data processing framework that support all phases of research, including event generation, detector simulation, data analysis, and visualization of simulated data. Such a comprehensive approach is necessary to explore and maximize the physics reach of the proposed detector solutions. 

The physics studies conducted during the Feasibility Study phase\cite{FCCMidTermReport} have been based mainly on a parameterized response of the detector, using Delphes, which provides a flexible tool for many purposes. However, several applications, like the study of the effects from the various backgrounds, require a more detailed simulation of the detector response. Within the Key4hep framework, the detector description is handled by DD4hep\cite{Gaede:2020tui} (Detector Description for High Energy Physics), which has been used to model the geometry of the sub-detectors composing the IDEA spectrometer. 
Once the full simulation of the detector is available, the digitization step needs to be implemented to allow the development of the reconstruction algorithms. 
Finally, a full event reconstruction with particle flow needs to be available to perform more complete physics performance studies. 
In the meantime, several approaches have been taken to be able to extract meaningful information with the available software tools. A summary of the most relevant performance estimates for the IDEA components follows. 

\subsection{IDEA tracking system performance and Particle Identification} 

\begin{figure}[htbp]
    \centering
    \includegraphics[width=1.0\textwidth]{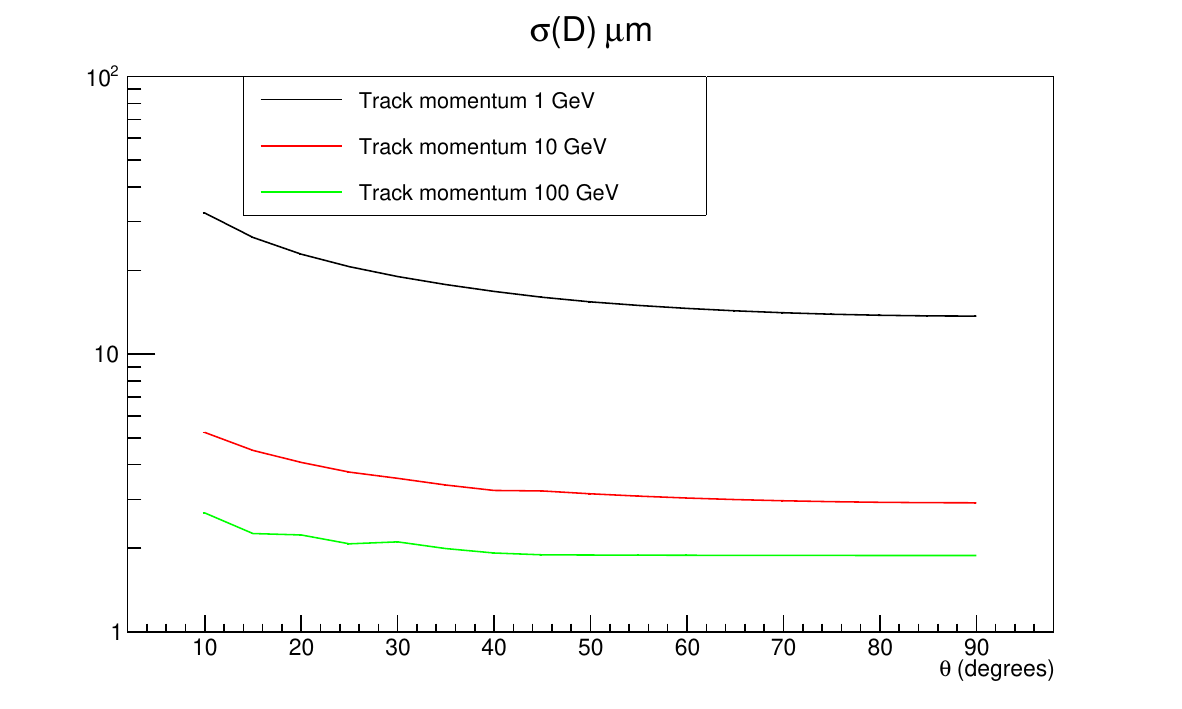}
    \caption{Transverse impact parameter resolution for different $\theta$ and muon momenta (Delphes) }
    \label{fig:idea_vertex_performance_comparison}
\end{figure}

\begin{figure}[htbp]
    \centering
    \includegraphics[width=1.0\textwidth]{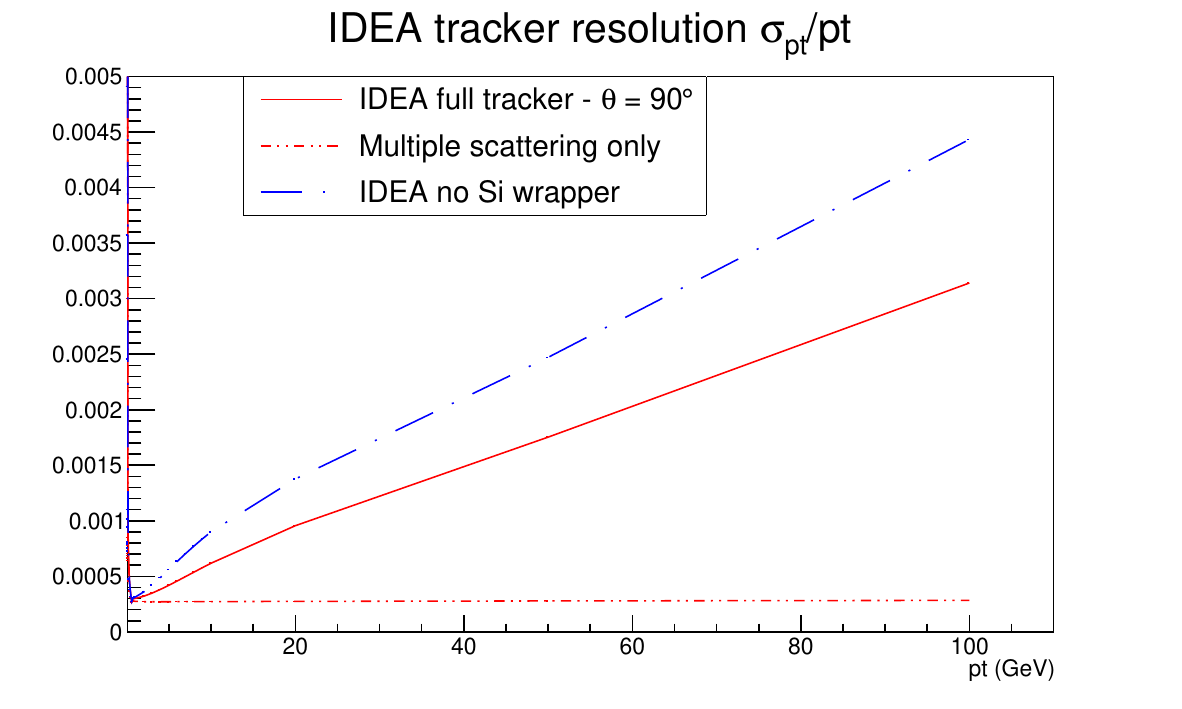}
    \caption{Transverse momentum resolution as a function of transverse momentum for the IDEA tracking system (Delphes). The (red) continuous line is the resolution of the baseline detector, while the (blue) dashed line shows the resolution without the silicon wrapper. The (red) dotted horizontal line shows the multiple scattering contribution.}
    \label{fig:idea_resplot}
\end{figure}

Since the IDEA tracking system, track reconstruction and track refit, still needs to be fully implemented in Key4hep,  a compromise solution has been chosen to estimate its performance. This includes modifying of the structure of the \textsc{DELPHES} Fast Simulation package in order to allow a full determination of the track covariance matrix that takes into account the detector geometry with the specific point resolutions and the presence of multiple scattering from the material contained in the tracking volume.  This fast tracking simulation is equivalent to performing a full fit of the tracks in the proposed detectors. The expected transverse impact parameter resolution as a function of polar angle, $\theta$, for some values of the transverse momentum, $p_t$, is shown in Fig.~\ref{fig:idea_vertex_performance_comparison}, indicating an asymptotic resolution for high momenta of $\sim 2\,\mu m$ over most of the polar angle range. The transverse momentum resolution is shown in Fig.~\ref{fig:idea_resplot} reaching $\sigma(p_t)/p_t \simeq 0.3\%$ for $p_t = 100$ GeV/c. The figure also shows how this resolution degrades if the silicon wrapper is removed. The high transparency of this tracking system is clearly understood from the small multiple scattering contribution as shown by the (red) dotted line.

The IDEA tracking system also provides excellent particle identification performance with cluster counting in the drift chamber and possibly timing in the silicon wrapper.

\begin{figure}[ht]
\centering
\includegraphics[width=0.55\linewidth]{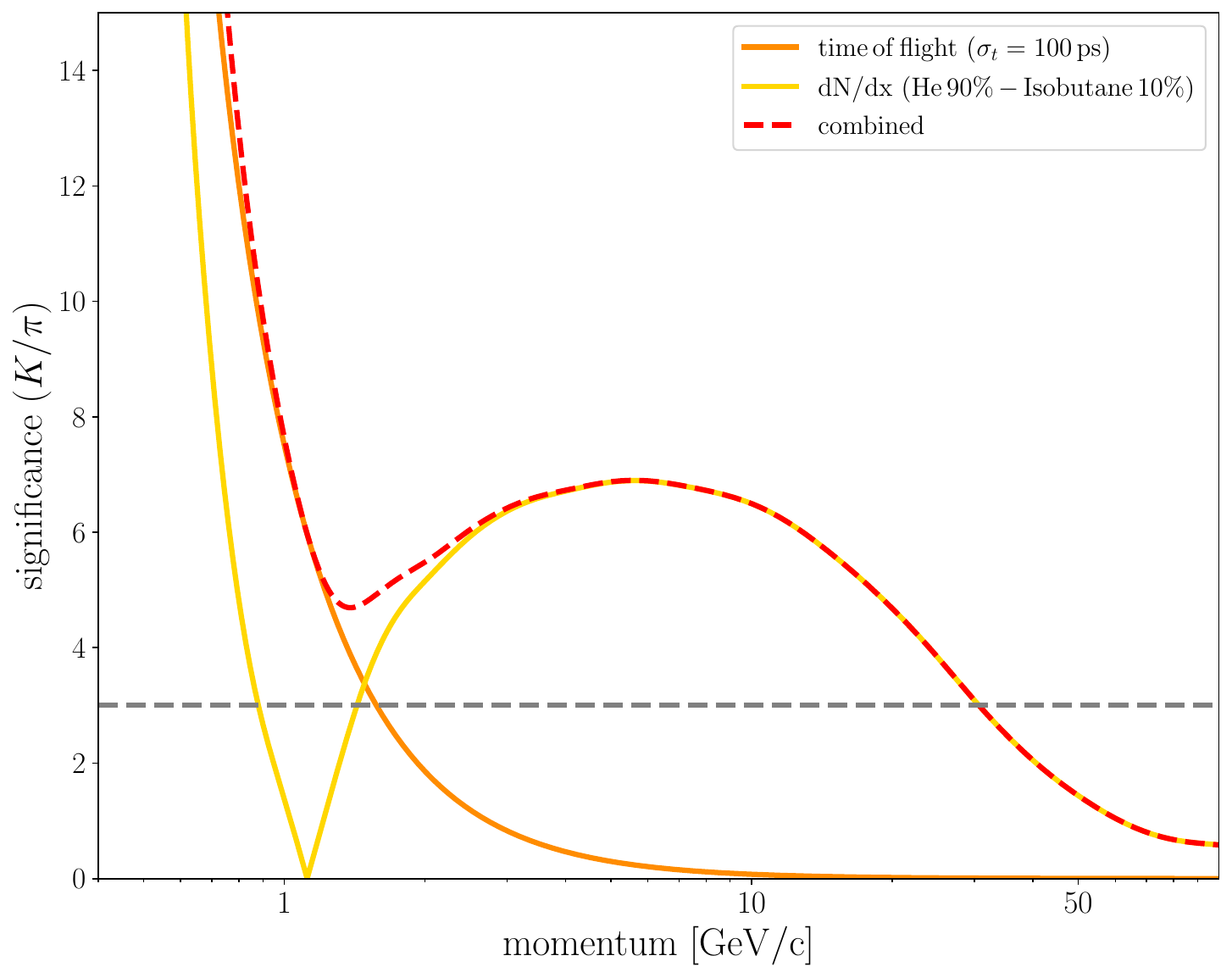}
\caption{$K/\pi$ separation significance (in standard deviations) of the IDEA drift chamber,
from a fast \textsc{Delphes} simulation study, with cluster counting and
using a drift chamber performance parametrised from \textsc{Garfield++} results.
A better than 3\,$\sigma$ separation is obtained up to momenta $\sim$\,30\,GeV. 
The $p<1.6$\,GeV region is covered by a time-of-flight system,  extending over a distance of 2\,m and having a timing resolution assumed to be 100\,ps.}
\label{fig:dch4}
\end{figure}

The cluster counting/timing technique provides a $\pi/K$ separation better than three standard deviations 
up to momenta about 30\,GeV, 
except in a narrow gap between 0.9 and 1.6\,GeV, 
where the Bethe--Bloch energy-loss curves for the two particle types cross. 
These values are obtained with a fast simulation study performed with \textsc{Delphes},
parametrizing \textsc{Garfield++}~\cite{Veenhof:1998tt} results. 
As illustrated in Fig.~\ref{fig:dch4}, 
the gap can be adequately covered with a time-of-flight measurement over a distance of 2\,m,
with a non-challenging resolution of $\mathcal{O}(100)$\,ps. 

Comprehensive studies of jet flavor tagging with the IDEA detector are shown in Ref~\cite{Bedeschi:2022rnj} along with detailed descriptions of these fast tracking and PID algorithms. 

\begin{figure}[H]
    \centering
    \includegraphics[width=0.5\textwidth]
    {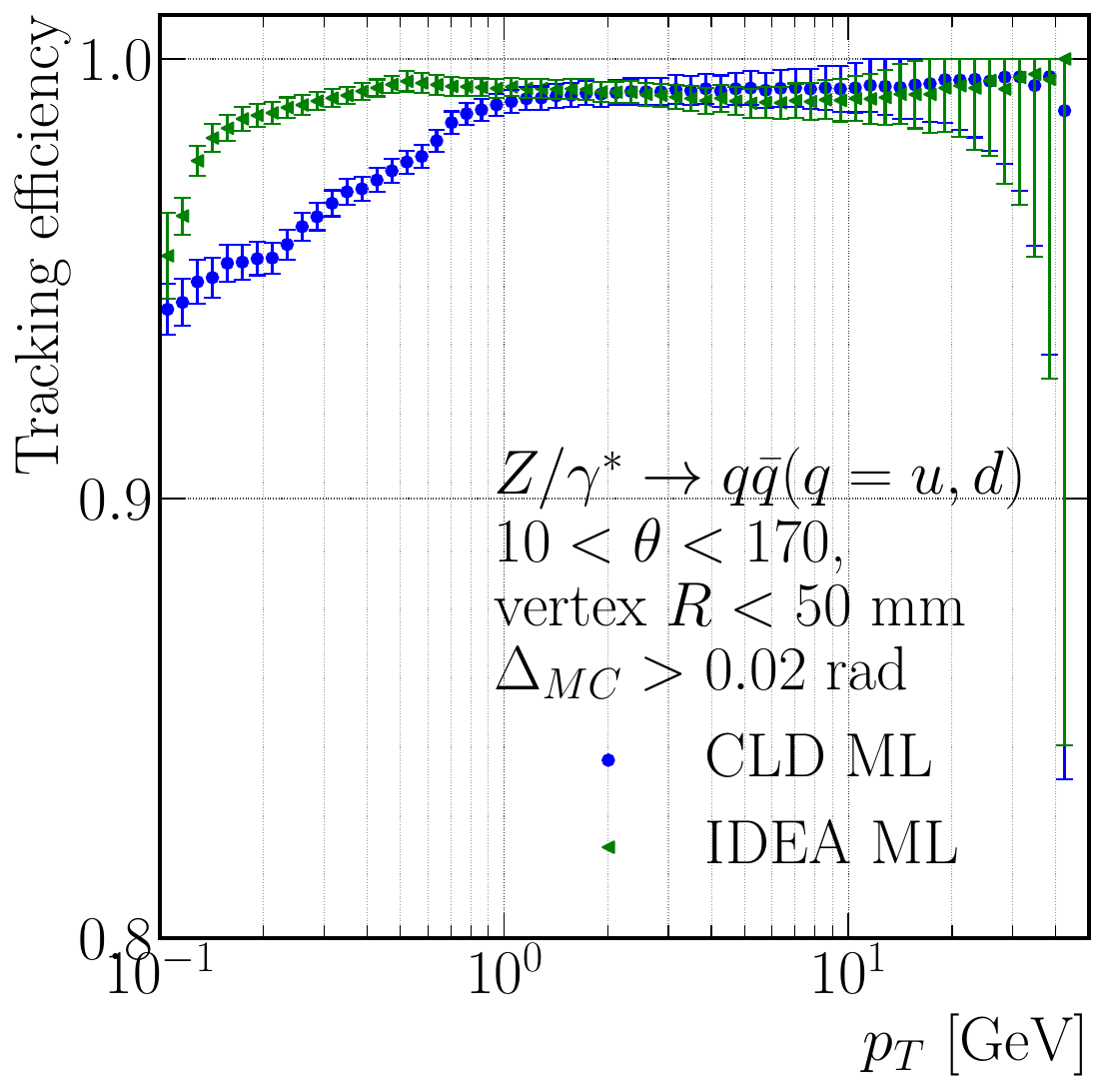}
    \caption{Tracking efficiency in IDEA as a function of p$_T$}
    \label{fig:MLTracking}
\end{figure}

%%%here introduce the FULL SIM 
In order to exploit the characteristics of the specific tracking system for the IDEA detector, a novel approach to tracking and full event reconstruction based on graph neural network (GNN) methods has been implemented~\cite{MLTRACKING_NOTE} in full simulation. 
A preliminary look at the tracking efficiency for the IDEA tracking system is shown in Fig.~\ref{fig:MLTracking} where we can appreciate how IDEA, with its superior number of track measurements, can reconstruct much lower transverse momentum tracks\footnote{The definition of efficiency used here is the percentage of reconstructable charged particles that match with a reconstructed track with at least four hits.}. This is also a crucial component for optimal Particle Flow reconstruction since a significant portion of the visible jet energy is carried by soft charged particles. 

\subsection{IDEA calorimeter combined performance}

%\begin{itemize}
%    \item brief outline of physics impact and gain in event reconstruction capabilities
%    \item energy resolution to electromagnetic particles (leading terms)
%    \item angular and spatial resolution
%   \item energy resolution to neutral hadrons (and applicability of the dual-readout method)
%    \item energy resolution to jets
%\end{itemize}

\begin{figure}[htbp]
    \centering
    \includegraphics[width=0.48\linewidth]{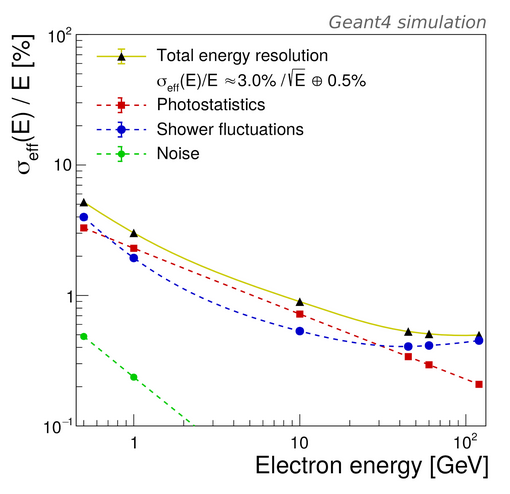}
    \includegraphics[width=0.495\linewidth]{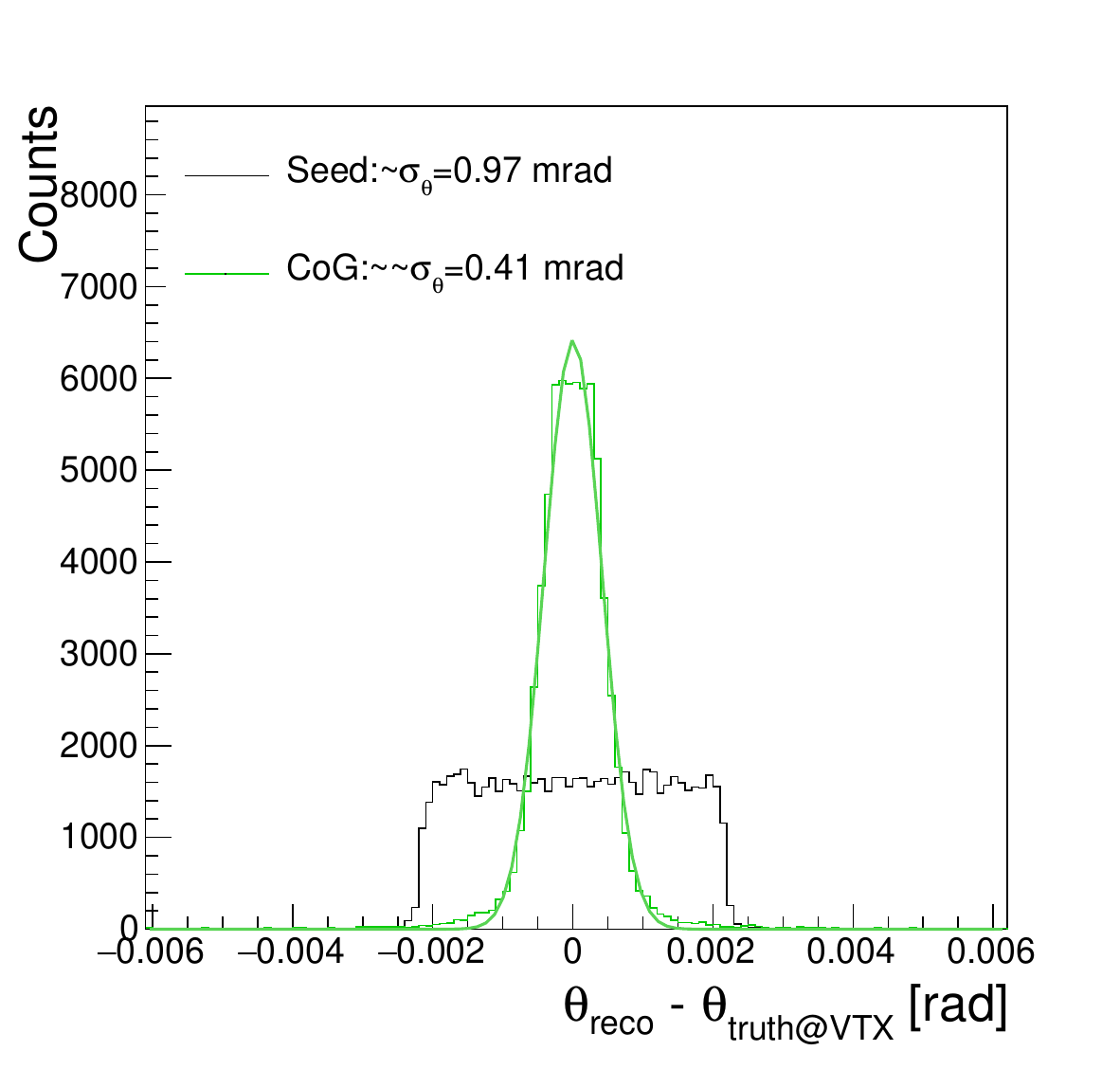}
        \caption{Left: Electromagnetic energy resolution as a function of the electron energy. Right: difference between the $\theta$ angle estimated with crystal barrel PWO segments and the true particle $\theta$ for photons with flat energy distribution in the 5-100 GeV range. The black distribution used the center of the crystal with the highest energy hit as position estimator while the green distribution calculated the center of gravity of the energy hits due to the electromagnetic shower developing in neighboring crystals.}
        \label{fig:performance}
\end{figure}

%physics impact from homogeneous em calorimetry
An electromagnetic energy resolution at the level of $3\%/\sqrt{E}$ is a unique asset for studies of heavy-flavor physics with low-energy photons in their final state \cite{RoyAleksan, Ciuchini_2011} and to improve the resolution of the $Z\rightarrow ee$ recoil mass in Higgsstrahlung events by recovering Bremsstrahlung photons. Furthermore, it enables an efficient clustering of photon pairs from $\pi^0$ decays which can effectively reduce $\pi^0$ photon splitting across jets in multi-jet events \cite{Lucchini_2020}.
%more on ALPS?
% em energy resolution
To meet its performance target on electromagnetic energy resolution a signal of at least 2000~phe/GeV is required to maintain the impact of photostatistics on the stochastic term below $3\%/\sqrt{E}$. Other contributions to the EM energy resolution are shown in Fig.~\ref{fig:performance} and consist of shower fluctuations (e.g. due to front and rear leakage or to dead material upstream of the calorimeter). Noise sources from either the SiPM dark counts or electronic noise are estimated to be subdominant contributions as discussed in \cite{Lucchini_2020}. Overall, the electromagnetic energy resolution can be reasonably parametrized as a function of the electron energy as $\sigma_{E}/E_{EM} = 3.0\%/\sqrt{E} \oplus 0.5\%$.

%The single hadron linearity and resolution of the energy response for the IDEA fiber sampling dual readout calorimeter hadronic section are shown in Figure~\ref{fig:fibre_single_hadron}. While the individual scintillation and Cherenkov readouts display a non-linear response (as expected for a non-compensating calorimeter), the combined readout fully recovers the measurement linearity. The dual-readout also improves the resolution at high hadron energy.  

%angular and spatial resolution
The two electromagnetic segments of PWO crystals provide a good angular resolution for photons as well as for electrons. The center of the gravity of the shower can be calculated to improve the actual resolution. The right panel in Fig.~\ref{fig:performance} show the $\theta$ angular resolution for photons with flat energy distribution in the 5-100 GeV range at the level of about 0.4 mrad (i.e. about 1~mm).

%single hadron and jet resolution with and without PFA
In addition to the state-of-the-art EM energy resolution, the IDEA crystal electromagnetic calorimeter provides an optimal integration with the fiber sampling dual-readout hadronic section of the calorimeter. %The use of the dual-readout method in the two calorimeter sections combined together provides a energy resolution for single hadrons of $\sigma_E/E_{had} \approx 30\%/\sqrt{E}\oplus 3\% $ as discussed in Section~\ref{sec:hybrid_calo_performance}.

%something on time resolution?
The crystal calorimeter can also provide a time resolution better than 30\,ps for EM showers with energy of 20~GeV or higher as demonstrated on beam tests with similar calorimeter prototypes \cite{FERRI2020162159} provided that adequate readout electronics are used. In addition, two thin and highly segmented layers of LYSO:Ce crystals (or a fast plastic scintillator) could be added in front of the calorimeter for time tagging of minimum ionizing particles with a time resolution of 20\,ps, exploiting a technology similar to that used by the CMS MTD \cite{CMS_MTD_TDR}. Such a timing detector could be used for time-of-flight measurements as well as to reconstruct the time of collision vertices with a precision at the level of 5~ps by combining charged tracks from the same vertex and offering a tool to mitigate the dispersion of center of mass energy in particular collision schemes \cite{AzziPerezSlides}. 
%Besides the time measurement such a scintillator based time of flight detector would measure the energy deposited by electrons or early showering photons to be combined with the calorimetric measurement.

%\begin{figure}[htbp]
%    \centering
%    \includegraphics[width=0.48\linewidth]{figs/fibreCalo/linearity_comparison.pdf}
%    \includegraphics[width=0.495\linewidth]{figs/fibreCalo/Simulation_Resolution_for_Energies_linear.pdf}
%        \caption{Linearity (left) and resolution (right) of the energy response to hadrons of the dual-readout fiber sampling calorimeter to single charged pions as obtained using the IDEA full simulation. The blue and red markers show the response of the Cherenkov and scintillating readouts, respectively. The gray markers show the combined response using the dual readout approach. The lines show fit results of parametrization of the results with usual analytic functions.}
%        \label{fig:fibre_single_hadron}
%\end{figure}

A complete simulation of the dual-readout crystal calorimeter integrated with the other IDEA sub-detectors within the Key4hep framework has recently been completed~\cite{chung_fullsim} to enable more detailed physics studies and overall optimization of the IDEA detector.

% \subsection{Combined crystal and fiber calorimeter performance}

\begin{figure}[htbp]
    \centering
    \includegraphics[width=0.495\linewidth]{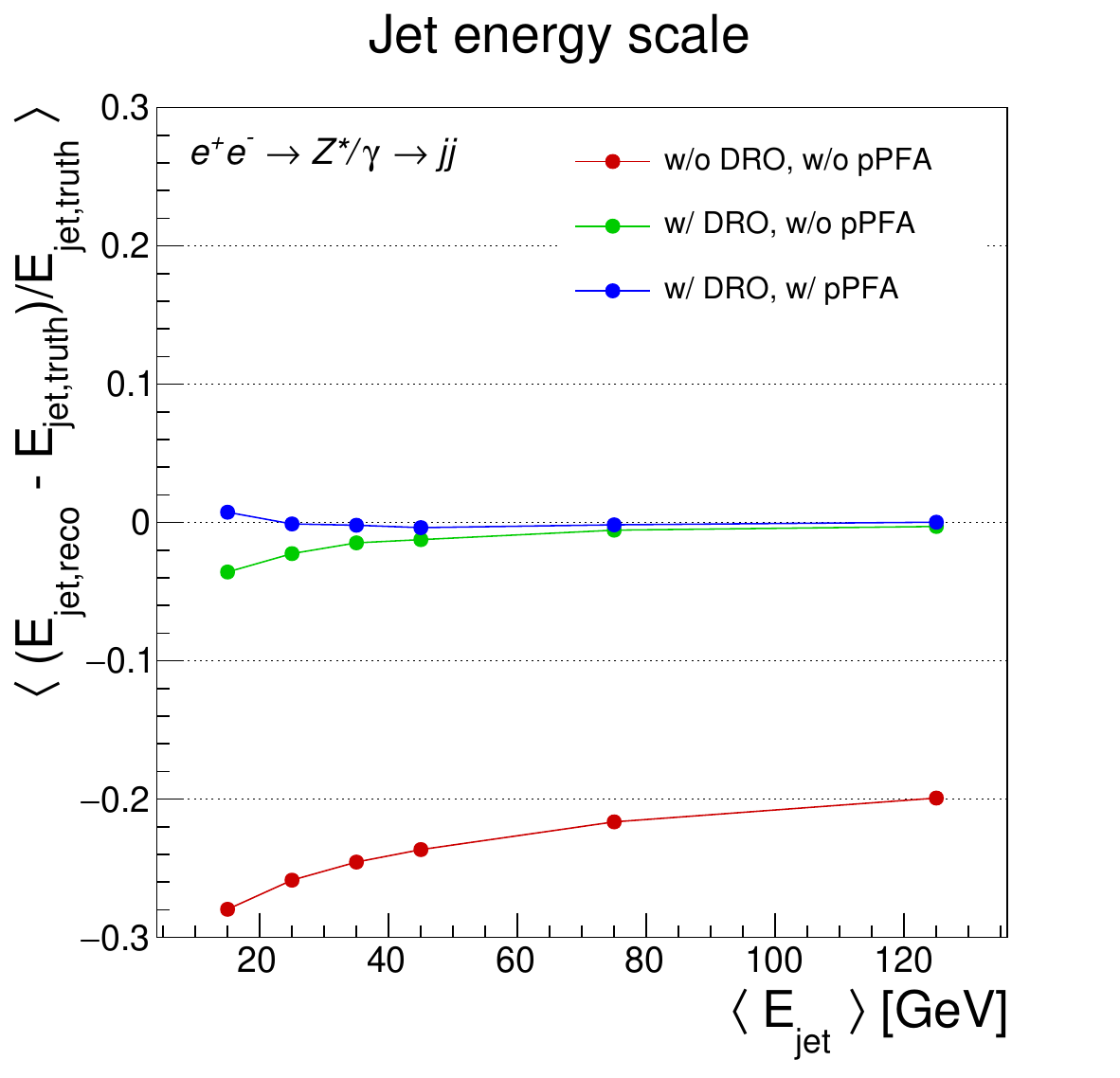}
    \includegraphics[width=0.495\linewidth]{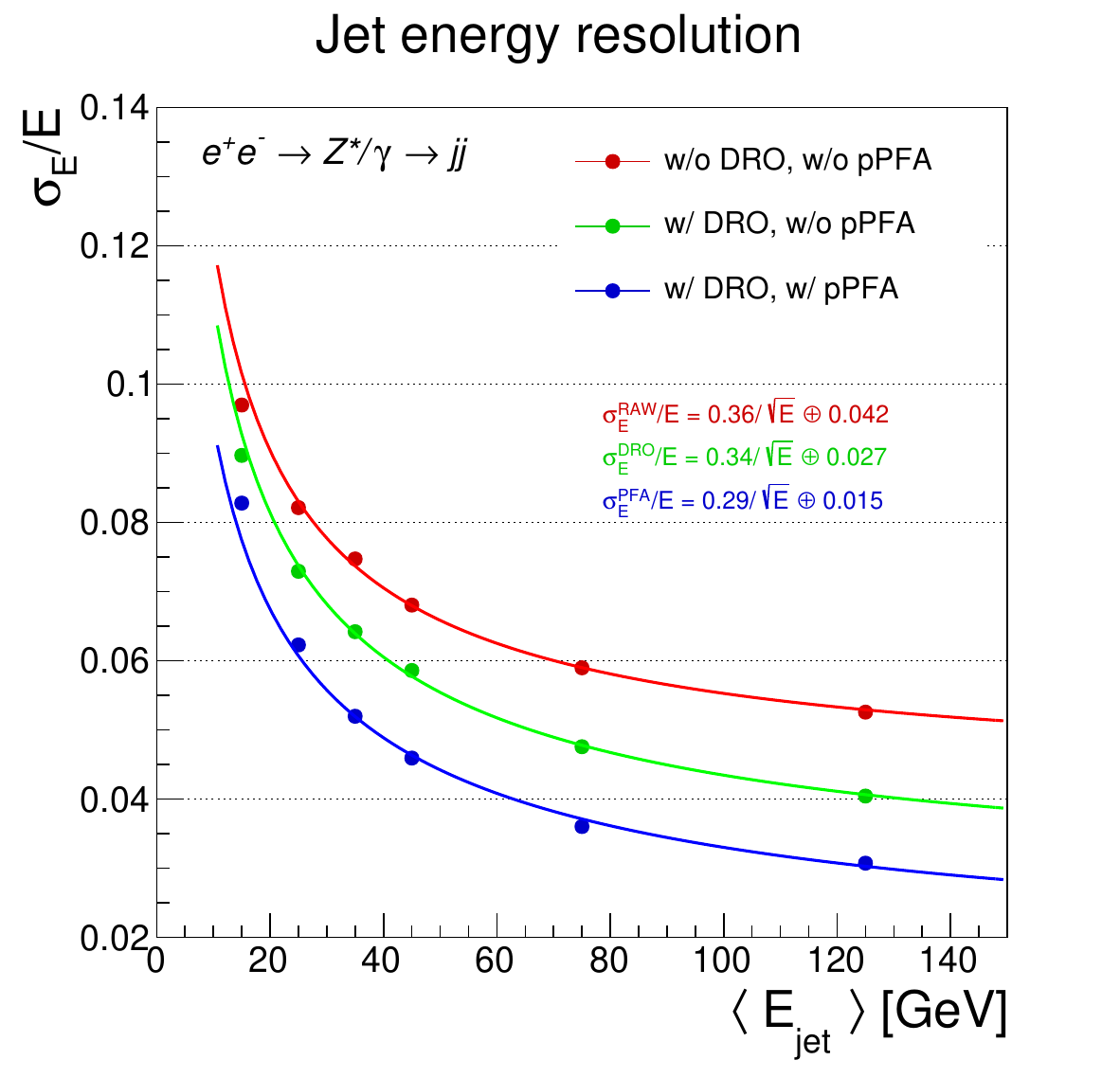}
        \caption{Hadronic jet energy linearity (left) and resolution (right) as measured with the scintillation signal only (red), the dual-readout corrected calorimeter signal (green), and the dual-readout plus particle-flow corrected estimation of the jet energy (blue). Figures from Ref~\cite{Lucchini_2022}.}
        \label{fig:combined_jet_performance}
\end{figure}

To study the calorimeter performance detailed full simulations of the dual-readout crystal section combined with the optical-fiber hadronic section have been performed before in standalone Geant4 simulations. The main objective was to study the possibility of performing particle-flow-like reconstruction of hadronic jets in such an innovative design. The results showed a good improvement in the hadronic jet energy concerning a standard configuration for which only the scintillating channel was used. We found a significant improvement in the energy and angular resolution when the dual-readout technique was used to compensate for the intrinsic shower fluctuations. An additional benefit by combining calo-clusters and tracks in a particle-flow-inspired reconstruction was achieved, as shown in Fig.~\ref{fig:combined_jet_performance}. The reconstruction algorithm was designed to exploit the correct estimation of the true hadron jet, on average, thanks to the dual-readout correction of hadronic signals on an event-by-event basis and is described in~\cite{Lucchini_2022}.

Recently, a new full simulation of the IDEA dual-readout fiber calorimeter exploiting the capillary-tubes technology has been developed within the FCC Key4hep software. This new simulation will allow us to repeat these studies with an updated description of all the IDEA subdetectors.

\subsection{An example of new capabilities: tau Particle Identification} 
\label{par:CP_tauparticleid}

Efficient and accurate reconstruction and categorization of particle decays constitute a pivotal undertaking within the ambit of measurement and searches strategies for the IDEA detector. To gauge dual-readout calorimeter capabilities in this respect we have developed innovative techniques for identifying tau decays using advanced neural network architectures based on geometrical deep learning, which aptly recognize leptonic and hadronic decay modes, whilst effectively discriminating them from QCD jets originating from Z decays. 

\begin{figure}[h]
	\centering
	\includegraphics[width=1\textwidth]{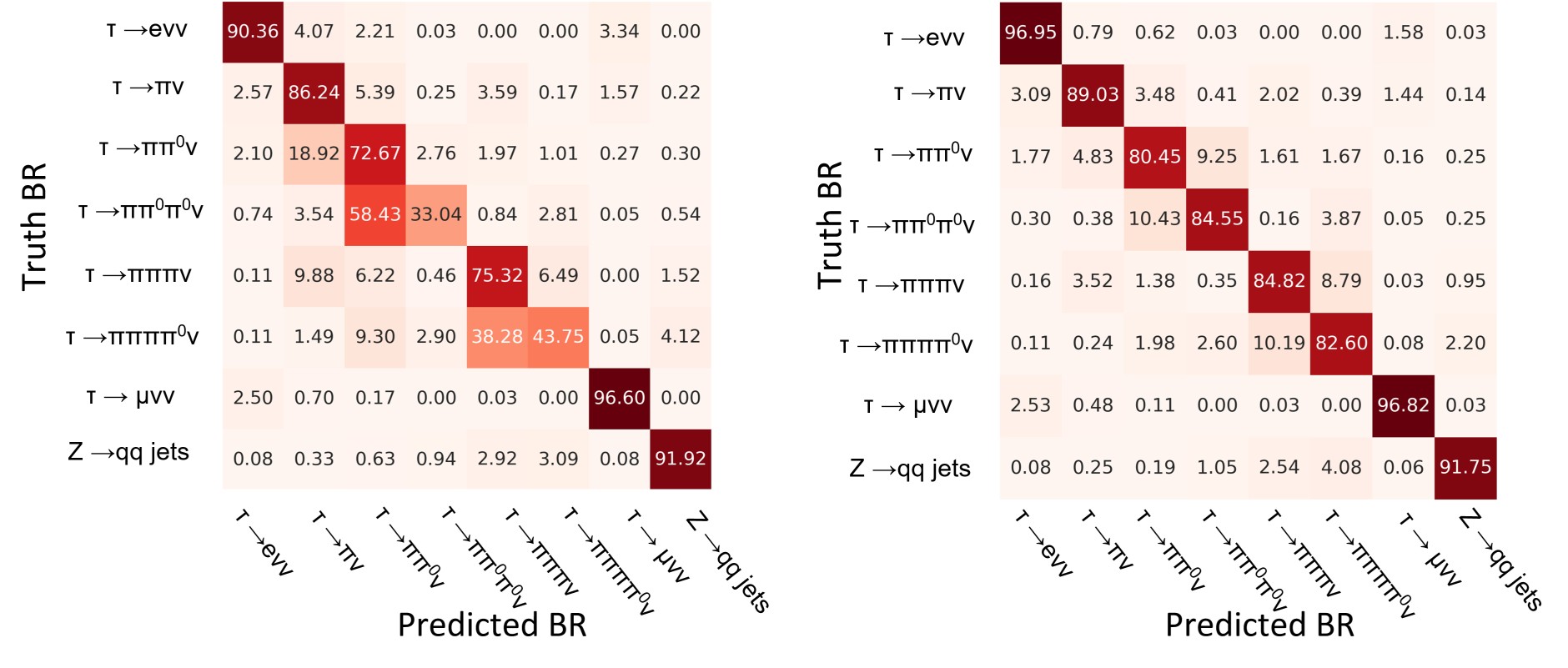}
	\caption{Confusion matrices of DGCNN on test dataset, using \emph{geometric only} (left) and \emph{geometric+fiber type} information (right). Matrices are normalized per row.}
	\label{fig:conf_geo_type}
\end{figure}

Starting from the ParticleNet idea~\cite{qu2020jet} developed for jet tagging, where jets are regarded as unordered sets of particles (Particle Cloud) and neural networks operate on dynamically constructed graphs, we have extended this approach for the task of tau decay identification \cite{Giagu_Tau}. 
In this study, the  deep neural network model, Dynamic Graph Convolutional Neural Network (DGCNN) proposed by Wang et al.~\cite{wang2019dynamic}, 
extended to provide an assessment of the uncertainty on the tau-identification prediction, has been trained on full simulation data of the IDEA detector of leptonic and hadronic tau lepton decays, and QCD jet events.
One assumed advantage of dual-readout technology is the distinctiveness of patterns generated in the calorimeter, owing to the differing properties of the constituent fibers, which is shown by an 88.3\% accuracy when incorporating fiber type information, compared to the 73.7\% of the geometrical information only. 
In Fig.~\ref{fig:conf_geo_type} the normalized confusion matrix obtained with and without the specific fiber type information is reported. 

This preliminary study highlights the enormous potential of dual-readout technology combined with advanced machine learning strategies for precise, event-by-event particle identification, while also contributing to the overall particle flow reconstruction. The synergy between innovative deep learning techniques and the unique information provided by the fiber structure significantly enhances discrimination power. However, even greater performance improvements can be achieved by incorporating information from the crystal electromagnetic calorimeter and the central tracker. This would further strengthen the capabilities of the IDEA detector in tau-identification, underscoring the critical role of tau leptons in the FCCee physics program, from precision tau polarization measurements to searches for new physics.

%\clearpage\newpage
\section{Summary \label{summary}}
Recent progress and design updates for the Innovative Detector for E$^+$e$^-$ Accelerator (IDEA) concept are presented.  Our design efforts aim to provide optimal performance for the majority of the physics goals at the proposed FCC-ee e$^+$e$^-$ collider.  A full complement of sub-detectors is described including a very low mass tracking system comprising a 
powerful vertex detector, a large drift chamber and outer a silicon wrapper, a high resolution dual-readout crystal electromagnetic calorimeter, an HTS based superconducting solenoid, a dual readout fiber calorimeter and three layers of muon chambers embedded in the magnet flux return yoke.  Simulation and R\&D efforts leading to the design choices for meeting the machine requirements and demanding precision of the physics drivers at the FCC-ee are summarized along with plans for future R\&D efforts advancing the technical design of each detector subsystem. Detailed simulations of the physics performance of the full detector concept are supported in the Key4hep framework with the geometry of the sub-detectors modeled using DD4hep.  The IDEA detector concept and associated R\&D programs build upon available technologies to satisfy the demanding performance requirements to exploit the unprecedented level of precision physics measurements made possible by the FCC-ee.

\clearpage\newpage
%\printbibliography %Prints bibliography
%\bibliography{mybib}% common bib file

\clearpage\newpage
\begin{center}
\begin{Large}
The IDEA Study Group
\end{Large}
\end{center}
\vskip 0.7cm
M. Abbrescia$^{1}$, S. Ajmal$^{2}$, N. Akchurin$^{3}$, M. Al-Thakeel$^{4}$, M. Alviggi$^{5}$, G. Ammirabile$^{7}$, A. Andreazza$^{74}$, B. Argiento$^{5}$, E. Auffray$^{36}$, P. Azzi$^{6}$
% ...
\vskip 1.0cm

$^{1}$ INFN Sezione di Bari and Università di Bari, Italy

$^{2}$ INFN Sezione di Perugia, Italy

$^{3}$ Texas Tech University, USA

$^{4}$ INFN Sezione di Bologna and Università di Bologna, Italy

$^{5}$ INFN Sezione di Napoli and Università di Napoli Federico II, Italy

$^{6}$ INFN Sezione di Padova, Italy

$^{7}$ INFN Sezione di Pisa, Italy

% ...
$^{36}$ CERN, Switzerland

% ...
$^{74}$ INFN Sezione di Milano and Università di Milano, Italy

% ...

\end{document}